\documentclass[transmag]{IEEEtran}

\usepackage[T1]{fontenc}
\usepackage{multirow}
\usepackage[table]{xcolor}
\usepackage{cite}
\usepackage{balance}
\usepackage{amsmath,amssymb,amsfonts}
\usepackage{graphicx}
\usepackage{textcomp}
\usepackage{xcolor}
\usepackage{bm}
\usepackage{subfigure}
\usepackage{algorithm}
\usepackage{algpseudocode}	
\usepackage[none]{hyphenat}	
\usepackage{stfloats}
\usepackage{flushend}

\begin{document}
	\bstctlcite{IEEEexample:BSTcontrol}

\title{Full-Duplex mmWave Massive MIMO Systems:\\
	A Joint Hybrid Precoding/Combining and\\
	Self-Interference Cancellation Design}

\author{
	Asil Koc,~\IEEEmembership{Student Member,~IEEE,}
	Tho Le-Ngoc,~\IEEEmembership{Life Fellow,~IEEE}
	\thanks{
		This work was supported in part by Huawei Technologies Canada and in part by the Natural Sciences and Engineering Research Council of Canada.
	}
	\thanks{
		A. Koc and T. Le-Ngoc are  with the Department of Electrical and Computer 	Engineering, McGill University, Montréal, QC H3A 0E9, Canada (e-mail:	asil.koc@mail.mcgill.ca; tho.le-ngoc@mcgill.ca). \textit{(Corresponding author: Asil Koc)}
	}
}

\IEEEtitleabstractindextext{
\begin{abstract}
Millimeter-wave (mmWave) massive multiple-input multiple-output (MIMO) systems have been considered as one of the primary candidates for the fifth generation (5G) and beyond 5G wireless communication networks to satisfy the ever-increasing capacity demands.
Full-duplex technology can further enhance the advantages of mmWave massive MIMO systems.
However, the strong self-interference (SI) is the major limiting factor in the full-duplex technology.
Hence, this paper proposes a novel angular-based joint hybrid precoding/combining (AB-JHPC) technique for the full-duplex mmWave massive-MIMO systems. Our primary goals are listed as: 
(i) 	improving the self-interference cancellation (SIC),
(ii) 	increasing the intended signal power,
(iii) 	decreasing the channel estimation overhead,
(iv) 	designing the massive MIMO systems with a low number of RF chains.
First, the RF-stage of AB-JHPC is developed via slow time-varying angle-of-departure (AoD) and angle-of-arrival (AoA) information. 
A joint transmit/receive RF beamformer design is proposed for covering (excluding) the AoD/AoA support of intended (SI) channel.
Second, the BB-stage of AB-JHPC is constructed via the reduced-size effective intended channel.
After using the well-known singular value decomposition (SVD) approach at the BB-stage, we also propose a new semi-blind minimum mean square error (S-MMSE) technique to further suppress the residual SI power by using AoD/AoA parameters.
Thus, the instantaneous SI channel knowledge is not needed in the proposed AB-JHPC technique.
Finally, we consider a transfer block architecture to minimize the number of RF chains.
The numerical results demonstrate that the SI signal is remarkably canceled via the proposed AB-JHPC technique. 
It is shown that AB-JHPC achieves $85.7$ dB SIC and the total amount of SIC almost linearly increases via antenna isolation techniques.
	We observe that the proposed full-duplex mmWave massive MIMO systems double the achievable rate capacity compared to its half-duplex counterpart as the antenna array size increases and the transmit/receive antenna isolation improves.
Moreover, the proposed S-MMSE algorithm provides considerably high capacity than the conventional SVD approach.
\end{abstract}


}

\maketitle

\section{Introduction}


\IEEEPARstart{M}{assive} multiple-input multiple-output (MIMO) systems operating in millimeter wave (mmWave) frequency bands are the primary candidates for fifth-generation (5G) and beyond 5G (B5G) wireless communication networks \cite{5G_Mas_MIMO_3,Mass_mmWave_REV,6G_prospective_look,Mass_mmWave_Survey}.
Massive MIMO systems with excessively large antenna arrays is a key technology to address the massive machine-type communications (mMTC) and enhanced mobile broadband (eMBB) requirements to support the development of various emerging applications (e.g., virtual reality, augmented reality, autonomous driving, internet of things, smart cities, etc.) \cite{6G_prospective_look}.
Furthermore, the wide range of available bandwidth in the mmWave frequency bands does not only address the bandwidth shortage issue in the current wireless networks but also provides greatly enhanced capacity. 
However, different from to the conventional rich scattering multipath channels, a limited-scattering propagation is experienced in the mmWave frequencies.
For ensuring the sufficient received signal power and compensating the severe path loss,
the massive MIMO technology is widely considered as a useful application in the mmWave communications \cite{Mass_mmWave_Survey,mmWave_Survey_2,mmWave_Antenna_Size,mmWave_Survey}. 
Because the high beamforming gain in the massive MIMO systems can focus the signal energy through the desired limited-scattering regions \cite{mmWave_Survey_2}.
Additionally, the shorter wavelengths in the mmWave frequency enable the utilization large antenna arrays under the area requirements in practical applications \cite{mmWave_Antenna_Size}.

Massive MIMO systems have been originally considered for the half-duplex communications (i.e., time-division duplex (TDD) \cite{MassiveMIMO_TDD,MassiveMIMO_TDD_2,MassiveMIMO_FDD} and frequency-division duplex (FDD) \cite{MassiveMIMO_FDD,MassiveMIMO_FDD_2,MassiveMIMO_FDD_3}).
The expected impacts of mmWave massive MIMO systems can be further extended via full-duplex communications, which theoretically doubles the capacity via more efficient utilization of the limited resources (i.e., time and frequency).
However, the performance of full-duplex communications is severely affected by the strong self-interference (SI) occurred due to the simultaneous transmission and reception over the same frequency band, which is not present in the half-duplex communications.
On the other hand, the recent developments in antenna technology and signal processing techniques make full-duplex communications more applicable via advanced SI cancellation (SIC) techniques, e.g., passive (propagation domain) antenna isolation, analog cancellation and digital cancellation 
\cite{FD_Survey, FD_Survey_Secrecy, FD_Survey_2019, FD_In_Band_Survey, FD_AntennaSep_60dB, FD_AntennaSep_74dB}.
The recent practical measurements in \cite{FD_AntennaSep_60dB} show that the passive antenna isolation achieves $60-65$ dB SIC.
Furthermore, in \cite{FD_AntennaSep_74dB}, the antenna isolation based SIC is enhanced up to $74$ dB via absorptive shielding, cross polarization and directional isolation.
The quality of SIC can be further improved via analog and digital cancellation techniques \cite{FD_Survey,FD_Survey_Secrecy, FD_Survey_2019, FD_In_Band_Survey}, however, the SI component cannot be completely removed.
Even though the residual SI limits the performance of full-duplex communications, the experimental and theoretical studies in the literature demonstrate that full-duplex transmission can outperform the conventional half-duplex as the quality of the SIC increases \cite{FD_Survey}. 
Additionally, the full-duplex massive MIMO systems enable enhanced estimation and cancellation of the strong SI channel by means of high degrees of freedom provided by the large antenna arrays \cite{FD_MIMO_SI_Canc_BF,FD_MIMO_mmWave,FD_MIMO_Beam_Domain,FD_MIMO_ANALOG,Satyanarayana2019,FD_MIMO_MU_Learning,FD_MIMO_Relay_mmWave_2,FD_MIMO_Relay_mmWave,FD_MIMO_1_bit_HPC}.

In order to ensure reliable transmission quality for the massive MIMO systems,
the precoding (combining) at the transmitter (receiver) is a crucial signal processing procedure for the downlink (uplink) transmission.
The conventional MIMO systems widely consider the single-stage fully-digital precoding/combining (FDPC) architecture \cite{Mass_MIMO_Precoding_Survey}.
However, the large antenna arrays in the massive MIMO systems induce two interesting challenges: 
(i) higher hardware cost/complexity as well as the power consumption (i.e., one dedicated power-hungry radio frequency (RF) chain per each antenna),
(ii) longer channel estimation overhead (i.e., the requirement of full instantaneous channel state information (CSI) in FDPC).
Hence, even though the large antenna arrays can improve the capacity, the energy-efficiency sharply decreases with the utilization of more power-hungry RF chains.
The single-stage fully analog precoding/combining (FAPC) technique significantly reduces the hardware cost/complexity by minimizing the RF chain utilization \cite{ANALOG_BF,ANALOG_BF_Heath,Mass_MIMO_Hyb_Survey}, however, FAPC achieves poor capacity compared to FDPC.
As a promising solution, the two-stage hybrid precoding/combining (HPC) technique has been proposed to address the aforementioned trade-off between FDPC and FAPC, while providing the comparable capacity as in FDPC \cite{Mass_MIMO_Hybrid_Survey,Mass_MIMO_Hyb_Survey,Mass_MIMO_Hybrid_Survey_2}.
In the HPC architecture, the analog RF-stage and digital baseband(BB)-stage are connected to each other via RF chains, where the number of RF chains is chosen between the number of antennas and the independent data streams.
Both half-duplex and full-duplex transmission schemes have been investigated for the HPC design in the massive MIMO systems in 
\cite{Mass_MIMO_Hyb_Low_Comp,MassMIMO_hybrid_NO_ADMA,SU_mMIMO_OFDM,OMP_Full_CSI,SU_mMIMO_Dynamic_SubArray,JSDM_LargeArray,MassMIMO_hybrid_JSDM_FA,ASIL_ABHP_Access,ASIL_Subconnected_GC,ASIL_PSO_PA_WCNC,ASIL_MC_PIMRC,ASIL_VTC_DAC_ADC}
and in
\cite{FD_MIMO_SI_Canc_BF,FD_MIMO_mmWave,FD_MIMO_Beam_Domain,FD_MIMO_ANALOG,Satyanarayana2019,FD_MIMO_MU_Learning,FD_MIMO_Relay_mmWave_2,FD_MIMO_Relay_mmWave,FD_MIMO_1_bit_HPC}, 
respectively.
\vspace{-2ex}

\subsection{Related Works}

Preliminary studies in the HPC technique employ the half-duplex communications.
There are two major approaches regarding the RF-stage design in the hybrid architectures: 
(i) utilizing full instantaneous CSI bringing in large channel estimation overhead \cite{Mass_MIMO_Hyb_Low_Comp,MassMIMO_hybrid_NO_ADMA,OMP_Full_CSI,SU_mMIMO_Dynamic_SubArray,SU_mMIMO_OFDM}, 
(ii) using the slow time-varying channel characteristics (e.g.,
channel covariance matrix, angle-of-departure (AoD), angle-of-arrival (AoA)) \cite{
	JSDM_LargeArray,
	MassMIMO_hybrid_JSDM_FA,
	ASIL_ABHP_Access,
	ASIL_MC_PIMRC,
	ASIL_Subconnected_GC,
	ASIL_PSO_PA_WCNC,
	ASIL_VTC_DAC_ADC}.
Thus, the latter approach only uses the reduced-size fast time-varying effective CSI at the BB-stage design.
Therefore, the hybrid architectures in 
	\cite{
		JSDM_LargeArray,
		MassMIMO_hybrid_JSDM_FA,
		ASIL_ABHP_Access,
		ASIL_Subconnected_GC,
		ASIL_PSO_PA_WCNC,
		ASIL_MC_PIMRC,
		ASIL_VTC_DAC_ADC}
can address two interesting challenges in the massive MIMO systems: (i) the reduced number of RF chains for low power consumption and the hardware cost/complexity, (ii) the low CSI overhead for shorter training sequence.
In \cite{ASIL_ABHP_Access}, a two-stage angular-based hybrid-precoding (HP) technique is proposed for the downlink massive MIMO systems, where the CSI overhead is greatly reduced by designing the RF-stage via AoD parameters.
The numerical results in \cite{ASIL_ABHP_Access} show that the angular-based HP significantly enhances the energy efficiency compared to the single-stage fully-digital precoding (FDP) by means of the reduced number of RF chains.
Moreover, the angular-based HP improves the sum-rate capacity with respect to its two-stage HP counterparts as eigen-beamforming based HP \cite{MassMIMO_hybrid_JSDM_FA}, block-diagonalization based HP \cite{JSDM_LargeArray}, 
non-orthogonal angle space based HP \cite{MassMIMO_hybrid_NO_ADMA}.
		In \cite{ASIL_Subconnected_GC}, the fully-connected and sub-connected hybrid architectures are analyzed for the angular-based HP technique.
	Then, a swarm intelligence based power allocation algorithm is proposed to further enhance the sum-rate capacity in \cite{ASIL_PSO_PA_WCNC}.
Afterwards, the angular-based HP technique is extended for the multi-cell systems in \cite{ASIL_MC_PIMRC}, where the downlink cooperation strategies are proposed to mitigate the inter-cell interference.
Considering both downlink and uplink transmission in the massive MIMO systems, an angular-based HPC technique is also proposed in \cite{ASIL_VTC_DAC_ADC}, where the effect of low-resolution digital-to-analog and analog-to-digital converters is also investigated.

Recently, the precoding/combining techniques are also studied for the full-duplex communication systems in \cite{FD_MIMO_SI_Canc_BF,FD_MIMO_mmWave,FD_MIMO_Beam_Domain,FD_MIMO_ANALOG,Satyanarayana2019,FD_MIMO_MU_Learning,FD_MIMO_Relay_mmWave_2,FD_MIMO_Relay_mmWave,FD_MIMO_1_bit_HPC}. 
First, the performance of full-duplex MIMO and massive MIMO systems is investigated for the FDPC architecture in \cite{FD_MIMO_SI_Canc_BF,FD_MIMO_mmWave,FD_MIMO_Beam_Domain}, where the promising results imply that full-duplex systems can outperform its half-duplex counterpart. 
Then, the FAPC architecture is designed for the full-duplex massive MIMO systems in \cite{FD_MIMO_ANALOG}, where the performance gap between the fully-digital and fully-analog structure is represented.
Afterwards, the HPC structure is built for the full-duplex massive MIMO systems in \cite{Satyanarayana2019,FD_MIMO_MU_Learning}, where the authors assume the availability of perfect SI channel knowledge for the SIC algorithm.
The full-duplex relay backhaul system is considered in \cite{FD_MIMO_Relay_mmWave_2}, where the SIC algorithm aims to mitigate the SI component via the HPC design based on the perfect SI channel knowledge.
Similarly, \cite{FD_MIMO_Relay_mmWave} also investigates the full-duplex relay system, whereas, the residual SI channel is modeled as Gaussian noise.
The effect of low-resolution phase-shifters at the RF-stage of HPC is analyzed in \cite{FD_MIMO_1_bit_HPC}, where the numerical results show that when 25 dB SIC is provided by antenna isolation, the HPC with 1-bit phase-shifters can improve the total SIC up to 44 dB.
It is worthwhile to note that the availability of full instantaneous CSI is needed for all above-mentioned FDPC, FAPC and HPC architectures developed for the full-duplex massive MIMO systems.
\vspace{-2ex}

\subsection{Contribution and Organization}

This paper proposes a novel angular-based joint HPC (AB-JHPC) technique for the full-duplex mmWave massive MIMO systems. The proposed AB-JHPC mainly targets to 
(i) 	improve the quality of SIC,
(ii) 	enhance the power of desired/intended signal,
(iii)	decrease the CSI overhead size,	
(iv) 	reduce the hardware cost/complexity and power consumption by employing less number of RF chains.
Table \ref{table_LiteratureReview} provides a detailed comparison of this work with respect to the state-of-the-art.
The main contributions of this paper are summarized as follows:

\begin{table*}
	\caption{State-of-the-art for the precoding/combining techniques in full-duplex massive MIMO systems.}
	\vspace{-1ex}
	\label{table_LiteratureReview}
	\footnotesize 
	\centering
	\renewcommand{\arraystretch}{1.7}
	\begin{tabular}{|c||c|c||c|c|c||c|c||c|}
		\hline
		\multirow{2}{*}{{\hspace{-1ex}Reference\hspace{-1ex}}} & 
		\multicolumn{2}{c||}{Transmission Mode}& 
		\multicolumn{3}{c||}{Precoder/Combiner Design}&
		\multicolumn{2}{c||}{CSI Overhead}&
		SIC without Instantaneous
		\\ \cline{2-8} 
		&Full-Duplex&Half-Duplex
		&Hybrid&Digital&Analog
		&Full&Low
		&SI Channel Knowledge
		\\ \hline\hline
		\cite{Mass_MIMO_Hyb_Low_Comp,MassMIMO_hybrid_NO_ADMA,OMP_Full_CSI,SU_mMIMO_Dynamic_SubArray,SU_mMIMO_OFDM}
		&&\checkmark&\checkmark&&&\checkmark&&\\ \hline
		\cite{
			JSDM_LargeArray,
			MassMIMO_hybrid_JSDM_FA,
			ASIL_ABHP_Access,
			ASIL_MC_PIMRC,
			ASIL_Subconnected_GC,
			ASIL_PSO_PA_WCNC,
			ASIL_VTC_DAC_ADC}
		&&\checkmark&\checkmark&&&&\checkmark&\\ \hline
		\cite{FD_MIMO_SI_Canc_BF,FD_MIMO_mmWave,FD_MIMO_Beam_Domain}
		&\checkmark&&&\checkmark&&\checkmark&&\\ \hline
		\cite{FD_MIMO_ANALOG}
		&\checkmark&&&&\checkmark&\checkmark&&\\ \hline
		\cite{Satyanarayana2019,FD_MIMO_MU_Learning,FD_MIMO_Relay_mmWave_2,FD_MIMO_Relay_mmWave,FD_MIMO_1_bit_HPC}
		&\checkmark&&\checkmark&&&\checkmark&&\\ \hline
		\cellcolor{yellow} This Paper
		&\cellcolor{yellow}\checkmark&&\cellcolor{yellow}\checkmark&&&&\cellcolor{yellow}\checkmark&\cellcolor{yellow}\checkmark
		\\ \hline
	\end{tabular}
	\vspace{-3ex}
\end{table*}

\begin{itemize}
	\item \textbf{Proposed Hybrid Precoding/Combining Technique:} Considering both transmission/reception and hybrid architecture, we aim to design four sub-blocks in the AB-JHPC technique: 
	(i)		BB precoder,
	(ii) 	BB combiner,
	(iii)	transmit RF beamformer,
	(iv) 	receive RF beamformer.
	The transmit/receive RF beamformer are built by the low-cost phase-shifters, which induces the constant modulus (CM) constraint at the RF-stage design.
	Therefore, the optimization problem for the total achievable-rate becomes non-convex due to the CM constraint. 
	Thus, the optimization problem in the proposed AB-JHPC is divided into two steps: (i) joint RF beamformer, (ii) BB precoder/combiner.

	\item \textbf{RF Beamformer:} In order to maximize the desired/intended signal power while suppressing the SI power, the transmit/receive RF beamformers are constructed via the slow time-varying AoD/AoA information.
	Hence, the RF-stage design does not require any instantaneous channel knowledge.
	For improving the quality of SIC, the joint RF beamformer generates the orthogonal transmit/receive beams spanning (excluding) the AoD/AoA support of intended (SI) channel.
	As presented in Section \ref{sec_Illustrative} and in Table \ref{table_RF_Chain_CSI}, the proposed RF-stage design is capable of decreasing the CSI overhead size by $99.8\%$.
	
	\item \textbf{BB Precoder \& BB Combiner:} After developing the transmit/receive RF beamformers, the BB-stage is designed via the reduced-size effective intended channel seen from the BB-stage. 
	First, the BB precoder and BB combiner are constructed via the well-known singular value decomposition (SVD).
	Also, a new semi-blind minimum mean square error (S-MMSE) algorithm is proposed for the BB combiner. It is classified as semi-blind because the S-MMSE algorithm uses only the fast time-varying effective intended channel coefficients.
	In other words, the S-MMSE algorithm does not depend on the fast time-varying SI channel coefficients, which might be impractical to be perfectly estimated. 
	Instead, similar to the RF-stage design, the S-MMSE-based BB combiner employs the AoA support of the SI channel as the slow time-varying channel parameters.
	Numerical results demonstrate that the proposed S-MMSE-based BB combiner enhances the quality of SIC and provides superior achievable-rate performance compared to the SVD-based BB combiner.

	\item \textbf{Self-Interference Cancellation:}  The illustrative results show that the SI channel is greatly suppressed via the proposed joint transmit/receive RF beamformer design. 
	It is seen that the power of far-field (near-field) SI channel is canceled by $81.5$ dB ($44.5$ dB) after the joint RF beamformer design, while it preserves most of the intended channel power with only $2$ dB degradation.
	Considering the SI occurred at a given data stream, we observe $85.7$ dB SIC is achieved via only the AB-JHPC technique without applying any antenna isolation. 
	Moreover, the total amount of SIC can be enhanced further enhanced via the advanced antenna isolation technique. After applying the proposed AB-JHPC technique, we monitor a near-linear relationship between the antenna isolation and the total amount of SIC.

	\item \textbf{Full-Duplex vs. Half-Duplex:} 
	The proposed full-duplex mmWave massive MIMO systems remarkably outperform its half-duplex counterpart in terms of total achievable rate capacity as the number of antennas and the antenna isolation based SIC increase.
	However, the full-duplex to half-duplex ratio gain slightly decreases as the number of data streams increases due to the enhanced SI power.
	
	
\end{itemize}

The rest of this paper is organized as follows. The system and channel models are introduced in Section \ref{sec_System} and Section \ref{sec_Channel}, respectively
In Section \ref{sec_AB_JHPC}, we develop the transmit/receive RF beamformer and BB precoder/combiner in the proposed AB-JHPC technique for the full-duplex mmWave massive MIMO systems. 
Section \ref{sec_RF_Reduction} expresses the transfer block architecture to further reduce the number of RF chains.
The illustrative results are provided in Section \ref{sec_Illustrative}.
Finally, the paper is concluded in Section \ref{sec_Conc}.
Table \ref{table_abbreviations} summarizes the various abbreviations used in this paper. 
For improving the clarity of mathematical exposition, the frequently-used symbols are listed in Table \ref{table_symbols}.


\begin{table*}
	\centering
	\caption{List of abbreviations.}
	\label{table_abbreviations}
	\setlength{\tabcolsep}{3pt}
	\renewcommand{\arraystretch}{1.3}
	\vspace{-0.5ex}
	\begin{tabular}{|l|l|}
		\hline
		2D    	& Two-dimensional \\ \hline
		3D    	& Three-dimensional                  \\ \hline
		3GPP 	& Third generation partnership project \\ \hline
		5G    	& Fifth generation                  \\ \hline
		AB-HPC	& Angular-based HPC \\ \hline
		AB-JHPC	& Angular-based joint HPC \\ \hline
		AoA 	& Angle-of-arrival \\ \hline
		AoD 	& Angle-of-departure \\ \hline
		B5G    	& Beyond 5G                      \\ \hline		
		BB    	& Baseband                      \\ \hline		
		CM     	& Constant modulus               \\ \hline
		CSI 	& Channel state information 	\\ \hline
		eMBB   	& Enhanced mobile broadband             \\ \hline
		FAPC    & Fully analog precoding/combining \\ \hline
		FDP    	& Fully digital precoding \\ \hline
		FDPC    & Fully digital precoding/combining \\ \hline
		HP    	& Hybrid precoding \\ \hline
	\end{tabular}
	\hspace{1ex}
	\begin{tabular}{|l|l|}
		\hline
		HPC    	& Hybrid precoding/combining \\ \hline
		ISI 	& Inter-symbol interference\\ \hline
		LoS 	& Line-of-sight \\ \hline
		MIMO   	& Multiple-input multiple-output \\ \hline
		MMSE 	& Minimum mean square error \\ \hline
		mMTC   	& Massive machine-type communications \\ \hline
		mmWave 	& Millimeter wave                \\ \hline
		MSE 	& Mean square error \\ \hline
		NLoS 	& Non-line-of-sight \\ \hline
		PSD		& Power spectral density		\\ \hline
		RF		& Radio frequency		\\ \hline
		S-MMSE 	& Semi-blind minimum mean square error \\ \hline
		SI 	  	& Self-interference					\\ \hline
		SIC 	& Self-interference cancellation \\ \hline
		SVD		& Singular value decomposition \\ \hline
		ULA		& Uniform linear array  \\ \hline
		URA		& Uniform rectangular array  \\ \hline
	\end{tabular}
\end{table*}

\begin{table*}
	\centering
	\caption{List of frequently-used symbols.}
	\label{table_symbols}
	\setlength{\tabcolsep}{3pt}
	\renewcommand{\arraystretch}{1.5}
	\vspace{-0.5ex}
	\begin{tabular}{|l|l|}
		\hline
		$i,j$ & Node index ($i,j \in \left\{ {1,2} \right\}$, $i \ne j$) \\ \hline
		${M_{t,i}}$& 
		\# of transmit antenna\\ \hline
		${M_{r,i}}$&
		\# of receive antenna\\ \hline
		$N_{t,i}$& \# of transmit RF chains \\ \hline
		$N_{r,i}$& \# of receive RF chains \\ \hline
		${S_i}$ & \# of data streams \\ \hline
		$P_T$	& Transmission power\\ \hline
		${{\bf{H}}_i} \in \mathbb{C}^{{M_{r,j}} \times M_{t,i}}$ & 
		Intended channel\\ \hline
		$\bm{\mathcal{H}}_i\in\mathbb{C}^{N_{r,j}\times N_{t,i}}$ & Effective intended channel\\ \hline
		${{\bf{H}}_{\textrm{SI},i}\in{\mathbb{C}^{M_{r,j} \times M_{t,j}}}}$&
		SI channel\\ \hline
		$\bm{\mathcal{H}}_{\textrm{SI},i}\in\mathbb{C}^{N_{r,i}\times N_{t,i}}$	& Effective SI channel\\ \hline
		${\bf{H}}_{\textrm{LoS},i}\in\mathbb{C}^{N_{r,i}\times N_{t,i}}$& Near-field SI channel\\ \hline
		${\bf{H}}_{\textrm{NLoS},i}\in\mathbb{C}^{N_{r,i}\times N_{t,i}}$ & Far-field SI channel\\ \hline
	\end{tabular}
	\hspace{1ex}
	\begin{tabular}{|l|l|}
		\hline
		${\bf{B}}_{t,i} \in {\mathbb{C}^{N_{t,i} \times {S_i}}}$&
		BB precoder\\ \hline
		${\bf{B}}_{r,i} \in {\mathbb{C}^{{S_j} \times N_{r,i}}}$&
		BB combiner\\ \hline
		${\bf{F}}_{t,i} \in {\mathbb{C}^{M_{t,i} \times N_{t,i}}}$&
		Transmit RF beamformer\\ \hline
		${\bf{F}}_{r,i} \in {\mathbb{C}^{N_{r,i} \times M_{r,i}}}$&
		Receive RF beamformer\\ \hline
		${\bf G}_i\in \mathbb{C}^{L_i\times L_i }$ & Diagonal path gain matrix \\ \hline
		${{\bf{\Phi }}_{t,i}}\in\mathbb{C}^{L_i \times M_{t,i}}$ & Transmit phase response matrix \\ \hline
		${{\bf{\Phi }}_{r,j}}\in\mathbb{C}^{M_{r,j}\times L_i}$ & Receive phase response matrix \\ \hline
		$L_i=\sum\nolimits_{c=1}^{C_i}L_{c,i}$ & \# of paths \\ \hline
		$C_i$ & \# of scattering-clusters\\ \hline
		${\theta _{r,c}^{\left(j\right)}}$, ${\psi _{r,c}^{\left(j\right)}}$ & Mean elevation and azimuth AoA\\ \hline
		${\theta _{t,c}^{\left(i\right)}}$, ${\psi _{t,c}^{\left(i\right)}}$ & Mean elevation and azimuth AoD\\ \hline
		${\delta _{r,c}^{\theta,\left(j\right)}}$, ${\delta _{r,c}^{\psi,\left(j\right)}}$ & Elevation and azimuth AoA spread\\ \hline
		${\delta _{t,c}^{\theta,\left(i\right)}}$, ${\delta _{t,c}^{\psi,\left(i\right)}}$ & Elevation and azimuth AoD spread\\ \hline
	\end{tabular}
\end{table*}

\textit{Notation:} 
Bold upper/lower case letters denote matrices/vectors. 
$\left( \cdot \right)^*$, $\left( \cdot \right)^T$, $\left( \cdot \right)^H$ $\left\| \cdot \right\|_2$ and $\left\| \cdot \right\|_F$ represent the complex conjugate, the transpose, the conjugate transpose, the $2$-norm and the Frobenius norm of a vector or matrix, respectively. 
$\mathbf{I}_K$, ${\mathbb{E}}\left\lbrace \cdot\right\rbrace $, $\rm{tr}\left(\cdot\right)$ and $\angle \left(\cdot\right)$ stand for $K\times K$ identity matrix, the expectation operator, the trace operator and the argument of a complex number, respectively. 
${\bf X}\left(m,n\right)$	denotes the element at the intersection of $m^{th}$ row  and $n^{th}$ column.
$\mathbf{X}\otimes\mathbf{Y}$ denotes the Kronecker product of two matrices $\mathbf{X}$ and $\mathbf{Y}$.
We use  ${{x}}\sim{\cal C}{\cal N}\left( {0 ,{\sigma}} \right)$ when ${x}$ is a complex Gaussian random variable with zero-mean and variance ${\sigma}$.
We use the ramp function as $\left(x\right)^+ = \max\left(0,x\right)$.


\section{System Model}\label{sec_System}
\begin{figure*}[!t]
	\centering
	\includegraphics[width=0.85\textwidth]
	{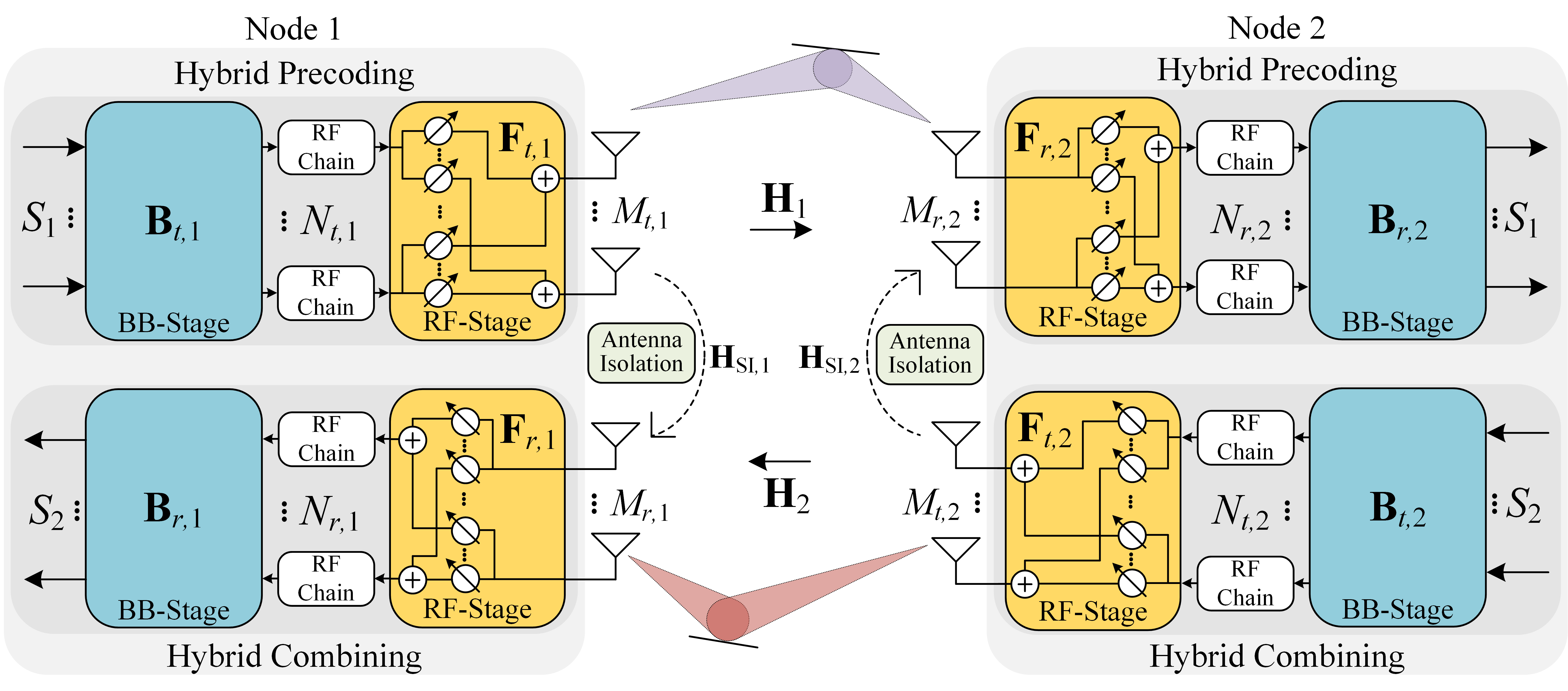}
	\vspace{-2ex}
	\caption{Full-duplex mmWave massive MIMO systems with HPC architecture.}
	\vspace{-0.5ex}
	\label{fig_1_SystemModel}
\end{figure*}

	Fig. \ref{fig_1_SystemModel} illustrates the system model for the full-duplex mmWave massive MIMO systems, where both nodes employ the HPC architecture and operate in full-duplex transmission mode. 
Hence, two-way data transmission and reception operations are simultaneously performed over the same frequency band.
As one of the practical applications of the proposed system model, the wireless backhaul links can be considered for connecting the multiple macro-cell base stations with large antenna arrays \cite{HPC_Wireless_Backhaul,FD_MIMO_Relay_mmWave_2}.
	The wireless backhaul links can also connect the macro-cell and small-cell base stations to enhance the cell-edge performance \cite{MassiveMIMO_Backhaul}.
Moreover, as specified in 3GPP standards, the massive MIMO cellular systems have the ability to switch between point-to-point (single-user) and multi-user transmission \cite{dahlman20164g}.
Therefore, the proposed system model has also practical applicability considering the point-to-point massive MIMO cellular systems.

As seen from Fig. \ref{fig_1_SystemModel}, when the node $i$ transmits data to the node $j$ ($i,j \in \left\{ {1,2} \right\}$ and $i \ne j$), the corresponding intended channel matrix between them is denoted as 
${{\bf{H}}_i} \in \mathbb{C}^{{M_{r,j}} \times M_{t,i}}$.
Moreover, ${{\bf{H}}_{\textrm{SI},i}\in{\mathbb{C}^{M_{r,j} \times M_{t,j}}}}$ represents the SI channel at the node $i$ due to the full-duplex transmission.
According to the HPC architecture, four sub-blocks are placed at each node: 
(i) the BB precoder 
${\bf{B}}_{t,i} \in {\mathbb{C}^{N_{t,i} \times {S_i}}}$,
(ii) the BB combiner   
${\bf{B}}_{r,i} \in {\mathbb{C}^{{S_j} \times N_{r,i}}}$,
(iii) the transmit RF beamformer 
${\bf{F}}_{t,i} \in {\mathbb{C}^{M_{t,i} \times N_{t,i}}}$
and 
(iv) the receive RF beamformer  
${\bf{F}}_{r,i} \in {\mathbb{C}^{N_{r,i} \times M_{r,i}}}$.
At the node $i$,  the transmitter and receiver are equipped with a uniform rectangular array (URA)\footnote{
	It is worthwhile to note that although uniform linear arrays (ULA) is widely investigated in the literature due to its simple structure (e.g., in 
	\cite{FD_MIMO_mmWave,FD_MIMO_Beam_Domain,FD_MIMO_ANALOG,FD_MIMO_MU_Learning,FD_MIMO_Relay_mmWave_2,FD_MIMO_Relay_mmWave,FD_MIMO_1_bit_HPC,Mass_MIMO_Hyb_Low_Comp,MassMIMO_hybrid_NO_ADMA,SU_mMIMO_OFDM,Satyanarayana2019}),
	URA is more advantageous in practical deployment by fitting antennas on two-dimensional (2D) grid as shown in
	\cite{SU_mMIMO_Dynamic_SubArray,MassMIMO_hybrid_JSDM_FA,ASIL_ABHP_Access,
		ASIL_Subconnected_GC,
		ASIL_PSO_PA_WCNC,ASIL_MC_PIMRC,ASIL_VTC_DAC_ADC,ASIL_Mobeen_2D_Array}. 
	Moreover, the URA structure enables three-dimensional (3D) beamforming by utilizing both azimuth and elevation angles, which is not possible for ULA \cite{3D_Beamforming_ElevAzim}.
} having 
${M_{t,i}} = {M_{t,i}^{\left(x\right)}} \times {M_{t,i}^{\left(y\right)}}$
and 
${M_{r,i}} = {M_{r,i}^{\left(x\right)}} \times {M_{r,i}^{\left(y\right)}}$
antennas, respectively. 
Here, 
${M_{t,i}^{\left(x\right)}}~\big( {{M_{r,i}^{\left(x\right)}}} \big)$
and 
${M_{t,i}^{\left(y\right)}}~\big( {{M_{r,i}^{\left(y\right)}}} \big)$
represent the transmit (receive) antennas along $x$-axis and $y$-axis, respectively.
As shown in Fig. \ref{fig_1_SystemModel}, an antenna isolation block is placed between the transmit and receive URAs for the propagation domain SIC techniques\footnote{
	The antenna isolation based SIC techniques have been investigated in the literature (e.g.,  in \cite{FD_Survey,FD_Survey_Secrecy, FD_Survey_2019, FD_In_Band_Survey, FD_AntennaSep_60dB, FD_AntennaSep_74dB}). 
	For instance, the numerical results in \cite{FD_AntennaSep_60dB} and \cite{FD_AntennaSep_74dB} show that the antenna isolation can provide $60$ dB and $74$ dB SIC, respectively.
	The design  of antenna isolation block is beyond the scope of this paper.
}.
The BB precoder and transmit RF beamformer are connected via $N_{t,i}$ RF chains at the node $i$. 
Similarly, $N_{r,i}$ RF chains connect the BB combiner and receive RF beamformer.	
Therefore, there are
${M_i} = M_{t,i} + M_{r,i}$
antennas and 
$N_i = N_{t,i} + N_{r,i}$
RF chains at the node $i$.
Additionally, ${S_i}$ denotes the number of parallel data streams transmitted from the node $i$ to the node $j$. 
In order to support ${S_i}$ parallel data streams, it is necessary to satisfy that 
${S_i} \le \min \big( {N_{t,i},N_{r,j}} \big)$.
On the other hand, to reduce the hardware cost/complexity of the full-duplex mmWave massive MIMO systems with large antenna arrays, the number of RF chains at each node is significantly smaller than the number of antennas (i.e., 
$N_{t,i}\ll M_{t,i}$ 
and 
$N_{r,i}\ll M_{r,i}$).
	It is important to note that when the point-to-point massive MIMO cellular system is considered for the proposed system model, the user node with few antenna elements has the flexibility of using either FDPC or HPC. In this special case, the user node using FDPC simply chooses the transmit/receive RF beamformers as identity matrices (i.e., 
		${\bf{F}}_{t,i} = {\bf{I}}_{M_{t,i}}$
		and 
		${\bf{F}}_{r,i} = {\bf{I}}_{M_{r,i}}$).

According to the HPC architecture, the precoded signal vector at the node $i$ is defined as ${{{\bf{s}}_i} = {\bf{F}}_{t,i}{\bf{B}}_{t,i}{{\bf{d}}_i} \in \mathbb{C}^{M_{t,i}}}$, where ${{\bf{d}}_i} \in \mathbb{C}^{S_{i}}$ is the data vector encoded by i.i.d. Gaussian codebooks (i.e., i.i.d. entries of ${{\bf{d}}_i}$ follow the distribution of $\mathcal{CN}\left(0,1\right)$, so we have
$\mathbb{E}\left\{{\bf d}_i {\bf d}_i^H\right\}= {\bf I}_{S_i}$).
The precoded signal is transmitted by the node $i$, which satisfies a transmit power constraint $P_T$ (i.e., $\mathbb{E}\left\{\left\|{\bf{s}}_i\right\|_2^2\right\}\le P_T$).
Then, the received signal by the node $j$ is given by:\vspace{-0.5ex}
\begin{equation}\label{eq_r_init}
	\begin{aligned}
		{{\bf{r}}_j} &= {{\bf{H}}_i}{{\bf{s}}_i} + {\bf{H}}_{\textrm{SI},j}{{\bf{s}}_j} + {{\bf{w}}_j}\\
		&=\underbrace {{{\bf{H}}_i}{\bf{F}}_{t,i}{\bf{B}}_{t,i}{{\bf{d}}_i}}_{{\rm{Intended}}{\rm{~Signal}}}+ 
		\underbrace {{\bf{H}}_{\textrm{SI},j}{\bf{F}}_{t,j}{\bf{B}}_{t,j}{{\bf{d}}_j}}_{{\rm{Self - Interference}}({\rm{SI}})} + \underbrace {{{\bf{w}}_j}}_{{\rm{Noise}}},
	\end{aligned}
\end{equation}
where 
${{\bf{w}}_j} \sim \mathcal{CN} \big( {\bf 0} , \sigma_w^2{\bf{I}}_{M_{r,j}}\big)$
is the complex circularly symmetric Gaussian noise vector.	
After the hybrid combiner at the node $j$, the combined signal can be written as follows:
\begin{equation}\label{eq_r}
	\begin{aligned}
		{{{\tilde{\bf r}}}_j} 
		&\hspace{-0.25ex}=\hspace{-0.25ex} {\bf{B}}_{r,j}{{\bf{F}}_{r,j}{{\bf{r}}_j}} \\
		&\hspace{-0.25ex}=\hspace{-0.25ex} \underbrace {{\bf{B}}_{r,j}\bm{\mathcal{H}}_i{\bf{B}}_{t,i}{{\bf{d}}_i}}_{{\rm{Intended~Signal}}} 
		+
		\underbrace {{\bf{B}}_{r,j}\bm{\mathcal{H}}_{\textrm{SI},j}{\bf{B}}_{t,j}{{\bf{d}}_j}}_{{\rm{Self - Interference}}({\rm{SI}})}
		+
		\underbrace {{\bf{B}}_{r,j}{\bf{F}}_{r,j}{{\bf{w}}_j}}_{{\rm{Modified~Noise}}},
	\end{aligned}
\end{equation}
where 
$\bm{\mathcal{H}}_i={\bf F}_{r,j} {\bf H}_i {\bf F}_{t,i}\in\mathbb{C}^{N_{r,j}\times N_{t,i}}$ 
and
$\bm{\mathcal{H}}_{\rm{SI},j}={{\bf{F}}_{r,j}{\bf{H}}_{\rm{SI},j}{\bf{F}}_{t,j}}\in\mathbb{C}^{N_{r,j}\times N_{t,j}}$
are respectively the reduced-size effective intended and SI channels seen from the BB-stages.
Particularly, the combined signal at the $k^{th}$ data stream with $k=1,\cdots,S_i$ is given by:
\begin{equation}\label{eq_r_k}
	\begin{aligned}
		{{{\tilde{r}}}_{j,k}} 
		&= \underbrace {{\bf{b}}_{r,j,k}^H\bm{\mathcal{H}}_i{\bf{b}}_{t,i,k}{{{d}}_{i,k}}}_{{\rm{Intended~Signal}}} 
		+
		\underbrace {\sum\nolimits_{q\ne k}^{S_i} {\bf{b}}_{r,j,k}^H\bm{\mathcal{H}}_i{\bf{b}}_{t,i,q}{{{d}}_{i,q}}}_{{\rm{Inter-Symbol~Interference(ISI)}}} \\
		&+		
		\underbrace {{\bf{b}}_{r,j,k}^H\bm{\mathcal{H}}_{\textrm{SI},j}{\bf{B}}_{t,j}{{\bf{d}}_j}}_{{\rm{Self - Interference}}({\rm{SI}})}
		+
		\underbrace {{\bf{b}}_{r,j,k}^H{\bf{F}}_{r,j}{{\bf{w}}_j}}_{{\rm{Modified~Noise}}},
	\end{aligned}
\end{equation}
where 
${\bf{b}}_{t,i,k}\in\mathbb{C}^{N_i}$ is the $k^{th}$ column of BB precoder ${\bf{B}}_{t,i}$ and
${\bf{b}}_{j,r,k}\in\mathbb{C}^{N_j}$ is the $k^{th}$ row of BB combiner ${\bf{B}}_{j,r}$.
Here, the inter-symbol interference	(ISI) can be well mitigated via the BB-stage design \cite{SU_mMIMO_OFDM}. However, the strong SI term given in \eqref{eq_r} and \eqref{eq_r_k} can be classified as the major limiting factor for the full-duplex communications.
Thus, during the HPC design, we aim to both maximize the intended signal power (i.e., 
$\big\| {{\bf{B}}_{r,j}{\bf{F}}_{r,j}{{\bf{H}}_i}{\bf{F}}_{t,i}{\bf{B}}_{t,i}{{\bf{d}}_i}} \big\|_2^2$) and suppress the SI power (i.e., $\big\| {{\bf{B}}_{r,j}{\bf{F}}_{r,j}{\bf{H}}_{\textrm{SI},j}{\bf{F}}_{t,j}{\bf{B}}_{t,j}{{\bf{d}}_j}} \big\|_2^2$). Therefore, a joint HPC design is needed to improve the performance of full-duplex mmWave massive MIMO systems. 
Regarding the SIC algorithm, the RF beamformers should satisfy the following condition as:
\begin{equation}\label{eq_approx_zero}
	\bm{\mathcal{H}}_{\textrm{SI},j}={{\bf{F}}_{r,j}{\bf{H}}_{\textrm{SI},j}{\bf{F}}_{t,j}}\approx{\bf 0}.
\end{equation}
The above approximate zero-condition can be achieved via the joint transmit/receive RF beamformer design.
It is worthwhile to remark that the SI terms given in \eqref{eq_r} and \eqref{eq_r_k} do not exist, when the half-duplex communications is employed at each node. 
By using the combined signal at the node $j$ given in \eqref{eq_r}, the covariance matrix of the combined SI and modified noise is given by:
\begin{equation}\label{eq_C}
	\begin{aligned}
		{{\bf{C}}_{tot,j}} 
		&\hspace{-0.5ex}=\hspace{-.5ex}
		\mathbb{E}\hspace{-.5ex}\left\{\hspace{-.5ex}
		\big(
		{\tilde {\bf{r}}_j} \hspace{-.5ex}-\hspace{-.5ex} {\bf{B}}_{r,j}{\bm{\mathcal{H}}_i}{\bf{B}}_{t,i}{{\bf{d}}_i}
		\big)
		\big(
		{\tilde {\bf{r}}_j} \hspace{-.5ex}-\hspace{-.5ex} {\bf{B}}_{r,j}{\bm{\mathcal{H}}_i}{\bf{B}}_{t,i}{{\bf{d}}_i}
		\big)^H
		\hspace{-.5ex}\right\}\\
		&\hspace{-0.5ex}=\hspace{-0.5ex}{\bf{B}}_{\hspace{-.15ex}r,j}\hspace{-.15ex}
		{\bm{\mathcal{H}}_{\hspace{-.15ex}\textrm{SI},j}\hspace{-0.15ex}{\bf{B}}_{\hspace{-.15ex}t,j}}\hspace{-.15ex}{\bf{B}}_{t,j}^H\hspace{-.25ex}\bm{\mathcal{H}}_{\hspace{-.15ex}\textrm{SI},j}^H\hspace{-.15ex} 
		{\bf{B}}_{\hspace{-.15ex}r,j}^H
		\hspace{-.5ex}+ \hspace{-.5ex}
		\sigma_{\hspace{-.35ex}w}^2{\bf{B}}_{\hspace{-.15ex}r,j}\hspace{-.15ex}{\bf{F}}_{\hspace{-.35ex}r,j}\hspace{-.15ex}{\bf{F}}_{\hspace{-.35ex}r,j}^H{\bf\hspace{-.15ex}{B}}_{\hspace{-.15ex}r,j}^H\hspace{-.15ex}.
	\end{aligned}
\end{equation}
By using \eqref{eq_r} and \eqref{eq_C}, the average achievable rate at the node $j$ for the proposed full-duplex mmWave massive MIMO systems is derived as follows:
\begin{IEEEeqnarray}{rCl}\label{eq_Rate_j}
	\IEEEyesnumber
	\IEEEyessubnumber*
	\hspace{-4ex}R_j\hspace{+2ex}&\hspace{-2ex}&\hspace{-2ex}=\mathbb{E}
	\left\{{\log _2} \left| 
	{{{\bf{I}}_{{S_j}}} 
		+ 
		{\bf{C}}_{tot,j}^{ - 1}{\bf{B}}_{r,j}{\bm{\mathcal{H}}}_i{\bf{B}}_{t,i}{\bf{B}}_{t,i}^H{\bm{\mathcal{H}}}_i^H{\bf{B}}_{r,j}^H}
	\right|\right\},\\	
	\hspace{-4ex}\rm{s.t.}&&~
	\bm{\mathcal{H}}_i\hspace{-0.25ex}=\hspace{-0.25ex}{\bf F}_{\hspace{-0.25ex}r,j} {\bf H}_i{\bf F}_{\hspace{-0.25ex}t,i}\hspace{-0.5ex}\in\hspace{-0.5ex}\mathbb{C}^{N_{r,j}\times N_{t,i}}, ~i,j \hspace{-0.5ex}\in\hspace{-0.5ex} \left\{ {1,2} \right\}, ~i \ne j,
	\label{eq_9b}\\
	\hspace{-4ex}& & ~ \left|\hspace{-0.15ex}{\bf{F}}_{\hspace{-0.25ex}t,i}\hspace{-0.35ex}\left(m,\hspace{-0.25ex}n\right)\hspace{-0.15ex}\right|\hspace{-0.5ex}=\hspace{-0.5ex}{\frac{1}{\sqrt{M_{t,i}}}},  
	\left|\hspace{-0.15ex}{\bf{F}}_{\hspace{-0.25ex}r,j}\hspace{-0.35ex}\left(m,\hspace{-0.25ex}n\right)\hspace{-0.15ex}\right|\hspace{-0.5ex}=\hspace{-0.5ex}{\frac{1}{\sqrt{M_{r,j}}}}, \forall m,n\label{eq_9c},\\
	\hspace{-4ex}& & ~
	\mathbb{E}\big\{ {\big\| {\bf{s}}_i \big\|_2^2} \big\} = 
	{\rm{tr}}\big( {{\bf{F}}_{t,i}}{{\bf{B}}_{t,i}}{{\bf{B}}_{t,i}^H}{{\bf{F}}_{t,i}^H} \big) \le {P_T}\label{eq_9d},
\end{IEEEeqnarray}
where \eqref{eq_9b} is the effective channel matrix from the node $i$ to the node $j$, \eqref{eq_9c} indicates the CM property due to the utilization of phase-shifters at the RF-stages as shown in Fig. \ref{fig_1_SystemModel},
\eqref{eq_9d} implies the maximum transmit power constraint $P_T$.
Then, the total achievable rate in the proposed full-duplex mmWave massive MIMO system is given by:
\begin{equation}\label{eq_Rate_total}
	R_{{total}} = R_1 + R_2 \textrm{~~[bps/Hz]}.
\end{equation}
As seen from \eqref{eq_C}, \eqref{eq_Rate_j} and \eqref{eq_Rate_total}, it is necessary to jointly construct the BB precoder, transmit RF beamformer, receive RF beamformer and BB combiner during the HPC design for enhancing the overall system capacity while improving the quality of SIC.
The proposed joint HPC technique is presented in Section \ref{sec_AB_JHPC}.

\section{Channel Model}\label{sec_Channel}

As illustrated in Fig. \ref{fig_1_SystemModel}, it is necessary to individually model both intended channel ${\bf{H}}_i$ and SI channel ${\bf{H}}_{\textrm{SI},j}$. Hence, we model both ot them throughout this section.

\subsection{Intended Channel}
Owing to the limited-scattering propagation environment, the mmWave channels experience only a few spatial paths different from the conventional rich-scattering channels \cite{mmWave_Survey}.
Thus, the intended channel from the node $i$ to the node $j$ is modeled by $C_i$ scattering-clusters and $L_{c,i}$ paths in the  $c^{th}$ cluster with $c=1,2,\cdots,C_i$.
In total, there are $L_i=\sum\nolimits_{c=1}^{C_i}L_{c,i}$ paths between the transmitter and receiver.
Considering the 3D geometry-based mmWave channel model \cite{ChannelModels} and the URA structure \cite{balanis2015antenna}, the channel matrix from the node $i$ to the node $j$ can be written as:\vspace{-0.5ex}
\begin{equation}\label{eq_H_mmWave}
\begin{aligned}
		{\bf{H}}_i 
	&= 
	\sum\limits_{c = 1}^{{C}} \sum\limits_{l = 1}^{{L_c^{\left(i\right)}}} 
	\tau_{i,c_l}^{-\eta}
	{{g_{i,c_l}}}  
	{{{\bm{\phi} }} _{r,j}}\left( {{\gamma _{r,c_l}^{\left(x\right)}},{\gamma _{r,c_l}^{\left(y\right)}}} \right)
	{\bm{\phi }} _{t,i}^H\big( {{\gamma _{t,c_l}^{\left(x\right)}},{\gamma _{t,c_l}^{\left(y\right)}}}\big)\\
	&=
	{{\bf{\Phi }}_{r,j}}{\bf{G}}_i{{\bf{\Phi }}_{t,i}},
\end{aligned}\vspace{-0.5ex}
\end{equation}
where 
$\tau_{i,c_l}$ 
and 
${g_{i,c_l}\sim\mathcal{CN}\big(0,\frac{1}{L_i}\big)}$
are respectively the distance and complex path gain of the $l^{th}$ path in the $c^{th}$ cluster with the path index of $c_l=l+\sum\nolimits_{q=1}^{c-1}L_q^{\left(i\right)}$,
$\eta$ 
is the path loss exponent,
${\bm{\phi} _{u,i}}\big( {\gamma _{u,c_l}^{\left(x\right)},\gamma _{u,c_l}^{\left(y\right)}} \big)\in \mathbb{C}^{M_{u,i}}$ with $u\in\left\{t,r\right\}$
is either the receive phase response vector for $u=r$ or the transmit phase response vector for $u=t$,
${\bf G}_i=\rm{diag}\big(\tau_{i,1}^{-\eta}g_{i,1},\cdots,\tau_{i,L_i}^{-\eta}g_{i,L_i}\big)\in \mathbb{C}^{L_i\times L_i }$ is the diagonal path gain matrix, 	
${{\bf{\Phi }}_{r,j}}\in\mathbb{C}^{M_{r,j}\times L_i}$ and ${{\bf{\Phi }}_{t,i}}\in\mathbb{C}^{L_i \times M_{t,i}}$ are the receive and transmit phase response matrices, respectively. 
Then, the phase response vector is defined as:
\begin{equation}\label{eq_phase_vector}
	\begin{aligned}
		\hspace{-1ex}{\bm{\phi} _{u,i}}\hspace{-0.5ex}\left( {{\gamma_x, \gamma_y}} \right) 
		&\hspace{-0.5ex}=\hspace{-0.5ex}
		\big[ \hspace{-0.25ex}{1\hspace{-0.15ex},\hspace{-0.25ex}{e^{ j2\pi d  {{\gamma_x }} }}\hspace{-0.25ex}, \hspace{-0.35ex}\cdots\hspace{-0.35ex},\hspace{-0.25ex}{e^{  j2\pi d( {{M_{u,i}^{\left(x\right)}} - 1} ) {{\gamma_x}} }}}\hspace{-0.25ex} \big]^T\\
		&\hspace{-0.5ex}\otimes\hspace{-0.5ex}
		\big[ \hspace{-0.25ex}{1\hspace{-0.15ex},\hspace{-0.25ex}{e^{ j2\pi d  {{\gamma_y }} }}\hspace{-0.25ex}, \hspace{-0.35ex}\cdots\hspace{-0.35ex},\hspace{-0.25ex}{e^{  j2\pi d( {{M_{u,i}^{\left(y\right)}} - 1} ) {{\gamma_y}} }}} \hspace{-0.25ex}\big]^T\hspace{-0.25ex}, ~\hspace{-0.25ex}u\hspace{-0.5ex}\in\hspace{-0.5ex}\left\{t,r\right\}\hspace{-0.5ex},
	\end{aligned}
\end{equation}
where $d$ is the distance between two antenna elements normalized by the wavelength.
By using \eqref{eq_H_mmWave} and \eqref{eq_phase_vector}, the receive phase response matrix at node $j$ and transmit phase response matrix at node $i$ are respectively given by:
\begin{equation}\label{eq_Phase_Matrix}
	\begin{aligned}
		{{\bf{\Phi }}_{r,j}} \hspace{-0.5ex}=\hspace{-1ex} \left[\hspace{-1.5ex} {\begin{array}{*{20}{c}}
				{{\bm{\phi} _{r,j}^H}\big( {{\gamma_{r,1}^{\left(x\right)}}},{{\gamma_{r,1}^{\left(y\right)}}} \big)}\\
				{\vdots}\\
				{{\bm{\phi} _{r,j}^H}\big( {{\gamma_{r,L_i}^{\left(x\right)}}},{{\gamma_{r,L_i}^{\left(y\right)}}} \big)}\\
		\end{array}}
		\hspace{-1.5ex}\right]^{\hspace{-.5ex}H}\hspace{-.5ex},~
		{{\bf{\Phi }}_{t,i}} \hspace{-0.5ex}=\hspace{-1ex} \left[\hspace{-1.5ex} {\begin{array}{*{20}{c}}
				{{\bm{\phi} _{t,i}^H}\big( {{\gamma_{t,1}^{\left(x\right)}}},{{\gamma_{t,1}^{\left(y\right)}}} \big)}\\
				{\vdots}\\
				{{\bm{\phi} _{t,i}^H}\big( {{\gamma_{t,L_i}^{\left(x\right)}}},{{\gamma_{t,L_i}^{\left(y\right)}}} \big)}\\
		\end{array}}
		\hspace{-1.5ex}\right]\hspace{-0.5ex}.
	\end{aligned}
\end{equation}
At the node $j$ as the receiver,  
${\gamma _{r,c_l}^{\left(x\right)}} = \sin \big( {{\theta _{r,c_l}^{\left(j\right)}}} \big)\cos \big( {{\psi _{r,c_l}^{\left(j\right)}}} \big)$ 
and 
${\gamma _{r,c_l}^{\left(y\right)}} = \sin \big( {{\theta _{r,c_l}^{\left(j\right)}}} \big)\sin \big( {{\psi _{r,c_l}^{\left(j\right)}}} \big)$ 
are the coefficients carrying the elevation AoA and azimuth AoA parameters of the $l^{th}$ path in the $c^{th}$ cluster, where 
${\theta _{r,c_l}^{\left(j\right)}} \in \big[ {{\theta _{r,c}^{\left(j\right)}} - {\delta _{r,c}^{\theta,\left(j\right)}}}, {{\theta _{r,c}^{\left(j\right)}} + {\delta _{r,c}^{\theta,\left(j\right)}}} \big]$
with mean elevation AoA 
${\theta _{r,c}^{\left(j\right)}}$
and elevation AoA spread 
${\delta _{r,c}^{\theta,\left(j\right)}}$, 
and   
${\psi _{r,c_l}^{\left(j\right)}} \in \big[ {{\psi _{r,c}^{\left(j\right)}} - {\delta _{r,c}^{\psi,\left(j\right)}}}, {{\psi _{r,c}^{\left(j\right)}} + {\delta _{r,c}^{\psi,\left(j\right)}}} \big]$
with mean azimuth AoA   
${\psi _{r,c}^{\left(j\right)}}$
and azimuth AoA spread  
${\delta _{r,c}^{\psi,\left(j\right)}}$.
Similarly, at the node $i$ as the transmitter, 
${\gamma _{t,c_l}^{\left(x\right)}} = \sin \big( {{\theta _{t,c_l}^{\left(i\right)}}} \big)\cos \big( {{\psi _{t,c_l}^{\left(i\right)}}} \big)$
and   
${\gamma _{t,c_l}^{\left(y\right)}} = \sin \big( {{\theta _{t,c_l}^{\left(i\right)}}} \big)\sin \big( {{\psi _{t,c_l}^{\left(i\right)}}} \big)$
reflect the elevation AoD and azimuth AoD information of the $l^{th}$ path in the $c^{th}$ cluster, where   
${\theta _{t,c_l}^{\left(i\right)}} \in \big[ {{\theta _{t,c}^{\left(i\right)}} - {\delta _{t,c}^{\theta,\left(i\right)}}}, {{\theta _{t,c}^{\left(i\right)}} + {\delta _{t,c}^{\theta,\left(i\right)}}} \big]$
with mean elevation AoD   
${\theta _{t,c}^{\left(i\right)}}$
and elevation AoD spread  
${\delta _{t,c}^{\theta,\left(i\right)}}$, 
and   
${\psi _{t,c_l}^{\left(i\right)}} \in \big[ {{\psi _{t,c}^{\left(i\right)}} - {\delta _{t,c}^{\psi,\left(i\right)}}}, {{\psi _{t,c}^{\left(i\right)}} + {\delta _{t,c}^{\psi,\left(i\right)}}} \big]$
with mean azimuth AoD   
${\psi _{t,c}^{\left(i\right)}}$
and azimuth AoD spread  
${\delta _{t,c}^{\psi,\left(i\right)}}$.

As in \cite{JSDM_LargeArray,MassMIMO_hybrid_JSDM_FA,ASIL_ABHP_Access,ASIL_MC_PIMRC,ASIL_VTC_DAC_ADC}, we consider that the intended channel has two parts: (i) fast time-varying instantaneous CSI as given in \eqref{eq_H_mmWave} and (ii) slow time-varying phase response matrices based on AoD/AoA information expressed in \eqref{eq_Phase_Matrix}.

\subsection{Self-Interference Channel}\label{secsec_SICchannel}

\begin{figure}[!t]
	\centering
	\includegraphics[width=0.7\columnwidth]{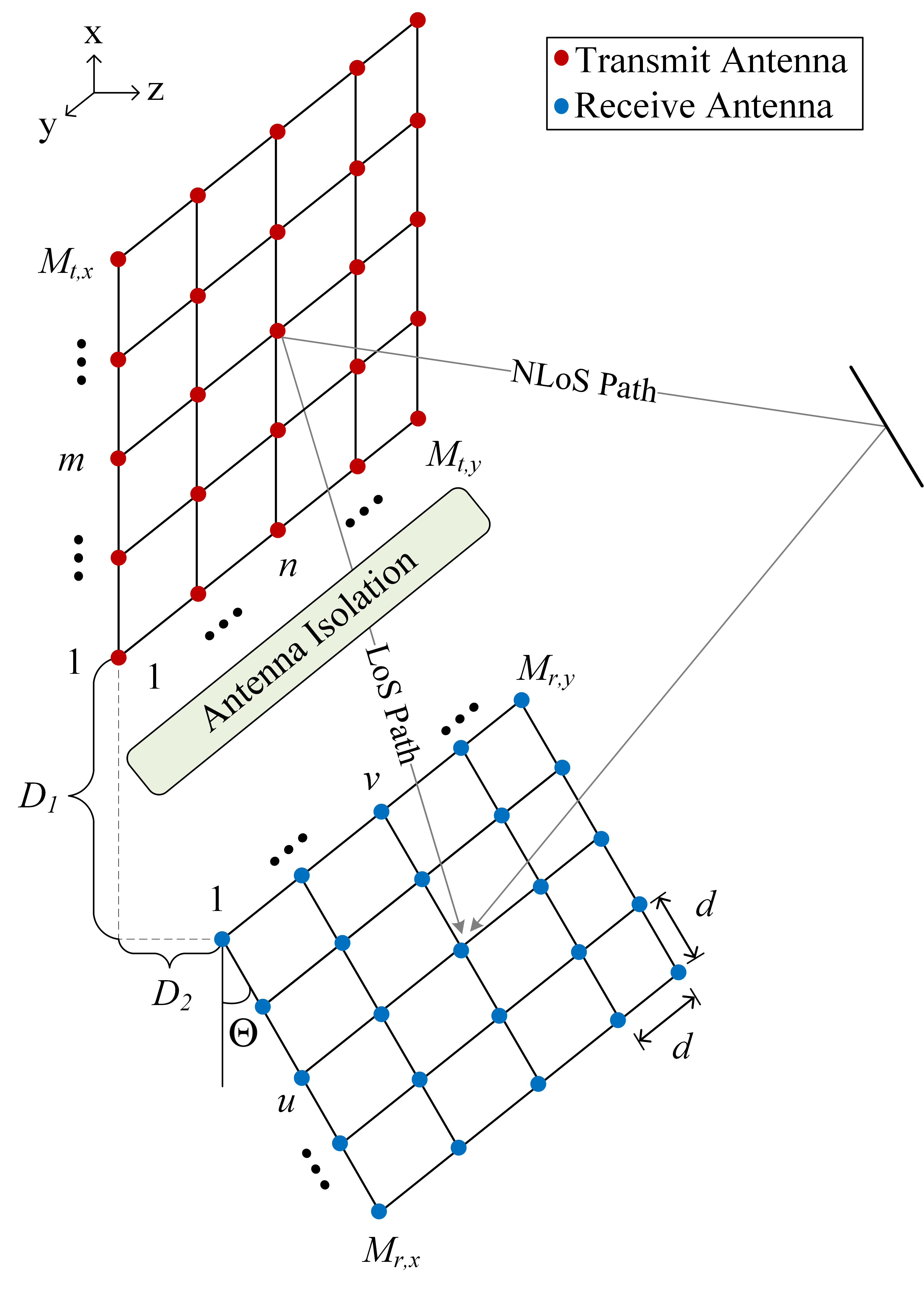}
	\vspace{-2ex}
	\caption{Illustration of transmit and receive URAs with the near-field and far-field SI channels based on LoS and NLoS paths.}
	\vspace{-2ex}
	\label{fig_2_TxRxAntennas}
\end{figure}

According to the two-way full-duplex transmission, the SI channel matrix ${\bf{H}}_{\textrm{SI}}^{\left(i\right)}$ is present at the node $i$. 
The transmit and receive URAs at each node are placed next to each other as illustrated in Fig. \ref{fig_2_TxRxAntennas}, where $D_1$ ($D_2$) is the distance between the transmit and receive URAs along $x$-axis ($z$-axis) normalized by the wavelength, $\Theta$ is the rotation angle of URA around $y$-axis.
As discussed earlier, the antenna isolation block is also included to suppress the SI channel power, especially the strong near-field effect (see Fig. \ref{fig_1_SystemModel}). However, it cannot completely remove the near-field SI channel \cite{FD_AntennaSep_74dB}.
Hence, as demonstrated in Fig. \ref{fig_2_TxRxAntennas}, the SI channel is composed of two parts:  (i)  the residual near-field SI channel representing the direct line-of-sight (LoS) paths after the antenna isolation based SIC,  (ii) the far-field SI  channel representing the reflected non-line-of-sight (NLoS) paths \cite{Satyanarayana2019,FD_MIMO_MU_Learning,FD_MIMO_Relay_mmWave_2}. Here, we first define the complete SI channel matrix at the node $i$  as:
\begin{equation}\label{eq_HSI}
	{\bf{H}}_{\textrm{SI},i}
	=
	{\bf{H}}_{\textrm{LoS},i}
	+
	{\bf{H}}_{\textrm{NLoS},i}
	\in\mathbb{C}^{M_{r,i}\times M_{t,i}},
\end{equation}
where 
${\bf{H}}_{\textrm{LoS},i}\in\mathbb{C}^{M_{r,i}\times M_{t,i}}$
is the residual near-field SI component and
${\bf{H}}_{\textrm{NLoS},i}\in\mathbb{C}^{M_{r,i}\times M_{t,i}}$
is the far-field SI component. 
The channel model given in \eqref{eq_H_mmWave} assumes the planar wavefront, which requires two communication nodes to be far apart. 
It makes the aforementioned channel model given in \eqref{eq_H_mmWave} an unrealistic model for ${\bf{H}}_{\textrm{LoS},i}$.
Based on the spherical wave propagation model, the near-field SI channel is modeled for the ULA structure in \cite{Satyanarayana2019,FD_MIMO_MU_Learning,FD_MIMO_Relay_mmWave_2}.
According to the transmit and receive URAs demonstrated in Fig. \ref{fig_2_TxRxAntennas}, the residual near-field SI channel between the $\left(m,n\right)^{th}$ transmit and $\left(u,v\right)^{th}$ receive antenna pair at the node $i$ is defined as in \eqref{eq_HLOS},
\begin{figure*}[bp]
	\normalsize
	\hrulefill
	\begin{equation}\label{eq_HLOS}
		{\bf{H}}_{\textrm{LoS},i}\big( {\big[ {\left( {u - 1} \right)M_{r,i}^{\left( y \right)} + v} \big],\big[ {\left( {m - 1} \right)M_{t,i}^{\left( y \right)} + n} \big]} \big) = \frac{\kappa }{{{\Delta_{\left( {m,n} \right)\rightarrow\left( {u,v} \right)}}}}{e^{ - j2\pi {{{\Delta_{\left( {m,n} \right)\rightarrow\left( {u,v} \right)}}}}}}.
	\end{equation}
	\begin{equation}\label{eq_delta_mn_uv}
		{\Delta _{\left( {m,n} \right)\rightarrow\left( {u,v} \right)}} \hspace{-0.5ex}=\hspace{-0.5ex} 
		\sqrt {
			{\left[ {\left( {u \hspace{-0.5ex}-\hspace{-0.5ex} 1} \right)\hspace{-0.25ex}d\sin \hspace{-.5ex}\left( \Theta  \right) \hspace{-0.5ex}+\hspace{-0.5ex} {D_2}} \right]}^2 
			+ 
			{\left[ {\left( {m \hspace{-0.5ex}-\hspace{-0.5ex} 1} \right)\hspace{-0.25ex}d \hspace{-0.5ex}+\hspace{-0.5ex} \left( {u \hspace{-0.5ex}- \hspace{-0.5ex}1} \right)\hspace{-0.25ex}d\cos \hspace{-0.5ex}\left( \Theta  \right) \hspace{-0.5ex}+\hspace{-0.5ex} {D_1}} \right]}^2
			+ 
			{\left[ {\left( {n \hspace{-0.5ex}-\hspace{-0.5ex} v} \right)\hspace{-0.25ex}d} \right]}^2
		}.
	\end{equation}
\end{figure*}
where 
$\kappa$ 
is the normalization scalar so that 
$10\log_{10}\left(\mathbb{E}\big\{\|{{\bf{H}}_{\textrm{LoS},i}}\|_F^2\big\}\right)=-P_{\textrm{IS,dB}}$, 
$P_{\textrm{IS,dB}}$ is the SIC in dB achieved by the antenna isolation\footnote{When $P_{\textrm{IS,dB}}=0$ dB inferring no antenna isolation, the near-field SI channel has the unity power on average as $\mathbb{E}\big\{\|{{\bf{H}}_{\textrm{LoS},i}}\|_F^2\big\}=1$. Additionally, if $P_{\textrm{IS,dB}}=30$ dB SIC is achieved by the antenna isolation, the average power of the residual near-field SI channel is calculated as $\mathbb{E}\big\{\|{{\bf{H}}_{\textrm{LoS},i}}\|_F^2\big\}=0.001$.},
${{\Delta_{\left( {m,n} \right)\rightarrow\left( {u,v} \right)}}}$
is the distance between the $\left(m,n\right)^{th}$ transmit and $\left(u,v\right)^{th}$ receive antenna pair  normalized by the wavelength. 
By using Fig. \ref{fig_2_TxRxAntennas}, the distance between the antenna elements is obtained as in \eqref{eq_delta_mn_uv}.

Regarding the far-field SI channel ${\bf H}_{\textrm{NLoS},i}$ given in \eqref{eq_HSI}, we utilize the 3D geometry-based mmWave channel model because the plane wave propagation model is suitable for the NLoS paths. By using \eqref{eq_H_mmWave} and \eqref{eq_Phase_Matrix}, the far-field SI channel at the node $i$ can be written as:
\begin{equation}\label{eq_H_NLoS}
	\begin{aligned}
		{\bf{H}}_{\textrm{{NLoS}},i}
		= {{\bf{\Phi }}_{\textrm{SI},r,i}}{\bf{G}}_{\textrm{SI},i}{{\bf{\Phi }}_{\textrm{SI},t,i}},
	\end{aligned}
\end{equation}
where 
${\bf G}_{\textrm{SI},i}
=
\rm{diag}\big(
\tau_{\textrm{SI},i,1}^{-\eta}g_{\textrm{SI},i,1},
\cdots,
\tau_{\textrm{SI},i,L_{\textrm{SI}}}^{-\eta}g_{\textrm{SI},i,L_{\textrm{SI}}}\big)\in 
\mathbb{C}^{
	L_{\textrm{SI}}
	\times 
	L_{\textrm{SI}}
}$ 
is the diagonal path gain matrix with $L_{\textrm{SI}}$ paths,
$\tau_{\textrm{SI},i,l}$ and $g_{\textrm{SI},i,l}\sim\mathcal{CN}\Big(0,\frac{1}{L_{\textrm{SI}}}\Big)$ 
are the distance and complex path gain of the $l^{th}$ NLoS path, respectively,
${{\bf{\Phi }}_{\textrm{SI},r,i}}\in\mathbb{C}^{M_{r,i}\times L_{\textrm{SI}}}$ 
and 
${{\bf{\Phi }}_{\textrm{SI},t,i}}\in\mathbb{C}^{L_{\textrm{SI}}\times M_{t,i}}$ 
are the receive and transmit phase response matrices for the far-field SI channel, respectively. 
In order to generate the phase response matrices as expressed in \eqref{eq_Phase_Matrix}, we first consider $C_{\textrm{SI}}$ scattering-clusters for the NLoS paths.
Then, the mean elevation and azimuth AoA (AoD) of paths inside the  $c^{th}$ cluster are defined as 	
${\theta _{\textrm{SI},r,c}^{\left(i\right)}}$ (${\theta _{\textrm{SI},t,c}^{\left(i\right)}}$)
and 
${\psi _{\textrm{SI},r,c}^{\left(i\right)}}$ (${\psi _{\textrm{SI},t,c}^{\left(i\right)}}$), respectively.
Also, the corresponding elevation and azimuth AoA (AoD) spread are given as 
$\delta_{\textrm{SI},r,c}^{\theta,\left(i\right)}$ ($\delta_{\textrm{SI},t,c}^{\theta,\left(i\right)}$)
and
$\delta_{\textrm{SI},r,c}^{\psi,\left(i\right)}$ ($\delta_{\textrm{SI},t,c}^{\psi,\left(i\right)}$),
respectively.

\section{Joint Hybrid Precoding/Combining}\label{sec_AB_JHPC}
In this section, the proposed angular-based joint hybrid precoding/combining (AB-JHPC) technique is developed for the full-duplex mmWave massive MIMO systems.
Here, we have four ultimate objectives:
(i) 	enhance the quality of SIC,
(i) 	maximize the intended signal power,
(iii) 	decrease the CSI overhead size,	
(iv) 	reduce the hardware cost/complexity by employing less number of RF chains.
Thus, we first introduce the transmit and receive RF beamformer design based on the slow time-varying AoD/AoA information, where we aim to maximize the beamforming gain for the intended signal and suppress the SI signal (please see \eqref{eq_approx_zero}).
Afterwards, the transmit BB precoder and receive BB combiner are built via the reduced-size effective CSI seen from the BB-stage, where we employ SVD and S-MMSE based algorithms.
\subsection{RF-Stage Design}

The primary target at the RF-stage design is to maximize the beamforming gain for the intended signal expressed in \eqref{eq_r}. 
By using \eqref{eq_H_mmWave}, the effective intended channel is rewritten as:\vspace{-1ex}
\begin{equation}
	\bm{\mathcal{H}}_i={\bf F}_{r,j} {\bf H}_i {\bf F}_{t,i}
	=
	{\bf F}_{r,j}{{\bf{\Phi }}_{r,j}}{\bf{G}}_i{{\bf{\Phi }}_{t,i}}{\bf F}_{t,i}.\vspace{-1ex}
\end{equation}
In order to maximize the transmit (receive) beamforming gain and exploit all degrees of freedom provided by the intended channel,
the columns of ${\bf F}_{t,i}$ (${\bf F}_{r,j}$) should belong to the subspace spanned
by ${{\bf{\Phi }}_{t,i}}$ (${{\bf{\Phi }}_{r,j}}$). 
Thus, we should satisfy $\rm{Span}\left({\bf F}_{t,i}\right) \subset \rm{Span}\left({{\bf{\Phi }}_{t,i}}\right)$ and 
$\rm{Span}\left({\bf F}_{r,j}\right) \subset \rm{Span}\left({{\bf{\Phi }}_{r,j}}\right)$.
Here, it is worthwhile to mention that the transmit (receive) phase response matrix ${{\bf{\Phi }}_{t,i}}$ (${{\bf{\Phi }}_{r,j}}$) is a function of slow time-varying AoD (AoA) information (please see \eqref{eq_Phase_Matrix}).
For assuring the primary target, AoD and AoA support for the intended channel from the node $i$ to the node $j$ are respectively defined as follows:\vspace{-0.5ex}
\begin{IEEEeqnarray}{rCl}\label{eq_AoD_AoA}
	\IEEEyesnumber
	\IEEEyessubnumber*
	\hspace{-4.5ex}{\rm{Ao}}{{\rm{D}}}_i \hspace{-0.5ex}&=&\hspace{-0.5ex} \Big\lbrace \hspace{-0.6ex}{{{\left[ {{\gamma _{x}}\hspace{-0.25ex},\hspace{-0.25ex}{\gamma _{y}}} \right]}} \hspace{-0.5ex}=\hspace{-0.5ex} \sin \hspace{-0.5ex}\left( \hspace{-0.25ex}\theta \hspace{-0.25ex} \right)\hspace{-0.5ex}{{\left[\hspace{-0.15ex} {\cos \hspace{-0.5ex}\left(\hspace{-0.25ex} \psi\hspace{-0.25ex}  \right)\hspace{-0.25ex},\hspace{-0.25ex}\sin \hspace{-0.5ex}\left( \hspace{-0.25ex}\psi \hspace{-0.25ex} \right)} \hspace{-0.15ex}\right]}}} \hspace{0.2ex}\Big|\hspace{0.2ex} 	\theta\hspace{-0.5ex}  \in\hspace{-0.5ex}{\bm {\theta }}_{t,i}\hspace{-0.25ex},\hspace{-0.25ex} \psi\hspace{-0.5ex}  \in\hspace{-0.5ex}	{\bm {\psi }}_{t,i}\hspace{-0.5ex}\Big\rbrace\hspace{-0.25ex},\label{eq_AoD_Supp}\\
	\hspace{-4.5ex}{\rm{Ao}}{{\rm{A}}}_j \hspace{-0.5ex}&=&\hspace{-0.5ex} \Big\lbrace \hspace{-0.6ex}{{{\left[ {{\gamma _{x}}\hspace{-0.25ex},\hspace{-0.25ex}{\gamma _{y}}} \right]}} \hspace{-0.5ex}=\hspace{-0.5ex} \sin \hspace{-0.5ex}\left( \hspace{-0.25ex}\theta \hspace{-0.25ex} \right)\hspace{-0.5ex}{{\left[\hspace{-0.15ex} {\cos \hspace{-0.5ex}\left(\hspace{-0.25ex} \psi\hspace{-0.25ex}  \right)\hspace{-0.25ex},\hspace{-0.25ex}\sin \hspace{-0.5ex}\left( \hspace{-0.25ex}\psi \hspace{-0.25ex} \right)} \hspace{-0.15ex}\right]}}} \hspace{0.2ex}\Big|\hspace{0.2ex} 	\theta\hspace{-0.5ex}  \in\hspace{-0.5ex}{\bm {\theta }}_{\hspace{-0.25ex}r,j}\hspace{-0.25ex},\hspace{-0.25ex} \psi\hspace{-0.5ex}  \in\hspace{-0.5ex}	{\bm {\psi }}_{\hspace{-0.25ex}r,j}\hspace{-0.5ex}\Big\rbrace\hspace{-0.25ex},\label{eq_AoA_Supp}
\end{IEEEeqnarray}
where 
${\bm {\theta }}_{t,i} \hspace{-3ex}=\hspace{-3ex}\cup_{c=1}^C  \big[ {{\theta _{t,c}^{\left(i\right)}} \hspace{1ex}-\hspace{1ex} {\delta _{t,c}^{\theta,\left(i\right)}}}, {{\theta _{t,c}^{\left(i\right)}} \hspace{1ex}+\hspace{1ex} {\delta _{t,c}^{\theta,\left(i\right)}}} \big]$ 
and
${\bm {\psi }}_{t,i} \hspace{-3ex}=\hspace{-3ex}\cup_{c=1}^C  \big[ {{\psi _{t,c}^{\left(i\right)}} \hspace{1ex} -\hspace{1ex}  {\delta _{t,c}^{\psi,\left(i\right)}}},   {{\psi _{t,c}^{\left(i\right)}} \hspace{1ex} +\hspace{1ex}  {\delta _{t,c}^{\psi,\left(i\right)}}} \big]$
denote the elevation and azimuth AoD supports of the intended channel at the node $i$, respectively.
Similarly, 
${\bm {\theta }}_{r,j} \hspace{-2ex}=\hspace{-2ex}\cup_{c=1}^C  \big[ {{\theta _{r,c}^{\left(j\right)}} \hspace{0.75ex}-\hspace{0.75ex} {\delta _{r,c}^{\theta,\left(j\right)}}}, {{\theta _{r,c}^{\left(j\right)}} \hspace{0.75ex}+\hspace{0.75ex} {\delta _{r,c}^{\theta,\left(j\right)}}} \big]$ 
and
${\bm {\psi }}_{r,j} \hspace{-1ex}=\hspace{-1ex}\cup_{c=1}^C  \big[ {{\psi _{r,c}^{\left(j\right)}} \hspace{0.25ex}-\hspace{0.25ex} {\delta _{r,c}^{\psi,\left(j\right)}}},   {{\psi _{r,c}^{\left(j\right)}} \hspace{0.25ex}+\hspace{0.25ex} {\delta _{r,c}^{\psi,\left(j\right)}}} \big]$
represent the elevation and azimuth AoA supports of the intended channel at the node $j$, respectively.

On the other hand, the secondary target at the RF-stage design is to suppress the effective SI channel power by satisfying the approximate zero condition given in \eqref{eq_approx_zero}. By substituting \eqref{eq_HLOS} and \eqref{eq_HSI} into \eqref{eq_approx_zero}, the effective SI channel at the node $j$ is also rewritten as:
\begin{equation}\label{eq_HSI_new_eff}
	\begin{aligned}
		\hspace{-1ex}\bm{\mathcal{H}}_{\hspace{-0.1ex}\textrm{SI},j}&
		\hspace{-0.5ex}=\hspace{-0.5ex}
		{{\bf{F}}_{r,j}{\bf{H}}_{\textrm{SI},j}{\bf{F}}_{t,j}}
		\hspace{-0.5ex}=\hspace{-0.5ex}
		{\bf{F}}_{r,j}
		\left(
		{\bf{H}}_{\textrm{NLoS},j}
		+
		{\bf{H}}_{\textrm{LoS},j},
		\right)
		{\bf{F}}_{t,j}\\
		&\hspace{-0.5ex}= \hspace{-1ex}
		\underbrace{
			{\bf{F}}_{\hspace{-0.25ex}r,j}
			{{\bf{\Phi }}_{\textrm{SI},r,j}}
		}_{\scriptstyle\rm{Receive}\atop \scriptstyle \rm{Beamforming}}
		\hspace{-1ex}{\bf{G}}_{\textrm{SI},j}\hspace{-1ex}
		\underbrace{
			{{\bf{\Phi }}_{\textrm{SI},t,j}}
			{\bf{F}}_{\hspace{-0.25ex}t,j}
		}_{\scriptstyle\rm{Transmit}\atop \scriptstyle \rm{Beamforming}}
		\hspace{-0.5ex}+\hspace{-1ex}
		\underbrace{
			{\bf{F}}_{\hspace{-0.25ex}r,j}
			{\bf{H}}_{\textrm{LoS},j}
			{\bf{F}}_{\hspace{-0.25ex}t,j}
		}_{\scriptstyle \rm{Suppressed~via}\atop \scriptstyle \rm{Antenna~ Isolation}}\hspace{-1ex}\approx\hspace{-.25ex}{\bf{0}},\\
	\end{aligned}
\end{equation}
where the approximate zero condition can be achieved via the joint design of the receive RF beamformer ${\bf{F}}_{r,j}$  and the transmit RF beamformer ${\bf{F}}_{t,j}$. 
\begin{figure*}[bp]
	\normalsize
	\hrulefill
	\addtocounter{equation}{+2}
	\begin{equation}\label{eq_steering_vector_vs_SI}
		\begin{aligned}
			{{\bf{\Phi }}_{\textrm{SI},t,j}}
			{{{\bf{e}} _{t,i}} \big(\lambda_{t,i}^{\left(x\right),k},\lambda_{t,i}^{\left(y\right),n}\big)}
			=
			\frac{1}{\sqrt {{M_{t,i}}}} \left[ {\begin{array}{*{20}{c}}
					{\sum\limits_{{m_x}}^{M_{t,i}^{\left( x \right)}} {\sum\limits_{{m_y}}^{M_{t,i}^{\left( y \right)}} {{e^{j2\pi d\left[ {m{  _x}\left( {\lambda _{t,i}^{\left( x \right),k} - \gamma _{t,1}^{\left( x \right)}} \right) +  {m_y}\left( {\lambda _{t,i}^{\left( y \right),k} - \gamma _{t,1}^{\left( y \right)}} \right)} \right]}}} } }\\
					{\vdots}\\
					{\sum\limits_{{m_x}}^{M_{t,i}^{\left( x \right)}} {\sum\limits_{{m_y}}^{M_{t,i}^{\left( y \right)}} \hspace{-0.75ex}{{e^{j2\pi d\left[\hspace{-0.25ex} {m{  _x}\hspace{-0.25ex}\left(\hspace{-0.25ex} {\lambda _{t,i}^{\left( x \right),k} \hspace{-0.25ex}- \gamma _{t,{L_{SI}}}^{\left( x \right)}} \hspace{-0.25ex}\right) +   {m_y}\hspace{-0.25ex}\left( \hspace{-0.25ex}{\lambda _{t,i}^{\left( y \right),k} \hspace{-0.25ex}- \gamma _{t,{L_{SI}}}^{\left( y \right)}} \hspace{-0.25ex}\right)} \hspace{-0.25ex}\right]}}} } }
			\end{array}} \right].
		\end{aligned}
	\end{equation}
\end{figure*}
Here, the effective SI channel has two components representing the LoS and NLoS paths. As discussed in Section \ref{secsec_SICchannel}, the LoS path reflecting the strong near-field SI channel is suppressed via the antenna isolation based SIC techniques \cite{FD_Survey,FD_Survey_Secrecy, FD_Survey_2019, FD_In_Band_Survey, FD_AntennaSep_60dB, FD_AntennaSep_74dB}. However, after the canceling strong near-field SI channel, the NLoS paths reflecting the far-field SI channel become more dominant\footnote{The experimental results in \cite{FD_AntennaSep_74dB} show that 74 dB SIC is achieved by antenna isolation measured in the anechoic chamber as a low-reflection environment. However, the SIC is reduced to 44 dB in the reflective room. The main reason considering 30 dB decrease of SIC is owing to the dominant reflected paths (i.e., NLoS paths).}.
Hence, the secondary target can be accomplished when the columns of ${\bf F}_{t,j}$ (${\bf F}_{r,j}$) should belong to the null space of ${{\bf{\Phi }}_{\textrm{SI},t,j}}$ 
(${{\bf{\Phi }}_{\textrm{SI},r,j}}$). 
In other words, the secondary target requires 
$\rm{Span}\left({\bf F}_{t,j}\right) \subset \rm{Null}\left({{\bf{\Phi }}_{\rm{SI},t,j}}\right)$
and 
$\rm{Span}\left({\bf F}_{r,j}\right) \subset \rm{Null}\left({{\bf{\Phi }}_{\textrm{SI},r,j}}\right)$.
For addressing the secondary target, we also define the AoD and AoA supports of the SI channel at the node $j$:
\addtocounter{equation}{-3}
\begin{IEEEeqnarray}{rCl}\label{eq_AoD_AoA_SI}
	\IEEEyesnumber
	\IEEEyessubnumber*
	\resizebox{0.83\hsize}{!}{$
	\hspace{-5ex}{\rm{Ao}}{{\rm{D}}}_{\textrm{SI},j} \hspace{-0.5ex}=\hspace{-0.5ex} \Big\lbrace \hspace{-0.6ex}{{{\left[ {{\gamma _{x}}\hspace{-0.25ex},\hspace{-0.25ex}{\gamma _{y}}} \right]}} \hspace{-0.5ex}=\hspace{-0.5ex} \sin \hspace{-0.5ex}\left( \hspace{-0.25ex}\theta \hspace{-0.25ex} \right)\hspace{-0.5ex}{{\left[\hspace{-0.15ex} {\cos \hspace{-0.5ex}\left(\hspace{-0.25ex} \psi\hspace{-0.25ex}  \right)\hspace{-0.25ex},\hspace{-0.25ex}\sin \hspace{-0.5ex}\left( \hspace{-0.25ex}\psi \hspace{-0.25ex} \right)} \hspace{-0.15ex}\right]}}} \hspace{0.2ex}\Big|\hspace{0.2ex} 	\theta\hspace{-0.5ex}  \in\hspace{-0.5ex}{\bm {\theta }}_{\textrm{SI},t,i}\hspace{-0.25ex},\hspace{-0.25ex} \psi\hspace{-0.5ex}  \in\hspace{-0.5ex}	{\bm {\psi }}_{\textrm{SI},t,i}\hspace{-0.5ex}\Big\rbrace\hspace{-0.25ex},\label{eq_AoD_Supp_SI}
	$}\\
	\resizebox{0.83\hsize}{!}{$
	\hspace{-5ex}{\rm{Ao}}{{\rm{A}}}_{\textrm{SI},j} \hspace{-0.5ex}=\hspace{-0.5ex} \Big\lbrace \hspace{-0.6ex}{{{\left[ {{\gamma _{x}}\hspace{-0.25ex},\hspace{-0.25ex}{\gamma _{y}}} \right]}} \hspace{-0.5ex}=\hspace{-0.5ex} \sin \hspace{-0.5ex}\left( \hspace{-0.25ex}\theta \hspace{-0.25ex} \right)\hspace{-0.5ex}{{\left[\hspace{-0.15ex} {\cos \hspace{-0.5ex}\left(\hspace{-0.25ex} \psi\hspace{-0.25ex}  \right)\hspace{-0.25ex},\hspace{-0.25ex}\sin \hspace{-0.5ex}\left( \hspace{-0.25ex}\psi \hspace{-0.25ex} \right)} \hspace{-0.15ex}\right]}}} \hspace{0.2ex}\Big|\hspace{0.2ex} 	\theta\hspace{-0.5ex}  \in\hspace{-0.5ex}{\bm {\theta }}_{\textrm{SI},r\hspace{-0.35ex},j}\hspace{-0.25ex},\hspace{-0.25ex} \psi\hspace{-0.5ex}  \in\hspace{-0.5ex}	{\bm {\psi }}_{\textrm{SI},r\hspace{-0.35ex},j}\hspace{-0.5ex}\Big\rbrace\hspace{-0.25ex},\label{eq_AoA_Supp_SI}
	$}
\end{IEEEeqnarray}
where 
${\bm {\theta }}_{\textrm{SI},t,j} \hspace{-2ex}=\hspace{-2ex}\cup_{c=1}^C  \big[ {{\theta _{\textrm{SI},t,c}^{\left(j\right)}}  \hspace{1ex}-\hspace{1ex} {\delta _{\textrm{SI},t,c}^{\theta,\left(j\right)}}}, {{\theta _{\textrm{SI},t,c}^{\left(j\right)}}  \hspace{1ex}+\hspace{1ex} {\delta _{\textrm{SI},t,c}^{\theta,\left(j\right)}}} \big]$ 
and
${\bm {\psi }}_{\textrm{SI},t,j} \hspace{-2ex}=\hspace{-2ex}\cup_{c=1}^C  \big[ {{\psi _{\textrm{SI},t,c}^{\left(j\right)}}  \hspace{1ex}-\hspace{1ex} {\delta _{\textrm{SI},t,c}^{\psi,\left(j\right)}}},   {{\psi _{\textrm{SI},t,c}^{\left(j\right)}}  \hspace{1ex}+\hspace{1ex} {\delta _{\textrm{SI},t,c}^{\psi,\left(j\right)}}} \big]$
denote the elevation and azimuth AoD supports of the SI channel at the node $j$, respectively.
Also, 
${\bm {\theta }}_{\textrm{SI},r,j} \hspace{-2ex}=\hspace{-2ex}\cup_{c=1}^C  \big[ {{\theta _{\textrm{SI},r,c}^{\left(j\right)}}  \hspace{1ex}-\hspace{1ex} {\delta _{\textrm{SI},r,c}^{\theta,\left(j\right)}}}, {{\theta _{\textrm{SI},r,c}^{\left(j\right)}} \hspace{1ex}+\hspace{1ex} {\delta _{\textrm{SI},r,c}^{\theta,\left(j\right)}}} \big]$ 
and
${\bm {\psi }}_{\textrm{SI},r,j} =\cup_{c=1}^C  \big[ {{\psi _{\textrm{SI},r,c}^{\left(j\right)}} - {\delta _{\textrm{SI},r,c}^{\psi,\left(j\right)}}},   {{\psi _{\textrm{SI},r,c}^{\left(j\right)}} + {\delta _{\textrm{SI},r,c}^{\psi,\left(j\right)}}} \big]$
represent the elevation and azimuth AoA supports of the SI channel at the node $j$, respectively.

For satisfying both the primary and secondary targets, the transmit/receive RF beamformers should cover the angular support of the intended channel given in \eqref{eq_AoD_AoA} while excluding the angular support of the SI channel given in \eqref{eq_AoD_AoA_SI}. For this purpose, according to the phase response vectors expressed in \eqref{eq_phase_vector}, we first define the transmit and receive steering vectors as 
${{\bf{e}}_{t,i}}\left( {{\gamma_x},{\gamma _y}} \right) = \frac{1}{\sqrt{M_{t,i}}}
{\bm{\phi} _{t,i}}\left( {{\gamma_x, \gamma_y}} \right)$
and 
${{\bf{e}}_{r,i}}\left( {{\gamma_x},{\gamma _y}} \right) = 
\frac{1}{\sqrt{M_{r,i}}}
{\bm{\phi} _{r,i}^*}\left( {{\gamma_x, \gamma_y}} \right)$, 
respectively.
Then, we define the orthogonal quantized angle-pairs $\big(\lambda_{u,i}^{\left(x\right),k},\lambda_{u,i}^{\left(y\right),n}\big)$ as:
\begin{IEEEeqnarray}{rCl}\label{eq_lambda}
	\IEEEyesnumber
	\IEEEyessubnumber*
	{\lambda _{u,i}^{\left(x\right),k}}
	&=&
	-1+\frac{2k-1}{{{M_{u,i}^{\left(x\right)}}}} 
	\textrm{,~~}\forall k = 1, \cdots,{M_{u,i}^{\left(x\right)}},\\
	{\lambda _{u,i}^{\left(y\right),n}}
	&=&
	-1+\frac{2n-1}{{{M_{u,i}^{\left(y\right)}}}}
	\textrm{,~~}\forall n = 1, \cdots,{M_{u,i}^{\left(y\right)}},
\end{IEEEeqnarray}
where $u\in\left\{t,r\right\}$
represents either the transmitter side for $u=t$ or the receiver side for $u=r$.
The orthogonality property of the quantized angle-pairs generates $M_{t,i}$ orthogonal transmit steering vectors and $M_{r,i}$ orthogonal receive steering vectors (i.e., 
${{{\bf{e}}_{u,i}^H}\big( 
	{\lambda _{u,i}^{\left(x\right),k}}
	{\lambda _{u,i}^{\left(x\right),n}}	 
	\big){{\bf{e}}_{u,i}}\big( 
	{\lambda _{u,i}^{\left(x\right),k'}}
	{\lambda _{u,i}^{\left(x\right),n'}}	 
	\big)=0}$
for
$u\in\left\{t,r\right\}$,
$\forall(k,n)\ne(k',n')$).
Furthermore, the quantized angle-pairs defined in \eqref{eq_lambda} produces the minimum number of angle-pairs in order to cover the complete 3D angular support \cite{ASIL_ABHP_Access}. It also reduces the number of RF chain utilization, while exploiting all degrees of freedom provided by the intended channel (i.e., $\textrm{Span}\left({\bf F}_{t,i}\right) \subset \textrm{Span}\left({{\bf{\Phi }}_{t,i}}\right)$ and 
$\textrm{Span}\left({\bf F}_{r,j}\right) \subset \textrm{Span}\left({{\bf{\Phi }}_{r,j}}\right)$). Hence, the primary target is satisfied, when we develop the transmit and receive RF beamformers by utilizing the corresponding steering vector with the quantized angle-pairs as $\big(\lambda_{t,i}^{\left(x\right),k},\lambda_{t,i}^{\left(y\right),n}\big)\in{\rm{AoD}}_i$ 
and
$\big(\lambda_{r,i}^{\left(x\right),k},\lambda_{r,i}^{\left(y\right),n}\big)\in{\rm{AoA}}_i$, respectively.

On the other hand, when the transmit RF beamformer is built via the transmit steering vector ${\bf {e}}_{t,i}\left(\cdot,\cdot\right)$ with any quantized angle-pairs satisfying both $\big(\lambda_{t,i}^{\left(x\right),k},\lambda_{t,i}^{\left(y\right),n}\big)\in{\rm{AoD}}_i$ 
and 
$\big(\lambda_{t,i}^{\left(x\right),k},\lambda_{t,i}^{\left(y\right),n}\big)\notin{\rm{AoD}}_{\textrm{SI},i}$,
one can show that each element of the multiplication given in \eqref{eq_steering_vector_vs_SI}
asymptotically converges to 0 for large transmit URAs as in the massive MIMO systems \cite[eq. (16)]{ASIL_ABHP_Access}. 
Similarly, this behavior is also valid for the receive RF beamformer design.
Thus, the approximate zero condition given in \eqref{eq_HSI_new_eff} can be satisfied by choosing the quantized angle-pairs outside of the AoD/AoA support of SI channel, which also ensures the secondary target.
In other words, 
as the antenna array size increases,
the corresponding quantized angle-pairs assures both
${\textrm{Span}\left({\bf F}_{t,j}\right) \subset \textrm{Null}\left({{\bf{\Phi }}_{\textrm{SI},t,j}}\right)}$
and 
${\textrm{Span}\left({\bf F}_{r,j}\right) \subset \textrm{Null}\left({{\bf{\Phi }}_{\textrm{SI},r,j}}\right)}$, which are earlier defined as the necessary conditions for the secondary target. 

In order to meet the primary and secondary targets, a joint design of transmit/receive RF beamformers is required based on the AoD/AoA support of the intended and SI channels. For the transmit RF beamformer at the node $i$, the quantized angle-pairs covering the AoD support of the intended channel ${\bf H}_i$ given in \eqref{eq_AoD_Supp} and excluding the AoD support of the SI channel ${\bf H}_{\textrm{SI},i}$ given in \eqref{eq_AoD_Supp_SI} are found as:
\addtocounter{equation}{+1}
\begin{equation}\label{eq_TxAnglePairs}
	\big(
	\lambda_{t,i}^{\left(x\right),k},
	\lambda_{t,i}^{\left(y\right),n}
	\big)
	\hspace{0.5ex}\left| \begin{array}{l}
		{\gamma _x} \in {\boldsymbol{\lambda }}_{t,i}^{\left(x\right),k},
		{\gamma _y} \in {\boldsymbol{\lambda}}_{t,i}^{\left(y\right),n}, \\ \\
		\left( {{\gamma _x},{\gamma _y}} \right) \hspace{-0.5ex}\in\hspace{-0.5ex} {\rm{Ao}}{{\rm{D}}_i},
		\left( {{\gamma _x},{\gamma _y}} \right) \hspace{-0.5ex}\notin\hspace{-0.5ex} {\rm{Ao}}{{\rm{D}}_{\textrm{SI},i}},
	\end{array} \right.
\end{equation}
where 
${\boldsymbol{\lambda }}_{t,i}^{\left(x\right),k}\hspace{-2ex}=\hspace{-2ex}
\big[{
	\lambda_{t,i}^{\left(x\right),k}
	\hspace{1ex}-\hspace{1ex}
	\frac{1}{{{M_{t,i}^{\left(x\right)}}}},
	\lambda_{t,i}^{\left(x\right),k}
	\hspace{1ex}+\hspace{1ex}
	\frac{1}{{{M_{t,i}^{\left(x\right)}}}}} 
\big]$ 
and 
${{\boldsymbol{\lambda }}_{t,i}^{\left(y\right),n}\hspace{1ex}=\hspace{1ex}
	\big[{
		\lambda_{t,i}^{\left(y\right),n}
		-
		\frac{1}{{{M_{t,i}^{\left(y\right)}}}},
		\lambda_{t,i}^{\left(y\right),n}
		+
		\frac{1}{{{M_{t,i}^{\left(y\right)}}}}} 
	\big]}$ 
represent the boundaries of $\lambda_{t,i}^{\left(x\right),k}$ and $\lambda_{t,i}^{\left(y\right),n}$, respectively.
Here, \eqref{eq_TxAnglePairs} implies that if any $\left(\gamma_x,\gamma_y\right)$ angle-pairs inside the boundaries of 
$\lambda_{t,i}^{\left(x\right),k}$ 
and 
$\lambda_{t,i}^{\left(y\right),n}$
is both within ${\rm{AoD}}_i$ and outside of ${\rm{AoD}}_{\textrm{SI},i}$, then the corresponding quantized angle-pair $\big(
\lambda_{t,i}^{\left(x\right),k},
\lambda_{t,i}^{\left(y\right),n}
\big)$ is utilized for the transmit RF beamformer design.
We consider that there are $N_{t,i}$ quantized angle-pairs satisfying \eqref{eq_TxAnglePairs}, which also represents the number of orthogonal-beams and the number of RF chains at the transmitter of the node $i$ (please see Fig. \ref{fig_1_SystemModel}). Finally, by using \eqref{eq_lambda} and \eqref{eq_TxAnglePairs}, the transmit RF beamformer at the node $i$ is obtained as follows:
\begin{equation}\label{eq_TX_RF}
	{{\bf{F}}_{t,i}}\hspace{-0.5ex}=\hspace{-0.5ex} \Big[\hspace{-0.15ex} {
		{\bf{e}}_{t,i}\hspace{-0.25ex}\big(
		\lambda_{t,i}^{\hspace{-0.35ex}\left(x\right),k_1}\hspace{-0.5ex},
		\lambda_{t,i}^{\hspace{-0.35ex}\left(y\right),n_1}
		\big), 
		\hspace{-0.5ex}\cdots\hspace{-0.5ex}, 	
		{\bf{e}}_{t,i}\big(
		\lambda_{t,i}^{\hspace{-0.5ex}\left(x\right),k_{N_{t,i}}}\hspace{-0.5ex},
		\lambda_{t,i}^{\hspace{-0.5ex}\left(y\right),n_{N_{t,i}}}
		\big)} \hspace{-0.35ex}\Big]\hspace{-0.25ex}.
\end{equation}
Similarly, regarding the receive RF beamformer at the node $j$, the quantized angle-pairs covering the AoA support of the intended channel ${\bf H}_i$ given in \eqref{eq_AoA_Supp} and excluding the AoA support of the SI channel ${\bf H}_{\textrm{SI},j}$ given in \eqref{eq_AoA_Supp_SI} are found as:
\begin{equation}\label{eq_RxAnglePairs}
	\big(
	\lambda_{r,j}^{\left(x\right),k},
	\lambda_{r,j}^{\left(y\right),n}
	\big)
	\hspace{0.5ex}\left| \begin{array}{l}
	{\gamma _x} \in {\boldsymbol{\lambda }}_{r,j}^{\left(x\right),k},
	{\gamma _y} \in {\boldsymbol{\lambda }}_{r,j}^{\left(y\right),n},
	\\ \\
	\left( {{\gamma _x},{\gamma _y}} \right) \hspace{-0.5ex}\in\hspace{-0.5ex} {\rm{Ao}}{{\rm{A}}_j},
	\left( {{\gamma _x},{\gamma _y}} \right) \hspace{-0.5ex}\notin\hspace{-0.5ex} {\rm{Ao}}{{\rm{A}}_{\textrm{SI},j}},
	\end{array} \right.
\end{equation}
where 
${\boldsymbol{\lambda }}_{r,i}^{\left(x\right),k}\hspace{-2ex}=\hspace{-2ex}
\big[{
	\lambda_{r,i}^{\left(x\right),k}
	\hspace{1ex}-\hspace{1ex}
	\frac{1}{{{M_{r,i}^{\left(x\right)}}}},
	\lambda_{r,i}^{\left(x\right),k}
	\hspace{1ex}+\hspace{1ex}
	\frac{1}{{{M_{r,i}^{\left(x\right)}}}}} 
\big]$ 
 	and
${{\boldsymbol{\lambda }}_{r,i}^{\left(y\right),n}\hspace{1ex}=\hspace{1ex}
	\big[{
		\lambda_{r,i}^{\left(y\right),n}
		-
		\frac{1}{{{M_{r,i}^{\left(y\right)}}}},
		\lambda_{r,i}^{\left(y\right),n}
		+
		\frac{1}{{{M_{r,i}^{\left(y\right)}}}}} 
	\big]}$ 
indicate the boundaries of $\lambda_{r,i}^{\left(x\right),k}$
and $\lambda_{t,i}^{\left(y\right),n}$, respectively.
By assuming $N_{r,j}$ quantized angle-pairs satisfying \eqref{eq_RxAnglePairs}, the receive RF beamformer at the node $j$ is derived as:
\begin{equation}\label{eq_RX_RF}
	{{\bf{F}}_{\hspace{-.5ex}r,j}}\hspace{-0.5ex}=\hspace{-0.75ex} \Big[\hspace{-0.25ex} {
		{\bf{e}}_{r,j}\hspace{-0.25ex}\big(\hspace{-0.25ex}
		\lambda_{r,j}^{\hspace{-0.35ex}\left(x\right)\hspace{-0.15ex},k_1}\hspace{-0.5ex},
		\lambda_{r,j}^{\hspace{-0.35ex}\left(y\right)\hspace{-0.15ex},n_1}
		\big)\hspace{-0.25ex}, 
		\hspace{-0.5ex}\cdots\hspace{-0.5ex}, \hspace{-0.25ex}	
		{\bf{e}}_{r,j}\hspace{-0.25ex}\big(\hspace{-0.25ex}
		\lambda_{r,j}^{\hspace{-0.5ex}\left(x\right),k_{N_{r,j}}}\hspace{-0.5ex},
		\lambda_{r,j}^{\hspace{-0.5ex}\left(y\right),n_{N_{r,j}}}
		\big)} \hspace{-0.35ex}\Big]^{\hspace{-0.25ex}T}\hspace{-0.5ex}.
\end{equation}
The RF-stage design  for the proposed AB-JHPC technique is summarized in Algorithm \ref{alg_RF}. It is important to mention that the transmit and receive RF beamformers developed in this section follow the CM property given in \eqref{eq_9c}, which makes it possible to design the RF-stage via the low-cost phase-shifters as illustrated in Fig. \ref{fig_1_SystemModel}.
By means of the quantized angle-pairs defined in \eqref{eq_lambda}, the transmit and receive RF beamformers at the node $i$ can be constructed via the phase-shifters with $\log_2\left(M_{t,i}\right)$ and $\log_2\left(M_{r,i}\right)$ bit resolution, respectively.
Furthermore, as expressed in Algorithm \ref{alg_RF}, the RF-stage design does not require the fast time-varying instantaneous CSI and it is only based on the slow time-varying AoD/AoA information. Particularly, the design of transmit (receive) RF beamformer requires only four angular parameters, which are the mean of elevation and azimuth AoD (AoA) and their spread\footnote{Standard angle estimation techniques requiring online channel sounding can be utilized to track the slow time-varying AoD/AoA parameters \cite{3D_Beamforming_ElevAzim,AoD_Est_2_Decades}. Recently, an offline channel estimation technique is proposed in \cite{ASIL_Xiaoyi_Channel_Estimation}, where the AoD/AoA parameters are tracked via the geospatial data, fuzzy c-means algorithm and ray tracing.}. Also, ${\bf F}_{t,i}$ and ${\bf F}_{r,i}^H$ are tall unitary matrices (i.e., ${\bf F}_{t,i}^H{\bf F}_{t,i}\hspace{-0.25ex}=\hspace{-0.25ex}{\bf I}_{N_{t,i}}$, ${\bf F}_{r,i}{\bf F}_{r,i}^H\hspace{-0.25ex}=\hspace{-0.25ex}{\bf I}_{N_{r,i}}$).

\begin{algorithm*}[!t]
	\caption{RF-stage design}
	\label{alg_RF}
	\begin{algorithmic}[1]
		\renewcommand{\algorithmicrequire}{\bf{Input:}}
		\renewcommand{\algorithmicensure}{\bf{Output:}}
		\Require $M_{t,i}$, $M_{r,i}$,
		$\big\lbrace
		\theta _{t,c}^{\left(i\right)}, 
		\psi _{t,c}^{\left(i\right)}, 
		\delta _{t,c}^{\theta,\left(i\right)},	
		\delta _{t,c}^{\psi,\left(i\right)},
		\big\rbrace$,
		$\big\lbrace
		\theta _{r,c}^{\left(i\right)}, 
		\psi   _{r,c}^{\left(i\right)}, 
		\delta _{r,c}^{\theta,\left(i\right)},
		\delta _{r,c}^{\psi,\left(i\right)}
		\big\rbrace$,
		for $i\hspace{-0.25ex}=\hspace{-0.25ex}1\hspace{-0.35ex}:\hspace{-0.35ex}2$ and $c\hspace{-0.25ex}=\hspace{-0.25ex}1\hspace{-0.35ex}:\hspace{-0.35ex}C$.
		\State Define the node index as $j \in \left\{ {1,2} \right\}$, $j \ne i$.
		\State {\bf{for}} {$i=1:2$}, {\bf{do}}
		\State ~~~Build ${\rm{AoD}}_i$ for the intended channel ${\bf H}_i$ via \eqref{eq_AoD_Supp}.
		\State ~~~Build ${\rm{AoA}}_i$ for the intended channel ${\bf H}_j$ via \eqref{eq_AoA_Supp}.
		\State ~~~Build ${\rm{AoD}}_{\textrm{SI},i}$ and ${\rm{AoA}}_{\textrm{SI},i}$ for the SI channel ${\bf H}_{\textrm{SI},i}$ via \eqref{eq_AoD_AoA_SI}.
		\State ~~~Construct the quantized angle-pairs $\big( 
		{\lambda _{t,i}^{\left(x\right),k}}
		{\lambda _{t,i}^{\left(x\right),n}}	 
		\big)$ and 
		$\big( 
		{\lambda _{r,i}^{\left(x\right),k}}
		{\lambda _{r,i}^{\left(x\right),n}}	 
		\big)$ 
		given in \eqref{eq_lambda}.
		\State ~~~Find $	\big(
		\lambda_{t,i}^{\left(x\right),k},
		\lambda_{t,i}^{\left(y\right),n}
		\big)$ pairs covering ${\rm{AoD}}_i$ and excluding ${\rm{AoD}}_{\textrm{SI},i}$ as in  \eqref{eq_TxAnglePairs}.
		\State ~~~Generate the transmit RF beamformer ${{\bf{F}}_{t,i}}$ via \eqref{eq_TX_RF}.
		\State ~~~Find $	\big(
		\lambda_{r,i}^{\left(x\right),k},
		\lambda_{r,i}^{\left(y\right),n}
		\big)$ pairs covering ${\rm{AoA}}_i$ and excluding ${\rm{AoA}}_{\textrm{SI},i}$ as in  \eqref{eq_RxAnglePairs}.
		\State ~~~Generate the receive RF beamformer ${{\bf{F}}_{r,i}}$ via \eqref{eq_RX_RF}.
		\State {\bf{end for}}
		\Ensure ${\bf{F}}_{t,1}$, ${\bf{F}}_{t,2}$, ${\bf{F}}_{r,1}$, ${\bf{F}}_{r,2}$
	\end{algorithmic} 
\end{algorithm*}

\vspace{-0.5ex}
\subsection{BB-Stage Design}

According to the proposed RF-stage design, the reduced-size effective intended channel $\bm{\mathcal{H}}_i$ is utilized for the BB-stage design. In contrast to \cite{Satyanarayana2019,FD_MIMO_MU_Learning}, the proposed BB-stage design does not require the instantaneous SI channel knowledge because the perfect estimation of SI channel could be impractical for the full-duplex communications\footnote{
	According to the received signal expression given in \eqref{eq_r_init}, when the full-size or reduced-size perfect SI channel knowledge (i.e., ${\bf H}_{\textrm{SI},j}$ or $\bm{\mathcal{H}}_{\textrm{SI},j}$) is available at the node $j$, any of passive antenna isolation, digital or analog SIC techniques is not necessary for the full-duplex communications. 
	Because the transmit RF beamformer ${\bf F}_{t,j}$, the BB precoder ${\bf B}_{t,j}$ and the data signal ${\bf d}_j$ is already available at the node $j$ in addition to the assumption of the perfect SI channel ${\bf H}_{\textrm{SI},j}$. Then, we could simply subtract the SI component ${\bf H}_{\textrm{SI},j}{\bf F}_{t,j}{\bf B}_{t,j}{\bf d}_j$ from the received signal given in \eqref{eq_r_init}, which could imply to doubling the capacity via the full-duplex communications with respect to the conventional half-duplex communications without any SIC techniques.
	However, the SI channel must be estimated through the pilot signals in practice, therefore, there exists the estimation errors due to the noise as well as the hardware imperfections \cite{FD_Survey,FD_Survey_Secrecy,FD_Survey_2019,AHMED_FD}.
	Thus, it motivates the development of advanced SIC techniques in the literature \cite{FD_Survey_Secrecy}.
}.
Here, we first develop the BB precoder and BB combiner via the well-known singular value decomposition (SVD).	
Then, we also propose a novel semi-blind minimum mean square error (S-MMSE) algorithm to further enhance the quality of SIC provided by the RF-stage. 

Let the SVD of $\bm{\mathcal{H}}_i$ be defined as
$\bm{\mathcal{H}}_i \hspace{-0.5ex}=\hspace{-0.5ex} {\bf U}_j{\bf \Sigma}_i{\bf V}_i^H$,
where 
${\bf U}_j\hspace{-0.5ex}\in\hspace{-0.5ex} \mathbb{C}^{N_{r,j}\times \textrm{rank}\left(\bm{\mathcal{H}}_i\right)}$
is a tall unitary matrix,
${\bf \Sigma}_n= \textrm{diag}\big(\sigma_{n,1}^2, \hspace{-0.5ex}\cdots\hspace{-0.5ex},\sigma_{n,\textrm{rank}\left(\bm{\mathcal{H}}_i\right)}^2\big)\in \mathbb{R}^{\textrm{rank}\left(\bm{\mathcal{H}}_i\right)\times \textrm{rank}\left(\bm{\mathcal{H}}_i\right)}$
is a diagonal matrix with the singular values in the decreasing order,
${\bf V}_i\in \mathbb{C}^{N_{t,i}\times \textrm{rank}\left(\bm{\mathcal{H}}\right)}$
is a tall unitary matrix. 
Assuming $\textrm{rank}\left(\bm{\mathcal{H}_i}\right)\ge S_i$, we can partition ${\bf V}_i$ and ${\bf U}_j$ as 
${\bf V}_i = \left[{\bf V}_{i,1}, {\bf V}_{i,2}\right]$ with ${\bf V}_{i,1}\in \mathbb{C}^{N_{t,i}\times S_i}$.
and
${\bf U}_j = \left[{\bf U}_{j,1}, {\bf U}_{j,2}\right]$ with ${\bf U}_{j,1}\in \mathbb{C}^{N_{r,j}\times S_i}$, respectively
According to the well-known SVD approach \cite{SU_mMIMO_Dynamic_SubArray,OMP_Full_CSI,SU_mMIMO_OFDM}, the BB precoder at the node $i$ and the BB combiner at the node $j$ are respectively derived as follows:
\begin{IEEEeqnarray}{rCl}\label{eq_BB_SVD}
	\IEEEyesnumber
	\IEEEyessubnumber*
	{\bf B}_{t,i}^{\textrm{SVD}} &=& {\bf V}_{i,1}{\bf P}_i^\frac{1}{2} \in\mathbb{C}^{N_{t,i}\times S_i},\label{eq_BB_PC_SVD}\\
	{\bf B}_{r,j}^{\textrm{SVD}} &=& {\bf U}_{j,1}^H \in\mathbb{C}^{S_i \times N_{r,j}},\label{eq_BB_CB_SVD}
\end{IEEEeqnarray}
where ${\bf P}_i\hspace{-0.5ex}=\hspace{-0.5ex}\textrm{diag}\left(P_{i,1},\hspace{-0.25ex}\cdots\hspace{-0.25ex},P_{i,S_i}\right)\hspace{-0.5ex}\in\hspace{-0.5ex}\mathbb{R}^{S_i \times S_i}$ is a diagonal power allocation matrix with $\sum\nolimits_{n=1}^{S_i}P_{i,n}\le {P_T}$ for the transmit power constraint given in \eqref{eq_9d}. By using the well-known water-filling technique \cite{SU_mMIMO_OFDM}, the allocated powers are obtained as:
\begin{equation}\label{eq_WF_PA}
	{P_{i,n}} \hspace{-0.5ex}=\hspace{-0.5ex} {\left( {\mu  - \frac{\sigma_w^2}{{P_T \sigma _n^2}}} \right)^ + },{\textrm{ s.t.}}\sum\limits_{n = 1}^{S_i} {\left( {\mu  - \frac{\sigma_w^2}{{P_T \sigma _n^2}}} \right)^ + }  \hspace{-0.5ex}=\hspace{-0.25ex} {P_T}.
\end{equation}
For maximizing the power of intended signal and mitigating the power of ISI expressed in \eqref{eq_r_k}, the BB precoder ${\bf{B}}_{t,i}^{\textrm{SVD}}$ is optimal as proven in \cite{SU_mMIMO_OFDM,SU_mMIMO_Dynamic_SubArray}.
	However, the BB combiner ${\bf{B}}_{r,j}^{\textrm{SVD}}$ is not optimal because it does not consider the effect of SI and noise present at the receiver node.
Hence, we aim to develop the optimal BB combiner by minimizing the mean square error (MSE). 
By using \eqref{eq_r}, the MSE expression at the node $j$ is written as follows:
\begin{equation}\label{eq_MSE}
	\begin{aligned}
		{{\rm{MSE}}}_j&=  \mathbb{E}\big\lbrace {{{\big\| {\tilde{\bf{r}}_j - {\bf{d}}_i} \big\|}_2^2}} \big\rbrace\\
		&\mathop  = \limits^{\left( a \right)} {\textrm{tr}}\big(\mathbb{E} {\big\{ {
				\tilde{\bf{r}}_j{{\tilde {\bf{r}}}_j^H} 
				- {\bf{d}}_i{{\tilde {\bf{r}}}_j^H} 
				- \tilde {\bf{r}}_j{{\bf{d}}_i^H} 
				+ {\bf{d}}_i{{\bf{d}}_i^H}} \big\}} 
		\big)\\
		&\mathop  = \limits^{\left( b \right)}
		\textrm{tr}\Big(\hspace{-0.5ex}  
		{{\bf{B}}_{r,j}}\hspace{-0.25ex}\big[ {{{\bm{\mathcal{H}}}_i}{{\bf{B}}_{t,i}}{\bf{B}}_{t,i}^H{\bm{\mathcal{H}}}_i^H \hspace{-0.5ex}+\hspace{-0.25ex} {{\bf{W}}_j} \hspace{-0.5ex}+\hspace{-0.25ex} \sigma _w^2{{\bf{I}}_{{N_{r,j}}}}} \big]{\bf{B}}_{r,j}^H
		\\
		&\Big. - {{\bf{B}}_{r,j}}{{\bm{\mathcal{H}}}_i}{{\bf{B}}_{t,i}} - {\bf{B}}_{t,i}^H{\bm{\mathcal{H}}}_i^H{\bf{B}}_{r,j}^H + {{\bf{I}}_{{S_i}}}
		\Big),\vspace{-1ex}
	\end{aligned}
\end{equation}
where ($a$) is the direct consequence of the linearity property of trace operator, ($b$) applies the unitary proper of the transmit RF beamformer (i.e., ${\bf F}_{t,i}^H{\bf F}_{t,i}\hspace{-0.25ex}=\hspace{-0.25ex}{\bf I}_{N_{t,i}}$) and $\mathbb{E}\left\{{\bf d}_i{\bf d}_i^H\right\}={\bf I}_{S_i}$.
Here, 
${{\bf{W}}_j} = \mathbb{E}\big\{ 
\bm{\mathcal H}_{{\textrm{SI}},j}{\bf{B}}_{t,j}			
{\bf{B}}_{t,j}^H\bm{\mathcal H}_{{\textrm{SI}},j}^H
\big\}$ 
is a function of the effective SI channel $\bm{\mathcal H}_{{\textrm{SI}},j}$. Nonetheless, the minimization problem for the current form of $\textrm{MSE}_j$ depends on the effective SI channel $\bm{\mathcal H}_{{\textrm{SI}},j}$,
which is not available at the node $j$.
According to the SI channel model given in \eqref{eq_HSI}, the SI channel has two parts indicating the LoS paths (i.e., the near-field SI channel) and NLoS paths (i.e., far-field SI channel). After applying the antenna separation based SIC, the NLoS paths become more dominant compared to the LoS paths (i.e., $\bm{\mathcal{H}}_{\textrm{SI},j}={{\bf{F}}_{r,j}{\bf{H}}_{\textrm{SI},j}{\bf{F}}_{t,j}}\approx {\bf{F}}_{r,j}{\bf H}_{\textrm{NLoS},j}{\bf{F}}_{t,j}$).
Hence, in order to remove the dependency of the instantaneous effective SI channel, we derive an approximation of ${{\bf{W}}_j}$ as:
\begin{equation}\label{eq_W_approx}
	\begin{aligned}
		{{\bf{W}}_j} 
		&= \mathbb{E}\Big\{ 
		\bm{\mathcal H}_{{\textrm{SI}},j}{\bf{B}}_{t,j}			
		{\bf{B}}_{t,j}^H\bm{\mathcal H}_{{\textrm{SI}},j}^H
		\Big\} \\
		&\mathop  \approx \limits^{\left( a \right)} \mathbb{E}\Big\{ 
		{\bf{F}}_{r,j}
		{\bf{H}}_{{\textrm{LoS}},j}
		{\bf{F}}_{t,j}
		{\bf{B}}_{t,j}
		{\bf{B}}_{t,j}^H
		{\bf{F}}_{t,j}^H
		{\bf{H}}_{{\textrm{LoS}},j}^H
		{\bf{F}}_{r,j}^H
		\Big\}\\
		&\mathop  = \limits^{\left( b \right)} \mathbb{E}\Big\{ 
		{\bf{F}}_{r,j}
		{{\bf{\Phi }}_{{\textrm{SI}},j,r}}
		{{\bf{G}}_{{\textrm{SI}},j}}
		{{\bf{\Phi }}_{{\textrm{SI}},j,t}}
		{\bf{F}}_{t,j}
		{\bf{B}}_{t,j}\Big.
		\\
		&\Big.
		\times
		{\bf{B}}_{t,j}^H
		{\bf{F}}_{t,j}^H
		{\bf{\Phi }}_{{\textrm{SI}},j,t}^H
		{\bf{G}}_{{\textrm{SI}},j}^H
		{\bf{\Phi }}_{{\textrm{SI}},j,r}^H
		{\bf{F}}_{r,j}^H
		\Big\}\\
		&\mathop  \approx \limits^{\left( c \right)}
		\frac{1}{\hat{\tau}_{\textrm{SI},j}^{2\eta}L_{\textrm{SI}}}
		{\bf{F}}_{r,j}
		{\hat{\bf{\Phi }}_{{\textrm{SI}},j,r}}
		{\hat{\bf{\Phi }}_{{\textrm{SI}},j,t}}
		{\bf{F}}_{t,j}
		{\bf{B}}_{t,j}\\
		&\times
		{\bf{B}}_{t,j}^H
		{\bf{F}}_{t,j}^H
		\hat{\bf{\Phi }}_{{\textrm{SI}},j,t}^H
		\hat{\bf{\Phi }}_{{\textrm{SI}},j,r}^H
		{\bf{F}}_{r,j}^H\\
		&\mathop  = \limits^{\left( d \right)}\hat{\bf W}_j,
	\end{aligned}
\end{equation}
where 
($a$) is obtained via the assumption of dominant NLoS paths after the antenna separation based SIC, 
($b$) is written by using \eqref{eq_H_NLoS},
($c$) is approximated by applying the expectation operator on the fast time-varying complex path gain matrix ${\bf G}_{\textrm{SI},j}$,
($d$) represents the approximation of ${\bf W}_j$ as $\hat{\bf W}_j$.
Here, $\hat{\tau}_{SI,j}$ is the average distance for the NLoS paths, $\hat{\bf{\Phi }}_{{\textrm{SI}},j,t}$ and $\hat{\bf{\Phi }}_{{\textrm{SI}},j,r}$ are the approximated transmit and receive phase response matrices, respectively.
It is worthwhile to remark that the approximated phase response matrices does not require the knowledge of exact AoD/AoA pairs. 
Instead, they require the knowledge of AoD/AoA support of the SI channel defined in \eqref{eq_AoD_AoA_SI} as in the RF-stage design (please see Algorithm \ref{alg_RF}). Therefore, we generate $\hat{L}_{SI}$ random AoD/AoA pairs inside the AoD/AoA support. Then, we build the approximated phase response matrices by using  \eqref{eq_Phase_Matrix} and the corresponding AoD/AoA pairs.
By replacing $\hat{\bf W}_j$ with ${\bf W}_j$ in \eqref{eq_MSE}, the MSE expression at the node $j$ can be also approximated as follows:
\begin{equation}\label{eq_MSE_approx}
	\begin{aligned}
		\hat{{\rm{MSE}}}_j &\hspace{-0.5ex}=\hspace{-0.5ex}
		\textrm{tr}\Big(  \hspace{-0.25ex}
		{{\bf{B}}_{r,j}}\big[ {{{\bm{\mathcal{H}}}_i}{{\bf{B}}_{t,i}}{\bf{B}}_{t,i}^H{\bm{\mathcal{H}}}_i^H \hspace{-0.25ex}+\hspace{-0.25ex} {\hat{\bf{W}}_j} \hspace{-0.25ex}+\hspace{-0.25ex} \sigma _w^2{{\bf{I}}_{{N_{r,j}}}}} \big]{\bf{B}}_{r,j}^H
		\\
		&\Big. \hspace{-0.5ex}-\hspace{-0.5ex} {{\bf{B}}_{r,j}}{{\bm{\mathcal{H}}}_i}{{\bf{B}}_{t,i}} - {\bf{B}}_{t,i}^H{\bm{\mathcal{H}}}_i^H{\bf{B}}_{r,j}^H + {{\bf{I}}_{{S_i}}}
		\Big).
	\end{aligned}
\end{equation}
Now, in addition to the effective intended channel $\bm{\mathcal{H}}_i$, the approximated $\hat{\textrm{MSE}}_j$ only depends the AoD/AoA support of SI channel.
In other words, it is blind to the effective SI channel $\bm{\mathcal{H}}_{\textrm{SI},j}$.
Therefore, the optimization problem for minimizing $\hat{\textrm{MSE}}_j$ is classified as semi-blind minimum mean square error (S-MMSE).
According to 
the expressed transmit RF beamformer ${\bf F}_{t,i}$ in \eqref{eq_TX_RF},
receive RF beamformer ${\bf F}_{r,j}$ in \eqref{eq_RX_RF}, and
BB precoder ${\bf B}_{t,i}$  in \eqref{eq_BB_PC_SVD},
the BB combiner problem for the proposed S-MMSE algorithm is defined as:
\begin{equation}\label{eq_SMMSE_Problem}
	{\bf B}_{r,j}^{\textrm{S-MMSE}}=\arg \min_{\left\{{\bf B}_{r,j}\right\}}  \hat{{\rm{MSE}}}_j.
\end{equation}
In order to find the optimal solution \eqref{eq_SMMSE_Problem}, we apply the differentiation rules for the complex-valid matrices. Then, the derivative of $\hat{\textrm{MSE}}_j$ with respect to ${\bf B}_{r,j}$ is obtained as:
\begin{equation}
	\begin{aligned}
		\frac{{\partial \hat{{\rm{MSE}}}_j}}{{\partial {{\bf{B}}_{r,j}}}} 
		&=2{{\bf{B}}_{r,j}}\big( {{{\bm{\mathcal{H}}}_i}{{\bf{B}}_{t,i}}{\bf{B}}_{t,i}^H{\bm{\mathcal{H}}}_i^H + {\hat{\bf{W}}_j} + \sigma _w^2{{\bf{I}}_{{N_{r,j}}}}} \big)\\
		&-  2{{\bf{B}}_{t,i}^H}{{\bm{\mathcal{H}}}_i^H}.
	\end{aligned}
\end{equation}
Finally, we derive the BB combiner ${\bf B}_{r,j}$ satisfying the S-MMSE criterion defined in \eqref{eq_SMMSE_Problem} (i.e., $\frac{{\partial \hat{{\rm{MSE}}}_j}}{{\partial {{\bf{B}}_{r,j}}}} =0$) as:
\begin{equation}\label{eq_SMMSE_Solution}
	{\bf B}_{r,j}^{\textrm{S-MMSE}}
	\hspace{-0.25ex}=\hspace{-0.25ex}
	{{\bf{B}}_{t,i}^H}
	{{\bm{\mathcal{H}}}_i^H}\hspace{-0.25ex}
	\big(\hspace{-0.25ex}
	{{\bm{\mathcal{H}}}_i}
	{{\bf{B}}_{t,i}}
	{\bf{B}}_{t,i}^H
	{\bm{\mathcal{H}}}_i^H 
	\hspace{-0.25ex}+ \hspace{-0.25ex}
	{\hat{\bf{W}}_j} 
	\hspace{-0.25ex}+ \hspace{-0.25ex}
	\sigma _w^2
	{{\bf{I}}_{{N_{r,j}}}} 
	\big)^{\hspace{-0.25ex}-1}.
\end{equation}
Algorithm \ref{alg_BB} outlines the BB-stage design for the proposed AB-JHPC technique. 
\begin{algorithm*}[!t]
	\caption{BB-stage design}
	\label{alg_BB}
	\begin{algorithmic}[1]
		\renewcommand{\algorithmicrequire}{\bf{Input:}}
		\renewcommand{\algorithmicensure}{\bf{Output:}}
		\Require 
		$\bm{\mathcal{H}}_i$,
		${\bf F}_{t,i}$,
		${\bf F}_{r,i}$,
		$\big\lbrace
		\theta _{t,c}^{\left(i\right)}, 
		\psi _{t,c}^{\left(i\right)}, 
		\delta _{t,c}^{\theta,\left(i\right)},	
		\delta _{t,c}^{\psi,\left(i\right)},
		\big\rbrace$,
		$\big\lbrace
		\theta _{r,c}^{\left(i\right)}, 
		\psi   _{r,c}^{\left(i\right)}, 
		\delta _{r,c}^{\theta,\left(i\right)},
		\delta _{r,c}^{\psi,\left(i\right)}
		\big\rbrace$,
		for $i\hspace{-0.25ex}=\hspace{-0.25ex}1\hspace{-0.35ex}:\hspace{-0.35ex}2$ and $c\hspace{-0.25ex}=\hspace{-0.25ex}1\hspace{-0.35ex}:\hspace{-0.35ex}C$.
		\State \textbf{for} $i=1:2$ \textbf{do}
		\State ~~~Perform SVD as $\bm{\mathcal{H}}_i \hspace{-0.5ex}=\hspace{-0.5ex} {\bf U}_j{\bf \Sigma}_i{\bf V}_i^H$ with 
		${\bf V}_i = \left[{\bf V}_{i,1}, {\bf V}_{i,2}\right]$.		
		and 
		${\bf U}_i = \left[{\bf U}_{i,1}, {\bf U}_{i,2}\right]$.
		\State ~~~Calculate the power allocation matrix ${\bf P}_i$ via \eqref{eq_WF_PA}.
		\State ~~~Generate SVD-based BB precoder ${\bf B}_{t,i}$ via \eqref{eq_BB_PC_SVD}.
		\State ~~~{\bf{if}} SVD-based BB combiner, {\bf{then}}
		\State ~~~~~~Generate BB combiner ${\bf B}_{r,i}$ via \eqref{eq_BB_CB_SVD}.
		\State {~~~\bf{else if}} S-MMSE-based BB combiner, {\bf then}
		\State ~~~~~~Find $\hat{\bf W}_j$ via \eqref{eq_W_approx}.
		\State ~~~~~~Form $\hat{\bf{\Phi }}_{{\textrm{SI}},j,t}$ and $\hat{\bf{\Phi }}_{{\textrm{SI}},j,r}$ via \eqref{eq_Phase_Matrix} with random AoD/AoA pairs according to \eqref{eq_AoD_AoA_SI}.
		\State ~~~~~~Generate BB combiner ${\bf B}_{r,i}$ via \eqref{eq_SMMSE_Solution}.
		\State ~~~{\bf{end if}}
		\State {\bf{end for}}
		\Ensure ${\bf{B}}_{t,1}$, ${\bf{B}}_{t,2}$, ${\bf{B}}_{r,1}$, ${\bf{B}}_{r,2}$
	\end{algorithmic} 
\end{algorithm*}	
Ultimately,  the RF-stage design explained in Algorithm \ref{alg_RF} requires only the slow time-varying AoD/AoA parameters of the intended and SI channels.
Then, the BB-stage design needs only the reduced-size effective intended channel $\bm{\mathcal{H}}_{t,i}$ in addition to the AoD/AoA parameters as seen in Algorithm \ref{alg_BB}. 
Therefore, the proposed AB-JHPC technique reduces the CSI overhead size from $M_{t,i}\times M_{r,j}$ to $N_{t,i}\times N_{r,j}$ in comparison to the conventional FDPC.


\section{RF Chain Reduction with Transfer Block Architecture}\label{sec_RF_Reduction}
\begin{figure}[!t]
	\centering
	\subfigure[]
	{\includegraphics[width=\columnwidth]{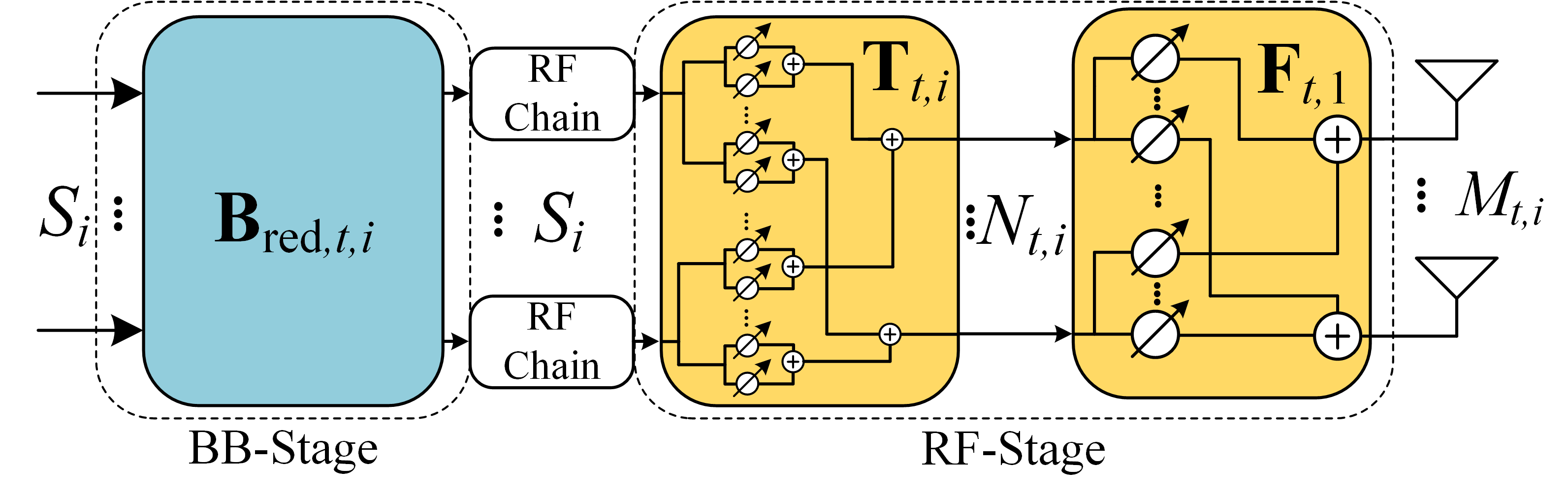}
		\label{fig_3a_HP_Transfer}}
	\subfigure[]
	{\includegraphics[width=\columnwidth]{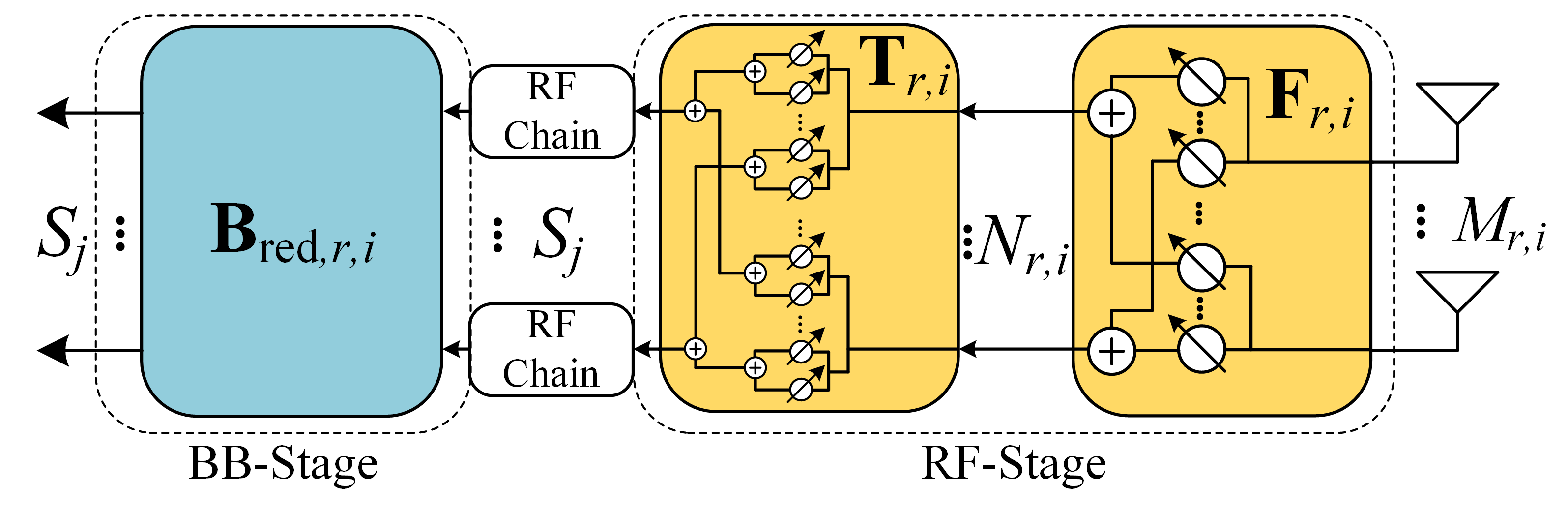}
		\label{fig_3b_HC_Transfer}}
	\vspace{-2ex}
	\caption{Transfer block architecture at node $i$: (a) hybrid precoding with $S_i$ RF chains, (b) hybrid combining with $S_j$ RF chains.}
	\label{fig_3_TransferBlock} 
\end{figure}

As shown in Fig. \ref{fig_1_SystemModel}, the proposed AB-JHPC technique utilizes $N_{t,i}+N_{r,i}$ RF chains at the node $i$ for exploiting all degrees of freedom provided by the intended channel while canceling the strong SI channel.
Thus, in comparison to the conventional FDPC, the number of RF chains is decreased from $M_{t,i}+M_{r,j}$ to $N_{t,i}+N_{r,j}$, where $M_{t,i}+M_{r,j}$ is a significantly large value for the massive MIMO systems (i.e., $N_{t,i}\ll M_{t,i}$ and $N_{r,i}\ll M_{r,i}$).
In order to further reduce the hardware cost/complexity and power consumption, we employ a transfer block architecture as demonstrated in Fig. \ref{fig_3_TransferBlock}, where the number of power-hungry RF chains is reduced to $S_i+S_j$ as the minimum value for simultaneously transmitting $S_i$ data streams and receiving $S_j$ data streams.
For the hybrid precoding shown in Fig. \ref{fig_3a_HP_Transfer}, the transmit transfer block ${\bf T}_{t,i}\in\mathbb{C}^{N_{t,i}\times S_i}$ is placed between the transmit RF beamformer ${\bf F}_{t,i}\in\mathbb{C}^{M_{t,i}\times N_{t,i}}$ and the reduced-size BB precoder ${\bf B}_{\textrm{red},t,i}\in\mathbb{C}^{S_i \times S_i}$.
Similarly, the hybrid combining design shown in Fig. \ref{fig_3b_HC_Transfer} utilizes the receive transfer block ${\bf T}_{r,i}\in\mathbb{C}^{S_j\times N_{r,i}}$ for connecting the receive RF beamformer ${\bf F}_{r,i}\in\mathbb{C}^{N_{r,i}\times M_{r,i}}$ and the reduced-size BB combiner ${\bf B}_{\textrm{red},r,i}\in\mathbb{C}^{S_j \times S_j}$.
It is seen that each input of ${\bf T}_{t,i}$ and ${\bf T}_{r,i}$ is passed through two-pair of phase-shifters and their summation is transferred to the corresponding output.
Therefore, the modulus of each element in ${\bf T}_{t,i}$ and ${\bf T}_{r,i}$ varies between $0$ and $2$.

When the BB precoder ${\bf B}_{t,i}$ is obtained by using \eqref{eq_BB_PC_SVD}, the optimization problem for the transmit transfer block ${\bf T}_{t,i}$ and the reduced-size BB precoder ${\bf B}_{\textrm{red},t,i}$ is written as:
\begin{equation}\label{eq_T_HP_Problem}
	\begin{aligned}
		\left[{\bf T}_{t,i},{\bf B}_{\textrm{red},t,i}\right]
		&=
		\arg\min_{\left\{{\bf T},{\bf B}_{\textrm{red}}\right\}}  \left\|{\bf B}_{t,i} - {\bf T} {\bf B}_{\textrm{red}}\right\|_F^2,\\
		\rm{s.t.}&~|{\bf{T}}\left(m,n\right)|\in\left[0,2\right],\forall m,n\\
		&~{\rm{tr}}\left({\bf B}_\textrm{red}^H{\bf T}^H{\bf T}{\bf B}_\textrm{red}\right) \le P_T,\\
	\end{aligned}
\end{equation}
where the constraints refer the modulus of the transmit transfer block and the maximum transmit power at the transmitter, respectively.
By using \cite[Lemma 1]{ASIL_ABHP_Access}, the optimal closed-form solutions are derived as follows:
\begin{equation}\label{eq_T_HP_Solution}
	\begin{aligned}
		{\bf T}_{t,i}\left(m,n\right)
		&=
		e^{j\angle {\bf B}_{t,i}\left(m,n\right)}\hspace{-0.5ex}
		\left[
		e^{j\beta_{t,i}^{\left(m,n\right)}}
		+
		e^{-j\beta_{t,i}^{\left(m,n\right)}}
		\right],\\
		{\bf B}_{\textrm{red},t,i}&=
		\left({\bf T}_{t,i}^H{\bf T}_{t,i}\right)^{-1}{\bf T}_{t,i}^H{\bf B}_{t,i},
	\end{aligned}
\end{equation}
where 
$\beta_{t,i}^{\left(m,n\right)}=\cos^{-1} \left(\frac{|{\bf B}_{t,i}\left(m,n\right)|}{\max_{u,v} |{\bf B}_{t,i}\left(u,v\right)|}\right)$.
Here, it is worthwhile to remark that the derived closed-form solutions guarantees the optimality by satisfying ${\bf B}_{t,i} = {\bf T}_{t,i} {\bf B}_{\textrm{red},t,i}$.

Similarly, after acquiring the BB combiner ${\bf B}_{r,i}$ via \eqref{eq_BB_CB_SVD} for the SVD approach or via \eqref{eq_SMMSE_Solution} for the proposed S-MMSE algorithm, the optimization problem for the receive transfer block architecture is defined as:
\begin{table*}
	\caption{AB-JHPC vs. FDPC: CSI Overhead Size, RF Chains, Phase-Shifters.}
	\vspace{-1ex}
	\label{table_RF_Chain_CSI}
	\footnotesize 
	\centering
	\renewcommand{\arraystretch}{1.4}
	\begin{tabular}{|c||c|c|c|}
		\hline
		& CSI Overhead Size & \# of RF Chains & \# of Phase Shifters \\ \hline\hline
		FDPC&
		$\sum\limits_{i=1}^{2}\sum\limits_{j\ne i}M_{r,j}\times M_{t,i}$&
		$\sum\limits_{i=1}^{2}M_{t,i}+M_{r,i}$&
		-\\ \hline
		\begin{tabular}[c]{@{}c@{}}AB-JHPC w/out \\ Transfer Block\end{tabular}&
		$\sum\limits_{i=1}^{2}\sum\limits_{j\ne i}N_{r,j}\times N_{t,i}$&
		$\sum\limits_{i=1}^{2}N_{t,i}+N_{r,i}$&
		$\sum\limits_{i=1}^{2}N_{t,i}\times M_{t,i} + N_{r,i}\times M_{r,i}$                      \\ \hline
		\begin{tabular}[c]{@{}c@{}}AB-JHPC using \\ Transfer Block\end{tabular}&
		$\sum\limits_{i=1}^{2}\sum\limits_{j\ne i}N_{r,j}\times N_{t,i}$                     &
		$\sum\limits_{i=1}^{2}2S_i$                 & 
		$\sum\limits_{i=1}^{2}\sum\limits_{j\ne i}N_{t,i}\times \left(M_{t,i}+2S_i\right) + N_{r,i}\times \left(M_{r,i} + 2S_j\right)$                      \\ \hline
	\end{tabular}
\end{table*}
\begin{table*}[!t]
	\caption{Simulation parameters.}
	\vspace{-1ex}
	\label{table_setup}
	\centering
	\renewcommand{\arraystretch}{1.5}
	\begin{tabular}{|l|c|}
		\hline
		\# of antennas\cite{Report_5G_UMi_UMa_Rel16} & $M_{t,i}=M_{r,i}=16\times 16 = 256$      \\ \hline
		Data Streams & $S_1=S_2=4$\\ \hline
		{Path loss exponent} \cite{Report_5G_Macro_PL_Rel_16}& $\eta=3.76$      \\ \hline
		{Noise PSD} \cite{Report_5G_Macro_PL_Rel_16}& $-174$ dBm/Hz      \\ \hline
		{Channel bandwidth} \cite{Report_5G_Macro_PL_Rel_16}& $10$ MHz      \\ \hline
		Antenna spacing (in wavelength)\cite{Report_5G_UMi_UMa_Rel16} & ${d\hspace{-0.25ex}=0.5}$\\ \hline
		{\# of network realizations} & $2.000$      \\\hline\hline
		${\bf H}_i$: Distance for paths& $\tau_{i}\in\left[35\rm{m},50\rm{m}\right]$\\ \hline
		${\bf H}_i$: \# of paths\cite{Report_5G_UMi_UMa_Rel16} & $L_i=20$\\ \hline
		${\bf H}_i$: Mean azimuth AoD/AoA& 
		$
		\psi_{t}^{\left(1\right)}=315^\circ,
		\psi_{t}^{\left(2\right)}=355^\circ~\left|\right|~
		\psi_{r}^{\left(1\right)}=205^\circ,
		\psi_{r}^{\left(2\right)}=245^\circ
		$\\ \hline
		${\bf H}_i$: Mean elevation AoD/AoA & 
		$
		\theta_{t}^{\left(i\right)}=
		\theta_{r}^{\left(i\right)}=40^\circ
		$\\ \hline
		${\bf H}_i$: Azimuth/elevation spread\cite{Report_5G_UMi_UMa_Rel16}  &
		$
		{\delta _{t}^{\psi,\left(i\right)}}=
		{\delta _{r}^{\psi,\left(i\right)}}=		
		{\delta _{t}^{\theta,\left(i\right)}}=		
		{\delta _{r}^{\theta,\left(i\right)}}=10^\circ
		$\\ \hline\hline
		${\bf H}_{\textrm{NLoS},i}$: Distance for paths & $\tau_{\textrm{SI},i}\in\left[5\rm{m},15\rm{m}\right]$\\ \hline
		${\bf H}_{\textrm{NLoS},i}$: \# of paths\cite{Report_5G_UMi_UMa_Rel16} & $L_{\textrm{SI}}=20$\\ \hline
		${\bf H}_{\textrm{NLoS},i}$: Mean azimuth AoD/AoA & 
		$
		\psi_{\textrm{SI},t}^{\left(i\right)}=150^\circ~\left|\right|~
		\psi_{\textrm{SI},r}^{\left(i\right)}=75^\circ
		$\\ \hline
		${\bf H}_{\textrm{NLoS},i}$: Mean elevation AoD/AoA& 
		$
		\theta_{\textrm{SI},t}^{\left(i\right)}=
		\theta_{\textrm{SI},r}^{\left(i\right)}=40^\circ
		$\\ \hline
		${\bf H}_{\textrm{NLoS},i}$: Azimuth/elevation spread\cite{Report_5G_UMi_UMa_Rel16}  &
		$
		{\delta _{\textrm{SI},t}^{\psi,\left(i\right)}}=
		{\delta _{\textrm{SI},r}^{\psi,\left(i\right)}}=		
		{\delta _{\textrm{SI},t}^{\theta,\left(i\right)}}=		
		{\delta _{\textrm{SI},r}^{\theta,\left(i\right)}}=10^\circ
		$\\ \hline
	\end{tabular}
\end{table*}
\begin{equation}\label{eq_T_HC_Problem}
	\begin{aligned}
		\left[{\bf B}_{\textrm{red},r,i},{\bf T}_{r,i}\right]
		&=
		\arg\min_{\left\{{\bf B}_{\textrm{red}},{\bf T} \right\}}  \left\|{\bf B}_{r,i} -  {\bf B}_{\textrm{red}}{\bf T}\right\|_F^2,\\	
		\rm{s.t.}&~
		|{\bf{T}}\left(m,n\right)|\in\left[0,2\right],\forall m,n,
	\end{aligned}
\end{equation}
By using \eqref{eq_T_HP_Problem} and \eqref{eq_T_HP_Solution}, we obtain the optimal closed-form solutions as follows:
\begin{equation}\label{eq_T_HC_Solution}
	\begin{aligned}
		{\bf T}_{r,i}\left(m,n\right)
		&=
		e^{j\angle {\bf B}_{r,i}\left(m,n\right)}\hspace{-0.5ex}
		\left[
		e^{j\beta_{r,i}^{\left(m,n\right)}}
		+
		e^{-j\beta_{r,i}^{\left(m,n\right)}}
		\right],\\
		{\bf B}_{\textrm{red},r,i}&=
		{\bf B}_{r,i}
		{\bf T}_{r,i}^H
		\left({\bf T}_{r,i}{\bf T}_{r,i}^H\right)^{-1},
	\end{aligned}
\end{equation}
where $\beta_{r,i}^{\left(m,n\right)} =\cos^{-1} \left(\frac{|{\bf B}_{r,i}\left(m,n\right)|}{\max_{u,v} |{\bf B}_{r,i}\left(u,v\right)|}\right)$. 
The obtained closed-form solutions also maintain the optimality by satisfying optimal solutions to \eqref{eq_T_HC_Problem} (i.e., ${\bf B}_{r,i} =  {\bf B}_{\textrm{red},r,i}{\bf T}_{r,i}$).
Therefore, the optimal solutions derived in \eqref{eq_T_HP_Solution} and \eqref{eq_T_HC_Solution} enable the proposed transfer block architecture to minimize the RF chain utilization without penalizing the achievable rate performance defined in \eqref{eq_Rate_j}.
Moreover, the reduced number of power-hungry RF chains enhances the energy efficiency, which is illustrated in Section \ref{sec_Illustrative}.

Finally, Table \ref{table_RF_Chain_CSI} 
summarizes the CSI overhead size, the number of RF chains and the number of phase-shifters requirements for the proposed full-duplex mmWave massive MIMO systems. 
We consider the proposed AB-JHPC technique without the transfer block as shown in Fig. \ref{fig_1_SystemModel} and with the transfer block as shown in Fig. \ref{fig_3_TransferBlock}.
Additionally, as a benchmark, the CSI overhead size and the number of RF chains for the conventional FDPC is given in Table \ref{table_RF_Chain_CSI} \cite{Mass_MIMO_Hybrid_Survey}.

\section{Illustrative Results} \label{sec_Illustrative}

In this section, we evaluate the performance of the proposed AB-JHPC technique for the full-duplex mmWave massive MIMO systems. 
Based on the recent 3GPP Release 16 specifications in \cite{Report_5G_UMi_UMa_Rel16,Report_5G_Macro_PL_Rel_16}, Table \ref{table_setup} summarizes the numerical values of the parameters used in the simulation setup, unless otherwise stated. 
Additionally, we assume that the transmit and receive URAs at each node are placed on the same surface with the separation by two wavelengths (i.e., $D_1=2$, $D_2=0$, $\Theta=0^\circ$ as shown in Fig. \ref{fig_2_TxRxAntennas}).
Here, we first demonstrate how the transmit/receive RF beamformers are generated according to the given AoD/AoA parameters.
Then, the amount of SIC achieved by the proposed AB-JHPC design is presented.
Afterwards, the total achievable-rate and energy efficiency results are provided to compare the advantages of the proposed full-duplex mmWave massive MIMO systems compared to its half-duplex counterparts.

\vspace{-1ex}
\subsection{RF-Stage Design via AoD/AoA Parameters}

\begin{figure}[!t]
	\centering
	\includegraphics[width=\columnwidth]{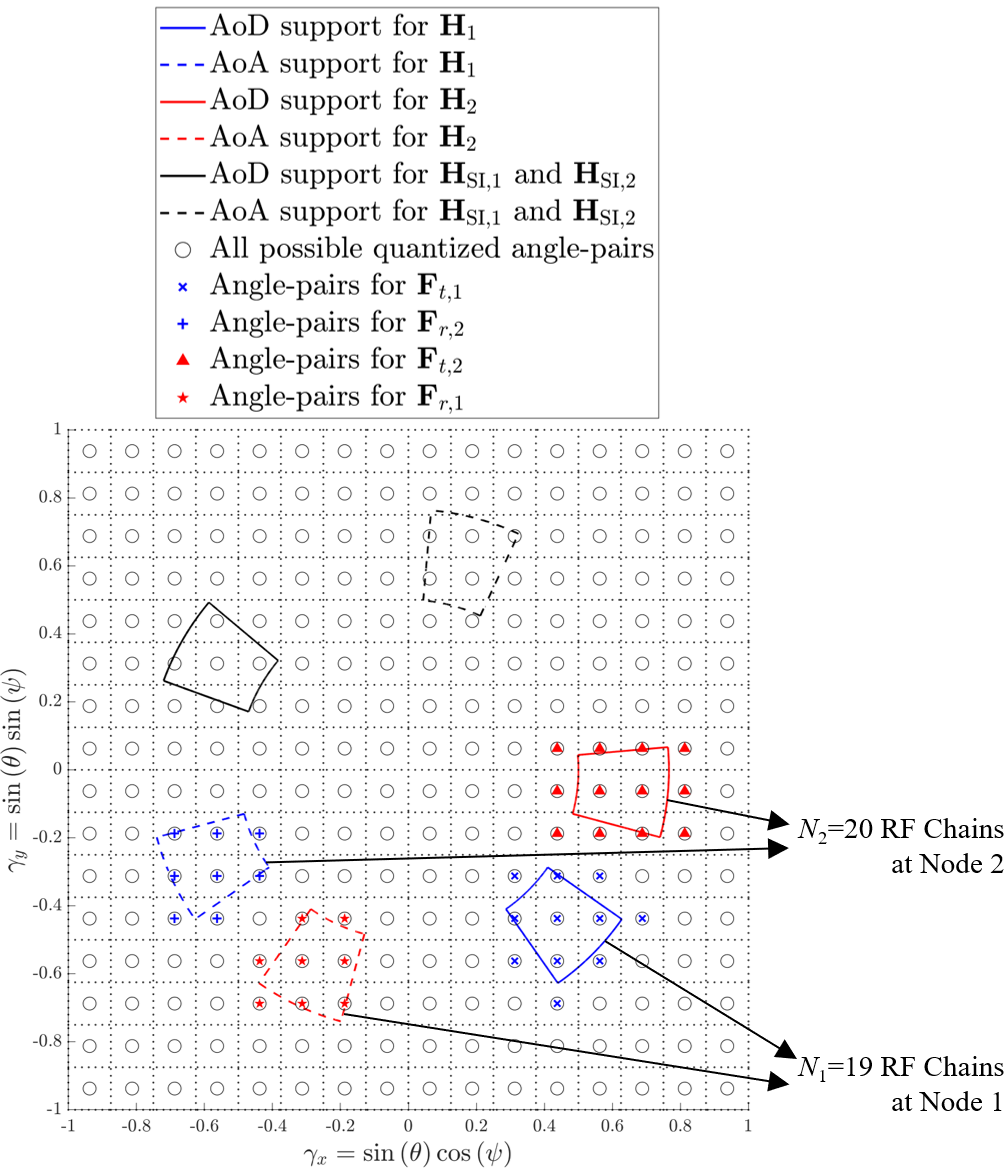}
	\caption{RF-stage design and AoD/AoA support for the intended and SI channels (${\bf H}_1$, ${\bf H}_2$, ${\bf H}_{\textrm{SI},1}$, ${\bf H}_{\textrm{SI},2}$).}
	\label{fig_4_RF_Design_AoA_AoD}
	\vspace{-2ex}
\end{figure}

Fig. \ref{fig_4_RF_Design_AoA_AoD} demonstrates the AoD/AoA supports for the intended channels (i.e., ${\bf H}_1$ and ${\bf H}_2$) and the SI channels (i.e., ${\bf H}_{\textrm{SI},1}$ and ${\bf H}_{\textrm{SI},2}$) based on the simulation parameters given in Table \ref{table_setup}, where we assume a single scattering cluster with $L_i=L_{\textrm{SI}}=20$ paths \cite{Report_5G_UMi_UMa_Rel16}.
Here, it is worthwhile to recall that $\gamma_x=\sin\left(\theta\right)\cos\left(\psi\right)$ and $\gamma_y=\sin\left(\theta\right)\sin\left(\psi\right)$ are the coefficients reflecting both elevation angle ($\theta$) and azimuth angle ($\psi$).
As summarized in Algorithm \ref{alg_RF}, the RF-stage design starts by building the AoD/AoA supports for each channel via \eqref{eq_AoD_AoA} and \eqref{eq_AoD_AoA_SI}. Then, we construct the quantized angle-pairs by using the transmit and receive antenna elements at each node. Considering transmit/receive URA size as $16\times16=256$ \cite{Report_5G_UMi_UMa_Rel16}, there are $256$ possible quantized angle-pairs represented via black circles in Fig. \ref{fig_4_RF_Design_AoA_AoD}.
Afterwards, we find the quantized angle-pairs covering the corresponding AoD/AoA support of the intended channel and excluding that of the SI channel.
For instance, $N_{t,1}=11$ quantized angle-pairs represented via blue cross symbols
are selected for the transmit RF beamformer ${\bf F}_{t,1}$, which are inside the AoD support of ${\bf H}_1$ and also orthogonal to the AoD support of ${\bf H}_{\textrm{SI},1}$.
Thus, in this simulation setup, the transmit RF beamformer ${\bf F}_{t,1}$ generates $N_{t,1}=11$ orthogonal-beams for sending data streams to the node $2$ and suppressing the SI signal power occurred at the node $1$.
Similarly, the receive RF beamformer ${\bf F}_{r,1}$ is generated by using $N_{r,1}=8$ quantized angle-pairs illustrated with red star symbols, which covers the AoA support of ${\bf H}_2$ and excludes that of ${\bf H}_{\textrm{SI},1}$.
Finally, the proposed hybrid architecture shown in ${\textrm{Fig. \ref{fig_1_SystemModel}}}$ requires only $N_1=N_{t,1}+N_{r,1}=19$ RF chains to develop transmit/receive RF beamformers at the node $1$.
Therefore, the hybrid architecture reduces the number of RF chains by $96.2\%$ compared to its fully-digital counterparts.
Furthermore, the number of RF chains can be further reduced via the propsoed transfer block architecture expressed in Section \ref{sec_RF_Reduction} (please see Table \ref{table_RF_Chain_CSI}).

\vspace{-1ex}
\subsection{Self-Interference Cancellation}
\begin{figure}[!t]
	\centering
	\includegraphics[width=\columnwidth]{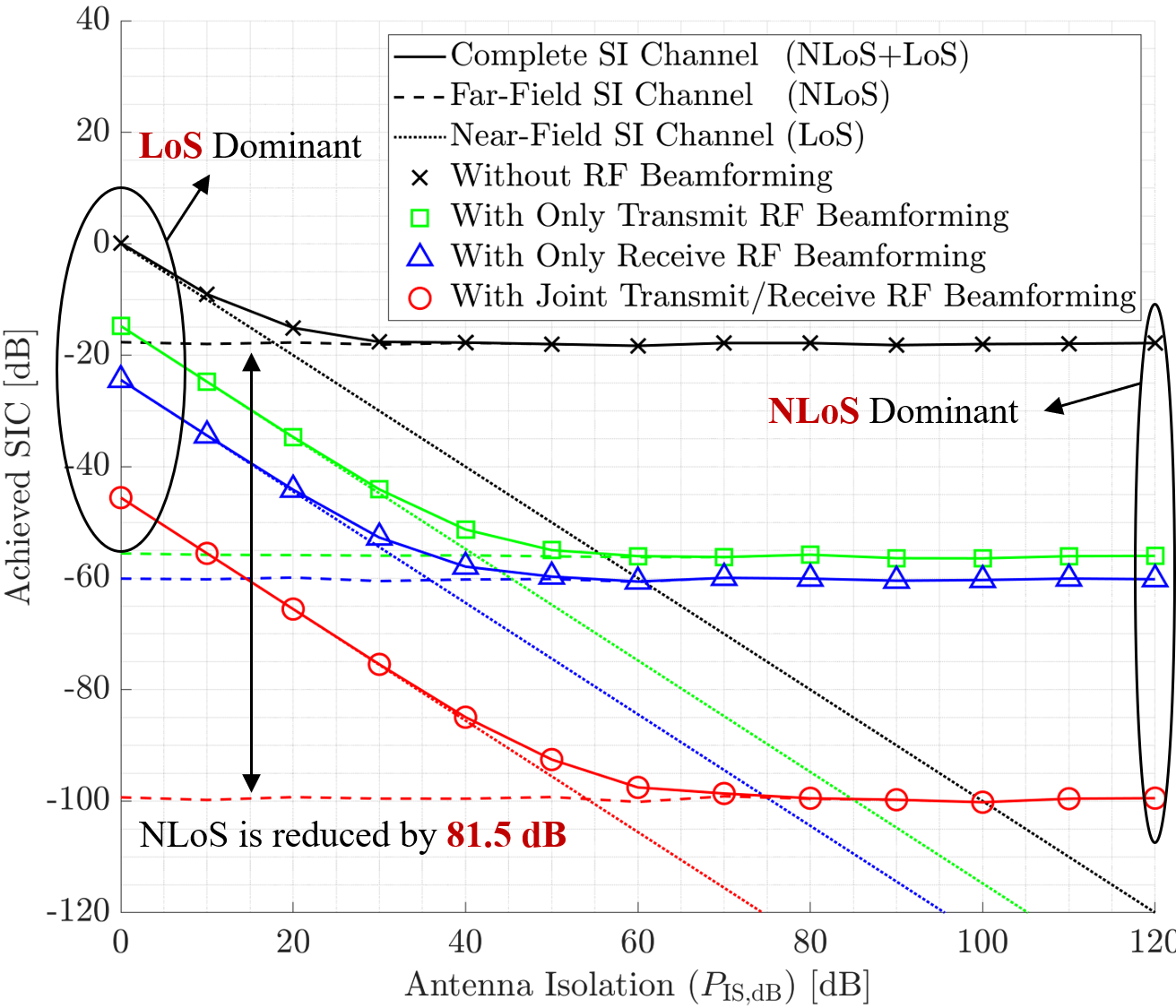}
	\caption{Self-interference cancellation (SIC) achieved by the joint transmit/receive RF beamforming versus antenna isolation.
	}
	\vspace{-2ex}
	\label{fig_5_RF_BF_SIC}
\end{figure}

After designing the RF-stage, Fig. \ref{fig_5_RF_BF_SIC} demonstrates the SIC achieved by the joint transmit/receive RF beamforming, where we consider that the power of near-field SI channel reflecting the strong LoS paths is reduced by 
$P_{\textrm{IS,dB}}\in\left[0\textrm{~dB}, 120\textrm{~dB}\right]$ via antenna isolation. 
According to the SI occurred at the node $2$, we compare 
the far-field SI  channel as ${\bf{H}}_{\textrm{NLoS},2}$, 
the near-field SI channel as ${\bf{H}}_{\textrm{LoS},2}$ and 
complete SI channel as 
${\bf{H}}_{\textrm{SI},2}
=
{\bf{H}}_{\textrm{LoS},2}
+
{\bf{H}}_{\textrm{NLoS},2}$
in terms of their powers\footnote{For the sake of simplicity, Fig. \ref{fig_5_RF_BF_SIC} only plots the SI channel powers at the node $2$. Equivalently, the same behaviors are also observed for the SI channel at the node $1$.}.
As expressed in Section \ref{sec_Channel}, the power of near-field SI channel is inversely proportional to the antenna isolation (i.e., $10\log_{10}\left(\mathbb{E}\big\{\|{{\bf{H}}_{\textrm{LoS},i}}\|_F^2\big\}\right)=-P_{\textrm{IS,dB}}$).
In order to precisely display the benefits of the proposed joint transmit/receive RF beamforming, the average SI channel power at the node $2$ is plotted for four cases: 
(i) without RF beamforming as $\mathbb{E}\left\{\left\|{\bf H}_{\textrm{SI},2}\right\|_F^2\right\}$
(ii) with only transmit RF beamforming as $\mathbb{E}\left\{\left\|{\bf H}_{\textrm{SI},2}{\bf F}_{t,2}\right\|_F^2\right\}$
(iii) with only receive RF beamforming as $\mathbb{E}\left\{\left\|{\bf F}_{r,2}{\bf H}_{\textrm{SI},2}\right\|_F^2\right\}$
(iv) with joint transmit/receive RF beamforming as $\mathbb{E}\left\{\left\|\bm{\mathcal{H}}_{\textrm{SI},2}\right\|_F^2\right\}=\mathbb{E}\left\{\left\|{{\bf{F}}_{r,2}{\bf{H}}_{\textrm{SI},2}{\bf{F}}_{t,2}}\right\|_F^2\right\}$.
\begin{itemize}
	\item 	\textbf{Case 1:} We observe that when $P_{\textrm{IS,dB}}$ is very poor (e.g., $P_{\textrm{IS,dB}}<15$ dB), the near-field SI channel represented with black dotted curve is dominant as expected. On the other hand, when $P_{\textrm{IS,dB}}$ is higher than $20$ dB, the far-field SI channel represented with black dashed curve becomes dominant. Moreover, even though the quality of SIC is enhanced via more advanced antenna isolation techniques, the complete SI channel remains constant due to the dominant NLoS paths. Because the power of far-field SI channel expressed in \eqref{eq_H_NLoS} is independent from $P_{\textrm{IS,dB}}$ and it is a function of the distance, path loss exponent and complex path gains. 
	Therefore, the complete SI channel power without RF beamforming is only reduced from $0$ dB to $-18$ dB, when $P_{\textrm{IS,dB}}$ is greater than 30 dB.	
	It implies that $18$ dB SIC is accomplished at most by only applying passive antenna isolation, which does not aim to suppress the far-field SI channel.		
	This behavior is also independently observed in the experiment results in \cite{FD_AntennaSep_74dB}, where the authors measure the SIC achieved by antenna isolation as $74$ dB in the anechoic chamber as a low-reflection environment. However, the measurements inside a reflective room in \cite{FD_AntennaSep_74dB} shows that the SIC achieved by antenna isolation is reduced to $44$ dB due to the dominant NLoS paths.
	
	\item 	\textbf{Cases 2 \& 3:} 	We apply only transmit (receive) RF beamforming for reducing the power of far-field SI channel and enhancing the quality of SIC. The numerical results show that the power of far-field SI channel is reduced from $-18$ dB to $-56$ dB ($-60$  dB) as represented with green (blue) dashed curves. Furthermore, when there is no antenna isolation (i.e., $P_{\textrm{IS,dB}}=0$ dB), only transmit (receive) RF beamforming reduces the power of near-field SI signal from $0$ dB to $-14$ dB ($-24$ dB). 
	Eventually, when $P_{\textrm{IS,dB}}\ge 50$ dB SIC is satisfied via antenna isolation, the power of the complete SI channel is $-56$ dB (${-60~\textrm{dB}}$), which means an additional $38$ dB ($42$ dB) reduction via only transmit (receive) RF beamforming compared to the first case without RF beamforming.
	\item \textbf{Case 4:} The proposed joint transmit/receive RF beamformer design in Section \ref{sec_AB_JHPC} further suppresses the complete SI channel power compared to the previous three cases as illustrated in Fig. \ref{fig_5_RF_BF_SIC}. The first important observation is that the far-field and near-field SI channel powers are reduced by $81.5$ dB and $45.5$ dB, respectively. In comparison to Cases 2 \& 3, the joint transmit/receive RF beamformer design achieves higher suppression than the summation of only transmit RF beamforming and only receive RF beamforming (i.e., $81.5 \textrm{ dB}> 38 \textrm{ dB} + 42 \textrm{ dB}$ for NLoS paths and $45.5 \textrm{ dB}> 14 \textrm{ dB} + 24 \textrm{ dB}$ for LoS paths).
	Moreover, when $P_{\textrm{IS,dB}}\le 50$ dB, there is a near-linear relationship between the complete SI channel power and $P_{\textrm{IS,dB}}$.
	Thus, according to the proposed joint transmit/receive RF beamforming, the complete SI channel power in this simulation setup can be modeled as follows:
	\begin{equation}
		\begin{aligned}
			&
			\resizebox{0.8\hsize}{!}{$
			\hspace{-1.5ex}10\hspace{-.25ex}\log_{10}\hspace{-0.75ex}\left(\hspace{-0.5ex}\mathbb{E}\hspace{-.5ex}\left\{\hspace{-.5ex}\left\|\hspace{-.15ex}\bm{\mathcal{H}}_{\textrm{SI},2}\hspace{-.25ex}\right\|_{\hspace{-.25ex}F}^{\hspace{-.25ex}2}\hspace{-.5ex}\right\}\hspace{-.5ex}\right)
			\hspace{-0.75ex}=\hspace{-0.75ex}
			10\hspace{-.25ex}\log_{10}\hspace{-.75ex}\left(\hspace{-.5ex}\mathbb{E}\hspace{-.5ex}\left\{\hspace{-.5ex}\left\|\hspace{-.15ex}{{\bf{F}}_{\hspace{-.35ex}r,2}{\bf{H}}_{\textrm{SI},2}{\bf{F}}_{\hspace{-.35ex}t,2}}\hspace{-.25ex}\right\|_{\hspace{-.25ex}F}^{\hspace{-.25ex}2}\hspace{-.5ex}\right\}\hspace{-.5ex}\right)\hspace{-1.5ex}
			$}
			\\
			&\hspace{4ex}\approx
			-P_{\textrm{IS,dB}}-45.5, ~\forall P_{\textrm{IS,dB}}\in\left[0\textrm{~dB},50\textrm{~dB}\right].
		\end{aligned}
	\end{equation}
	The above equation affirms the importance of utilizing both joint transmit/receive RF beamforming and advanced antenna isolation techniques in order to cancel the strong SI channel. In other words, they works well together. For example, in the case 1 (without RF beamforming), the residual SI channel power is saturated at $-18$ dB, even if we achieve ${74\textrm{~dB}}$ antenna isolation as in \cite{FD_AntennaSep_74dB}. On the other hand, in the case 4 (joint transmit/receive RF beamforming), we have $45.5$ dB SIC, when the antenna isolation is omitted (i.e., ${P_{\textrm{IS,dB}}=0\textrm{~dB}}$). However, when we combine both of the techniques, the complete SI channel power can be suppressed up to $99.5$ dB as shown in Fig. \ref{fig_5_RF_BF_SIC}.
\end{itemize}

\begin{figure}[!t]
	\centering
	\includegraphics[width=\columnwidth]{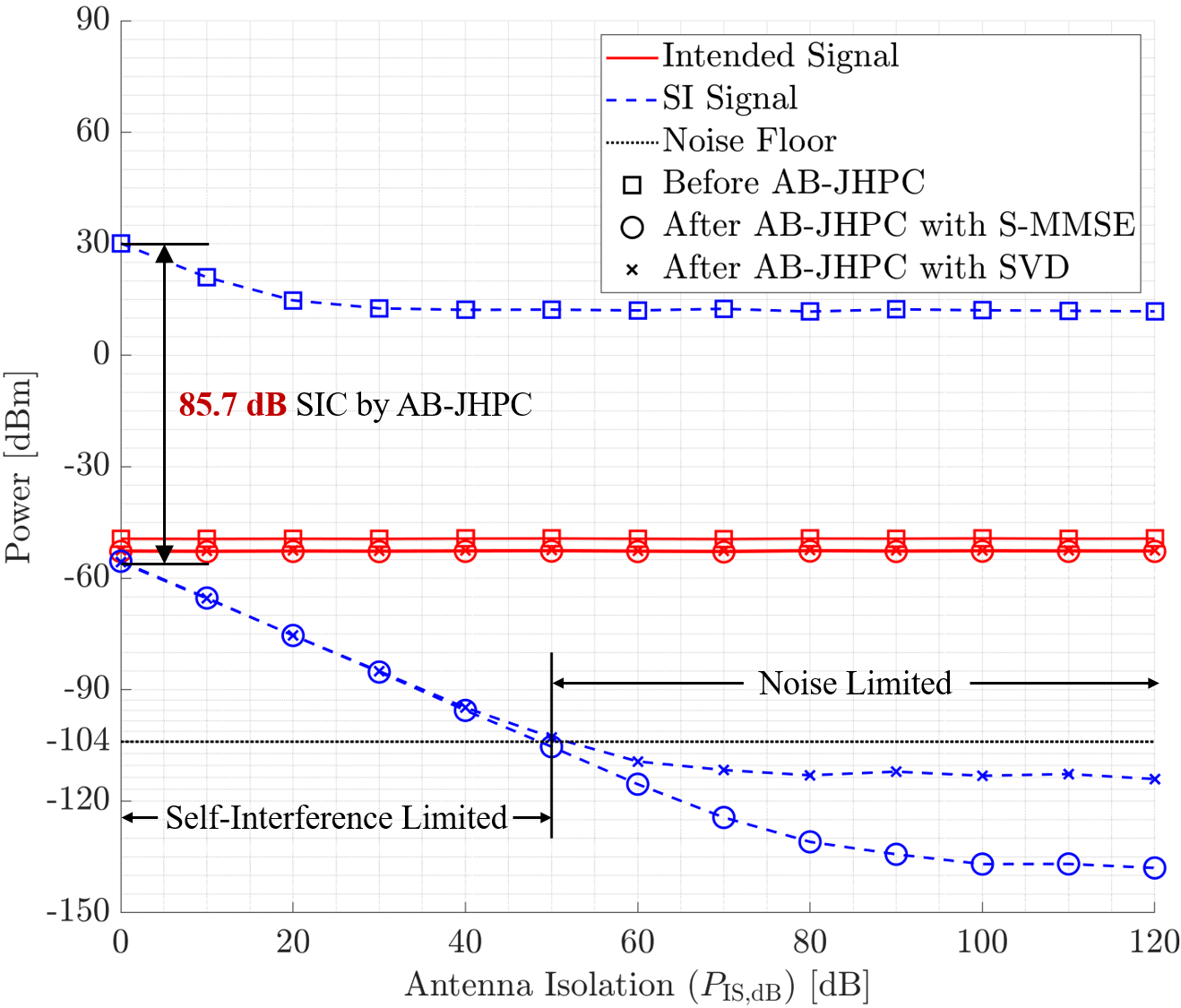}
	\caption{Self-interference cancellation (SIC) achieved by the proposed AB-JHPC technique versus antenna isolation, where the transmit power is $P_T=30$ dBm.
	}
	\label{fig_6_SIC_by_HPC}
\end{figure}

Fig. \ref{fig_6_SIC_by_HPC} analyzes the SIC achieved by the proposed AB-JHPC technique, where we consider the effects of both RF-stage and BB-stage design. The transmit power is chosen as ${P_T=30}$ dBm for the simultaneous transmission of $S_1=S_2=4$ data streams at both nodes.
By using the combined signal expression given in \eqref{eq_r_k}, Fig. \ref{fig_6_SIC_by_HPC} plots the powers of intended signal, SI signal and noise for the first data stream at the node 2\footnote{The power of ISI given in \eqref{eq_r_k} is not plotted in Fig. \ref{fig_6_SIC_by_HPC} because it is well mitigated via the BB-stage design. Hence, ISI is not a limiting factor for the full-duplex mmWave massive MIMO systems compared to the noise and strong SI.}. 
According to the given noise power spectral density (PSD) as $-174$ dBm/Hz and the channel bandwidth as $10$ MHz \cite{Report_5G_Macro_PL_Rel_16}, the noise power is calculated as $\sigma_w^2=-104$ dBm. Also, the power of modified noise given in \eqref{eq_r_k} is kept as $\sigma_w^2=-104$ dBm by normalizing the BB combiner\footnote{By using
	${{{\bf{F}}_{r,2}}{\bf{F}}_{r,2}^H = {{\bf{I}}_{{N_{r,2}}}}}$ and ${{\bf{w}}_2} \sim \mathcal{CN} \big( 0 , \sigma_w^2{\bf{I}}_{M_{r,2}}\big)$, the power of modified noise in \eqref{eq_r_k} is calculated as $\mathbb{E}\big\{{\big| {{\bf{b}}_{r,2,k}^H{{\bf{F}}_{r,2}}{{\bf{w}}_2}} \big|^2} \big\}=\sigma_w^2{\bf{b}}_{r,2,k}^H{{\bf{b}}_{r,2,k}}$.
	For keeping the modified noise power as $\sigma_w^2$, the $k^{th}$ row of BB combiner is normalized as ${{{\bar{\bf b}}}_{r,2,k}} = \frac{{{{\bf{b}}_{r,2,k}}}}{{\sqrt {{\bf{b}}_{r,2,k}^H{{\bf{b}}_{r,2,k}}} }}$.
}. 
Therefore, the noise floor represented with black dotted curve is placed at $-104$ dBm.
Regarding the intended signal received at the node $2$, we first consider the power of full-size intended channel ${\bf H}_1$ before applying AB-JHPC. In that case, the transmit power $P_T$ is assumed to be equally divided among four data streams, then the corresponding intended signal power at the first data stream is calculated as $\mathbb{E}\left\{\frac{P_T}{4}\left\|{\bf H}_1\right\|_F^2\right\}$.
After applying the proposed AB-JHPC with S-MMSE and SVD, we also plot the power of intended signal given in \eqref{eq_r_k}.
It is seen that both S-MMSE and SVD approaches used in the BB combiner of AB-JHPC provides almost the same intended signal power. Moreover, 	
approximately $2$ dB degradation is observed compared to the full-size intended channel power before applying AB-JHPC, which means that the proposed  AB-JHPC scheme preserves most of the power provided by the channel.
Considering the SI signal occurred at the node $2$, we also plot the power of full-size SI channel ${\bf H}_{\textrm{SI,2}}$ with the transmit power $P_T$ before applying AB-JHPC, which is calculated as $\mathbb{E}\left\{{P_T}\left\|{\bf H}_{\textrm{SI},2}\right\|_F^2\right\}$. As expected, when there is no antenna isolation, the power of corresponding received SI is $30$ dBm. For larger  antenna isolation, the corresponding received SI is reduced to $12$ dBm, when $P_{\textrm{IS,dB}}$ is greater than 30 dB. Therefore, the antenna isolation without applying AB-JHPC only provides up to $18$ dB SIC similar to Fig. \ref{fig_5_RF_BF_SIC}.
Then, we employ the proposed AB-JHPC technique to significantly enhance the achievable SIC. 
The residual SI is reduced from $30$ dBm to $-55.7$ dBm, even when there is no antenna isolation (i.e., ${P_{\textrm{IS,dB}}=0}$ dB).
This represents $85.7$ dB SIC provided for the first data stream by only employing the proposed AB-JHPC.
Moreover, when $P_{\textrm{IS,dB}}$ increases from $0$ dB to $70$ dB, the total SIC achieved via antenna isolation and AB-JHPC with S-MMSE is a near-linear function of $P_{\textrm{IS,dB}}$ (i.e., the total SIC is $P_{\textrm{IS,dB}}+85.7$ dB).
Additionally, the power of residual SI signal is reduced below noise floor level for $P_{\textrm{IS,dB}}\ge 50$ dB. 
On the other hand, the residual SI power is the major limiting factor for $P_{\textrm{IS,dB}}< 50$ dB.
Furthermore, regarding the BB combiner of AB-JHPC, the proposed S-MMSE algorithm further suppresses the SI signal power compared to the SVD approach for $P_{\textrm{IS,dB}}\ge 50$ dB, where the NLoS paths are dominant in the complete SI channel. 
For instance, when $P_{\textrm{IS,dB}}= 60$ dB, AB-JHPC with S-MMSE reduces the residual SI signal power to $-115.4$ dB, which is $6.1$ dB lower compared to AB-JHPC with SVD.

\begin{figure}[!t]
	\centering
	\includegraphics[width=\columnwidth]{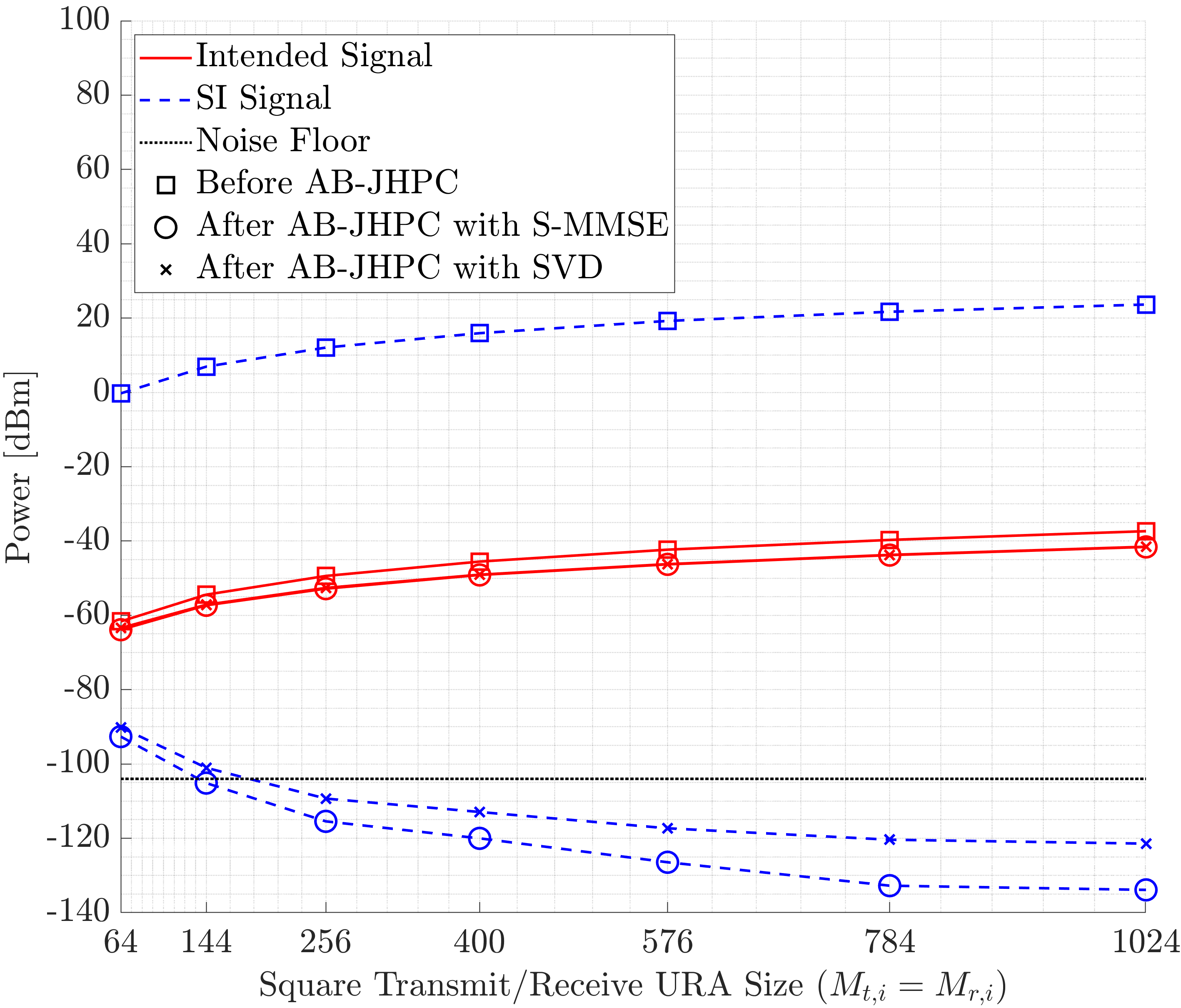}
	\caption{Self-interference cancellation (SIC) achieved by the proposed AB-JHPC versus the square transmit/receive URA sizes, where the transmit power is $P_T=30$ dBm and the antenna isolation is $P_{\textrm{IS,dB}}=60$ dB. 
	}
	\label{fig_7_SIC_by_HPC}
\end{figure}

Fig. \ref{fig_7_SIC_by_HPC} illustrates the signal power versus the square URA size, where the transmit power is $P_T=30$ dBm and the antenna isolation is $P_{\textrm{IS,dB}}=60$ dB. Before applying AB-JHPC, we observe that both the intended signal and SI powers increase as the larger URAs are utilized by means of the antenna array gain. 
On the other hand, when more antenna elements are utilized, the power of SI signal declines after applying the proposed AB-JHPC.
The main reason for this behavior is the enhancement of beamforming gain in the desired direction, which also suppressing the side lobes affecting the undesired directions, such as the SI signal.
Moreover, it is shown that the quality of SIC also improves for the larger URAs.
To illustrate, after applying AB-JHPC with S-MMSE,  the SI signal power is $29.8$ dB below the noise floor $M_{t,i}=M_{r,i}=1024$ antenna elements
However, when there are transmit and receive $M_{t,i}=M_{r,i}=64$ antenna elements, the SI power is $11.4$ dB above the noise floor.
Furthermore, the proposed AB-JHPC with S-MMSE algorithm offers higher amount of SIC for all URA sizes.

\vspace{-1ex}

\subsection{Total Achievable Rate: Full-Duplex vs. Half-Duplex}
\begin{figure}[!t]
	\centering
	\subfigure[]
	{\includegraphics[width=\columnwidth]{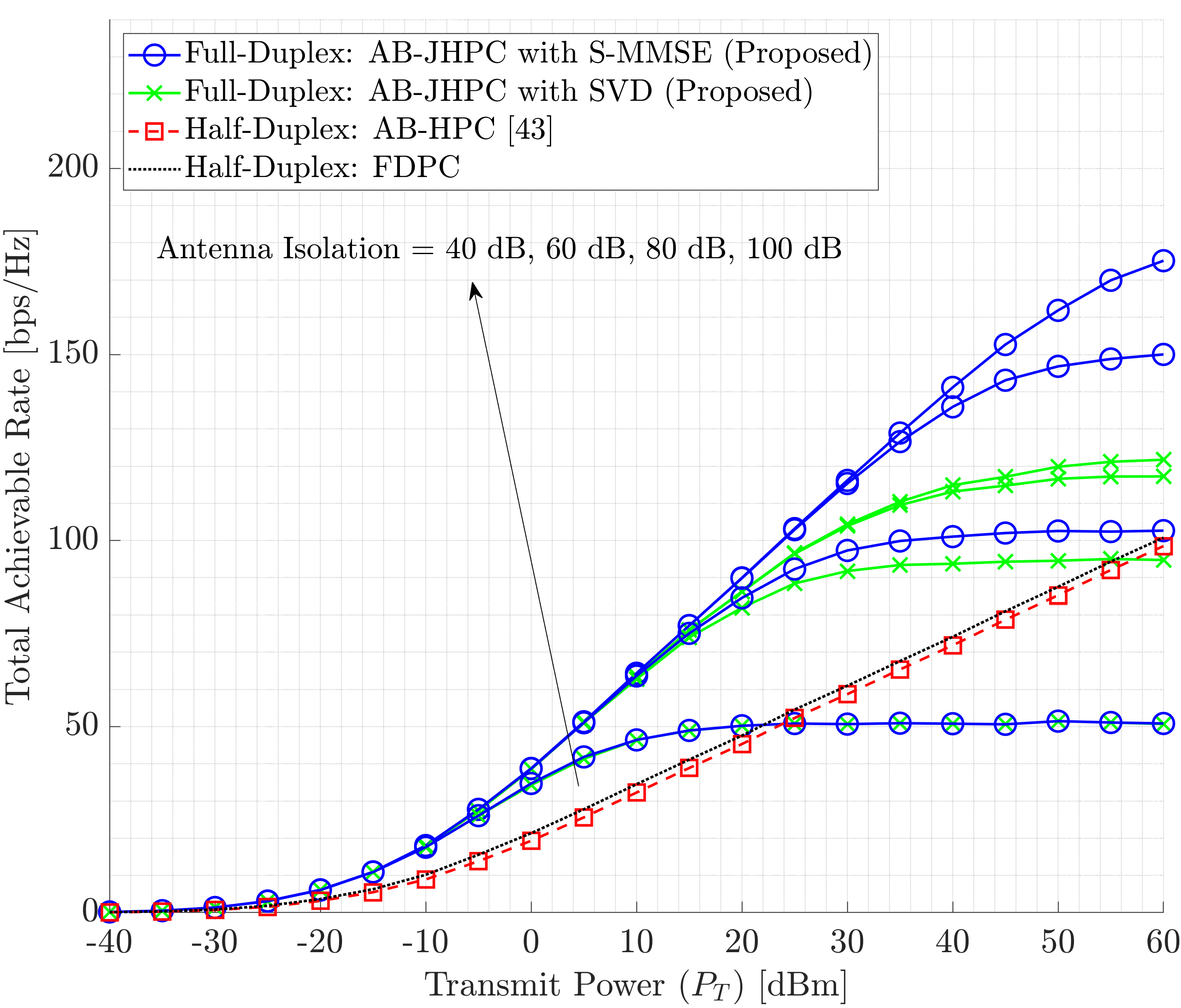}
		\label{fig_8a_Rate}}
	\vspace{2ex}\\
	\subfigure[]
	{\includegraphics[width=\columnwidth]{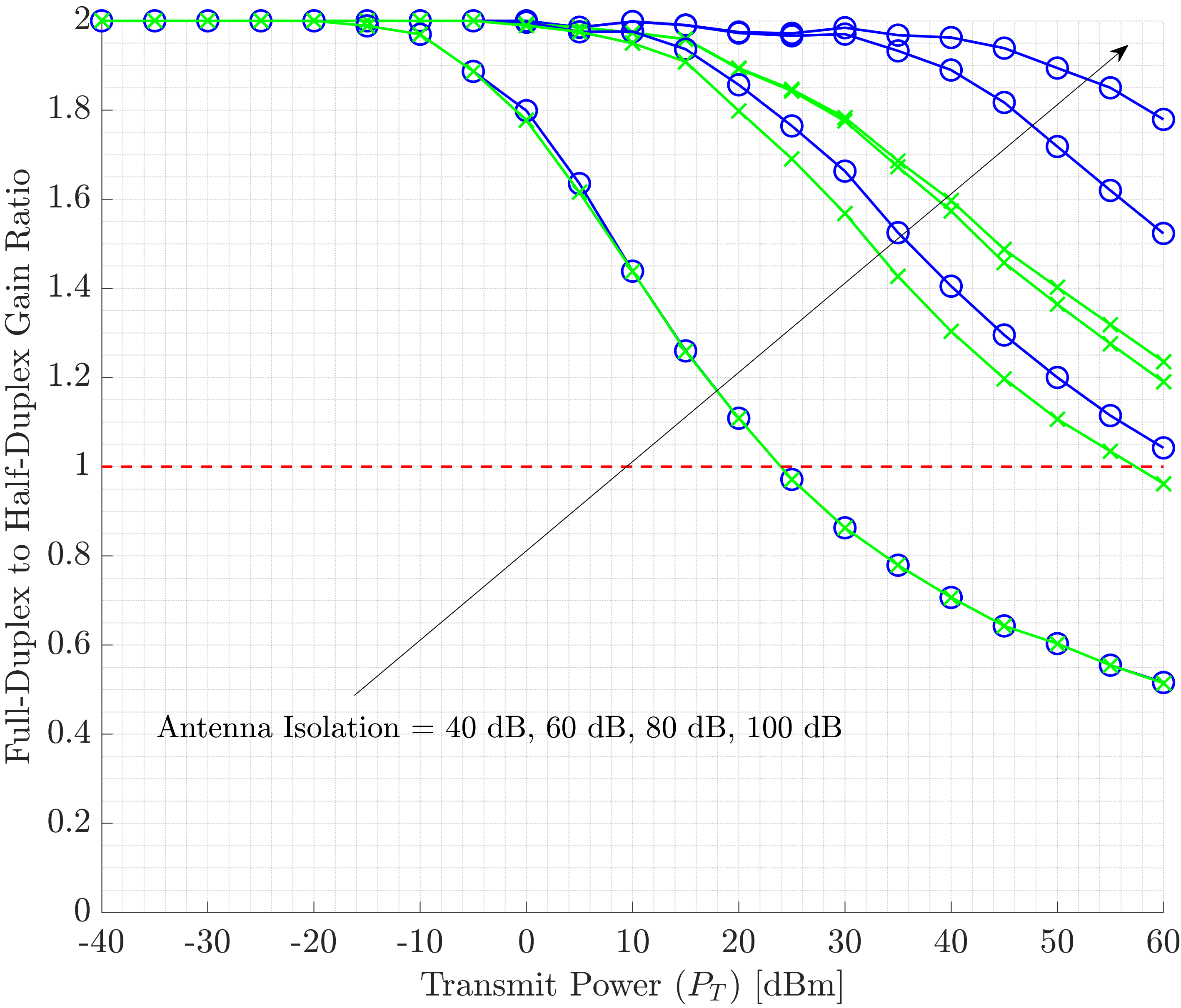}
		\label{fig_8b_Rate}}
	\caption{Achievable rate performance versus the transmit power ($P_T$) for the proposed AB-JHPC technique using full-duplex compared to AB-HPC \cite{ASIL_VTC_DAC_ADC} and FDPC using half-duplex:
		(a) total achievable rate,
		(b) full-duplex to half-duplex gain ratio.}
	\label{fig_8_Rate} 
	\vspace{2ex}
\end{figure}

The total achievable rate performance of the proposed full-duplex mmWave massive MIMO systems is investigated in Fig. \ref{fig_8_Rate}, where  the antenna isolation is assumed as 
${P_{\textrm{IS,dB}}=
	40\textrm{~dB},
	60\textrm{~dB},
	80\textrm{~dB},
	100\textrm{~dB}
}$. 
It is important to highlight that each node is equipped with $16\times16=256$ transmit and receive antenna elements  to simultaneously support $S_1=S_2=4$ data streams.
Also, we consider two half-duplex schemes as a benchmark to compare the performance of full-duplex communications via the proposed AB-JHPC technique: (i) AB-HPC as a hybrid architecture reducing the number of RF chains and CSI overhead size as in the proposed scheme \cite{ASIL_VTC_DAC_ADC}, (ii) FDPC as the fully-digital architecture requiring large CSI overhead and a single RF chain per each antenna element\footnote{It is worthwhile to note that the total achievable rate for the full-duplex communications is obtained by using \eqref{eq_Rate_j} and \eqref{eq_Rate_total}.
	On the other hand, when both communication nodes employ half-duplex, it is necessary to use separate time/frequency resources in order to establish two-way data transmission and reception operations. 
	Moreover, in the case of half-duplex, there is no SI component expressed in \eqref{eq_r_k}.
	Therefore, the total achievable rate for the half-duplex mmWave massive MIMO systems is calculated via 
	${R_{total}^{\left(\textrm{HD}\right)} = \frac{1}{2}R_1^{\left(\textrm{HD}\right)} + \frac{1}{2}R_2^{\left(\textrm{HD}\right)}}$, where $R_1^{\left(\textrm{HD}\right)}$ and $R_2^{\left(\textrm{HD}\right)}$ can be obtained by adopting \eqref{eq_r_k} and \eqref{eq_Rate_j} without the SI component.}.
As presented in Fig. \ref{fig_4_RF_Design_AoA_AoD}, the proposed AB-JHPC requires $N_1=19$ and $N_2=20$ RF chains at the node 1 and node 2, respectively. Therefore, in order to compare the hybrid architecture under the same hardware cost/complexity, AB-HPC is also designed by using the same number of RF chains.
In Fig. \ref{fig_8a_Rate}, when we first analyze the half-duplex communications, the performance gap between FDPC and AB-HPC is observed as only $1.4$ dB. On the other hand, the hybrid architectures significantly reduce the hardware cost/complexity by placing only $N_1=19$ ($N_2=20$) RF chains at the node $1$ (node $2$), whereas FDPC requires $M_1=M_2=512$ RF chains at each node.
It implies a remarkably reduction in the hardware cost/complexity by $96.2\%$. Also, the CSI overhead size is remarkably lowered by $99.8\%$ (please see Table \ref{table_RF_Chain_CSI} and Fig. \ref{fig_4_RF_Design_AoA_AoD}).
Regarding the achievable rate performance of the proposed AB-JHPC technique, we both utilize the S-MMSE and SVD approaches for the BB combiner design. 
Also, by using the total achievable rate curves for AB-JHPC using full-duplex and AB-HPC using half-duplex plotted in Fig. \ref{fig_8a_Rate}, the gain ratio shown in Fig. \ref{fig_8b_Rate} is simply calculated by taking the ratio of them (i.e., 
${
	R_{total}^{\left(\textrm{FD}\right)}}/
{R_{total}^{\left(\textrm{HD}\right)}
}$).
In Fig. \ref{fig_8a_Rate}, when the transmit power increases, the performance floor behavior is monitored for the full-duplex transmission due to the residual SI power. We observe that as the quality of antenna isolation improves, the total achievable rate capacity reaches to a higher performance floor.
For instance, as shown in Fig. \ref{fig_6_SIC_by_HPC}, the residual SI signal power is higher than the noise floor,  when $P_{\textrm{IS,dB}}=40$ dB and $P_T=30$ dBm. It is also visible in Fig. \ref{fig_8a_Rate}, where AB-JHPC reaches to interference floor at the transmit power of $P_T=30$ dBm for $P_{\textrm{IS,dB}}=40$ dB.
On the other hand, when the antenna isolation accomplishes $P_{\textrm{IS,dB}}=80$ dB SIC, the noise floor is above the residual SI power in Fig. \ref{fig_6_SIC_by_HPC}, especially for AB-JHPC with S-MMSE. It is also verified by the achievable rate results in Fig. \ref{fig_8a_Rate} (e.g., no performance floor at $P_{T}=30$ dBm and $P_{\textrm{IS,dB}}=80$ dB) and in Fig. \ref{fig_8b_Rate} (e.g., the gain ratio is found as $1.98$  at $P_{T}=30$ dBm and $P_{\textrm{IS,dB}}=80$ dB).
Then, Fig. \ref{fig_8b_Rate} shows that the proposed full-duplex scheme doubles the achievable rate capacity for the low and medium transmit power regime. 
Furthermore, the doubled-capacity property remains resistant for the high transmit power regime as the antenna isolation based SIC is enhanced. 
Regarding the BB combiner design of AB-JHPC, the proposed S-MMSE algorithm provides higher achievable rate and converges to the higher performance floor compared to the conventional SVD approach.

\begin{figure}[!t]
	\centering
	\subfigure[AB-JHPC with S-MMSE]
	{\includegraphics[width=\columnwidth]{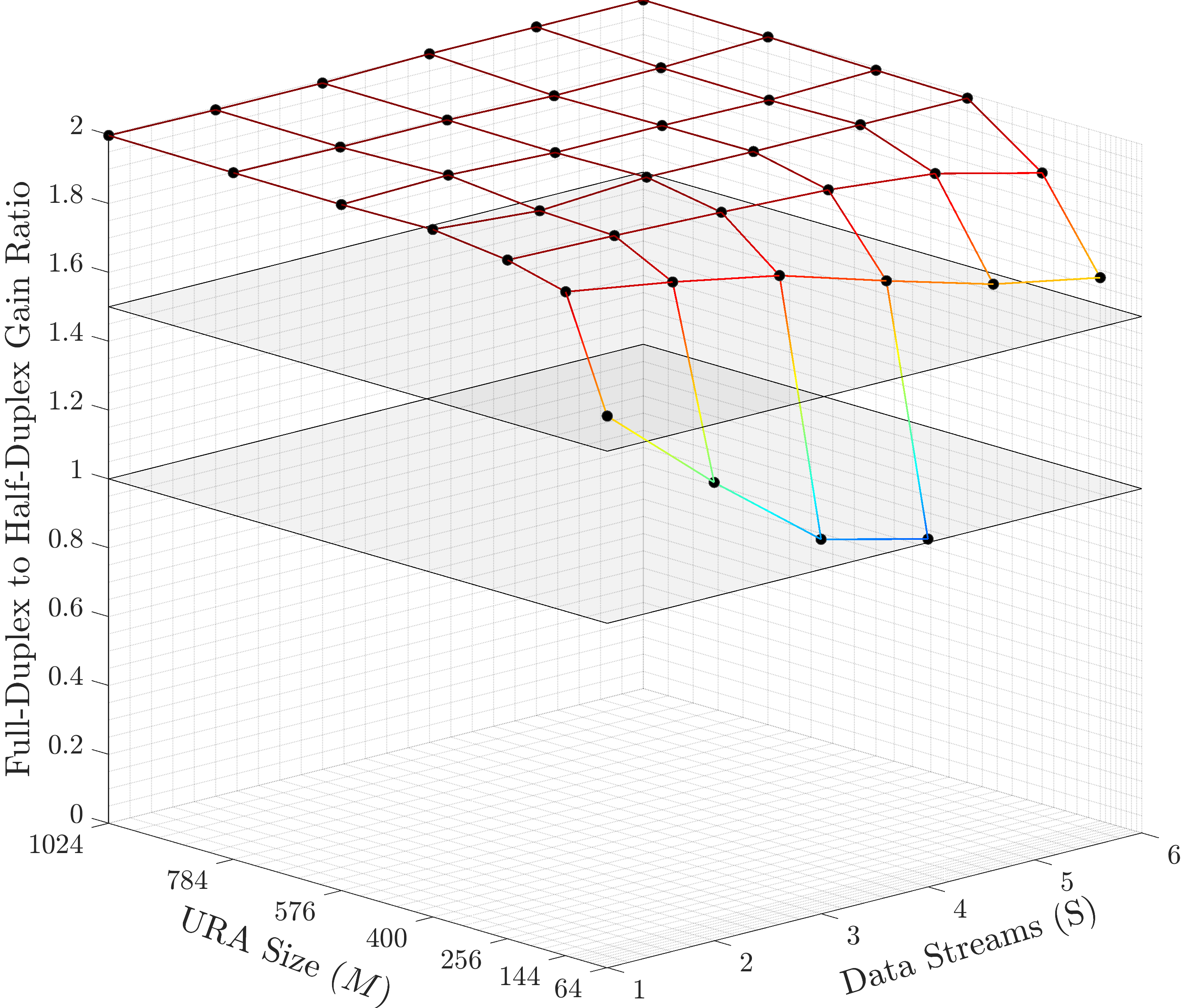}
		\label{fig_9a_DataStream}}
	\vspace{2ex}\\
	\subfigure[AB-JHPC with SVD]
	{\includegraphics[width=\columnwidth]{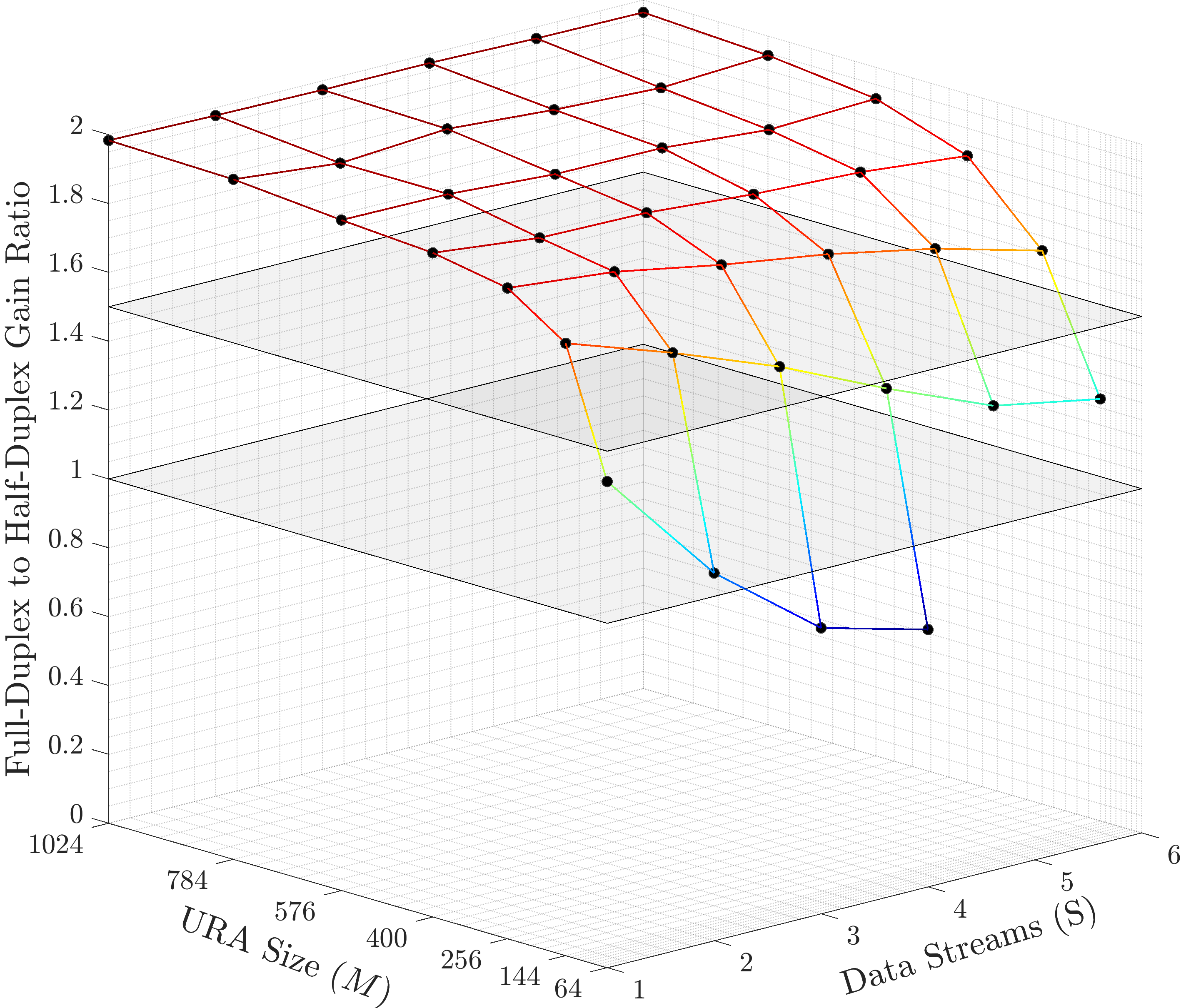}
		\label{fig_9b_DataStream}}
	\caption{Full-duplex to half-duplex gain ratio versus the square URA sizes ($M=M_{t,i}=M_{r,i}$) and number of data streams ($S=S_1=S_2$): (a) AB-JHPC with S-MMSE and (b) AB-JHPC with SVD, where the transmit power is $P_{T}=30$ dBm and the antenna isolation is $P_{\textrm{IS,dB}}=74$ dB \cite{FD_AntennaSep_74dB}.}
	\label{fig_9_DataStream} 
	\vspace{-1ex}
\end{figure}

Fig. \ref{fig_9_DataStream} illustrates the full-duplex to half-duplex gain ratio in terms of achievable rate versus the square URA sizes ($M=M_{t,i}=M_{r,i}$) and number of data streams ($S=S_1=S_2$),  where the transmit power is $P_{T}=30$ dBm and the antenna isolation is $P_{\textrm{IS,dB}}=74$ dB as in \cite{FD_AntennaSep_74dB}.
Although the number of data streams is selected between $S=1$ and $S=6$, the gain ratio for $M=64$ antennas is plotted until $S=4$ data streams because we obtain $\min\left(N_{t,1},N_{r,1},N_{t,2},N_{r,2}\right)=4$ for $M=64$.
It is seen that the gain ratio almost approaches to $2$ for the larger URAs utilized at the transmitter/receiver. However, when more data streams are transmitted, the gain ratio degrades due to the increased residual SI power.
Moreover, AB-JHPC with S-MMSE has always the gain ratio higher than unity as illustrated in Fig. \ref{fig_9a_DataStream}. By comparing Fig. \ref{fig_9a_DataStream} and Fig. \ref{fig_9b_DataStream}, we observe that the proposed S-MMSE technique outperforms SVD.
For instance, when the URA size is chosen as $M=144$ for serving $S=3$ data streams, AB-JHPC with S-MMSE provides the gain ratio of $1.83$, while it is only $1.55$ for AB-JHPC with SVD.

\subsection{Energy Efficiency}
In addition to improving the quality of SIC and reducing the hardware cost/complexity, another primary objective in the proposed hybrid architecture is decreasing the total power consumption for the massive MIMO systems. 
By using the models given in \cite{EnergyEff_1,EnergyEff_2,EnergyEff_3}, the total power consumption for the two-way data transmission is defined as:
\begin{equation}\label{eq_P_total}
	P_{total} = 2P_T + N_{RF}P_{RF} + N_{PS}P_{PS} \textrm{~~[W]},
\end{equation}
where 
$P_T$  is the total transmission power at each node,
$N_{RF}$ and $N_{PS}$ are the number of RF chains and phase-shifters, respectively,
$P_{RF}$ and $P_{PS}$ are the power consumption by each RF chain and phase-shifter, respectively.
Table \ref{table_RF_Chain_CSI} summarizes the number of hardware components for the proposed scheme using transfer block architecture (please see Fig. \ref{fig_3_TransferBlock}) or without transfer block architecture (please see Fig. \ref{fig_1_SystemModel}).
For instance, when no transfer block is utilized, there are $N_{RF}=39$ RF chains as shown in Fig. \ref{fig_4_RF_Design_AoA_AoD}. 
On the other hand, only $N_{RF}=16$ RF chains are required in the transfer block architecture to support $S_1=S_2=4$ data streams.	
Also, we consider ${P_{RF}}=250$ mW and ${P_{PS}}=1$ mW as in \cite{EnergyEff_1,EnergyEff_2}.
Afterwards, the energy efficiency is calculated by taking the ratio of total achievable rate given in \eqref{eq_Rate_total} with respect to the total power consumption given in \eqref{eq_P_total} (i.e., $\frac{R_{total}}{P_{total}}$).

\begin{figure}[!t]
	\centering
	\subfigure[]
	{\includegraphics[width=\columnwidth]{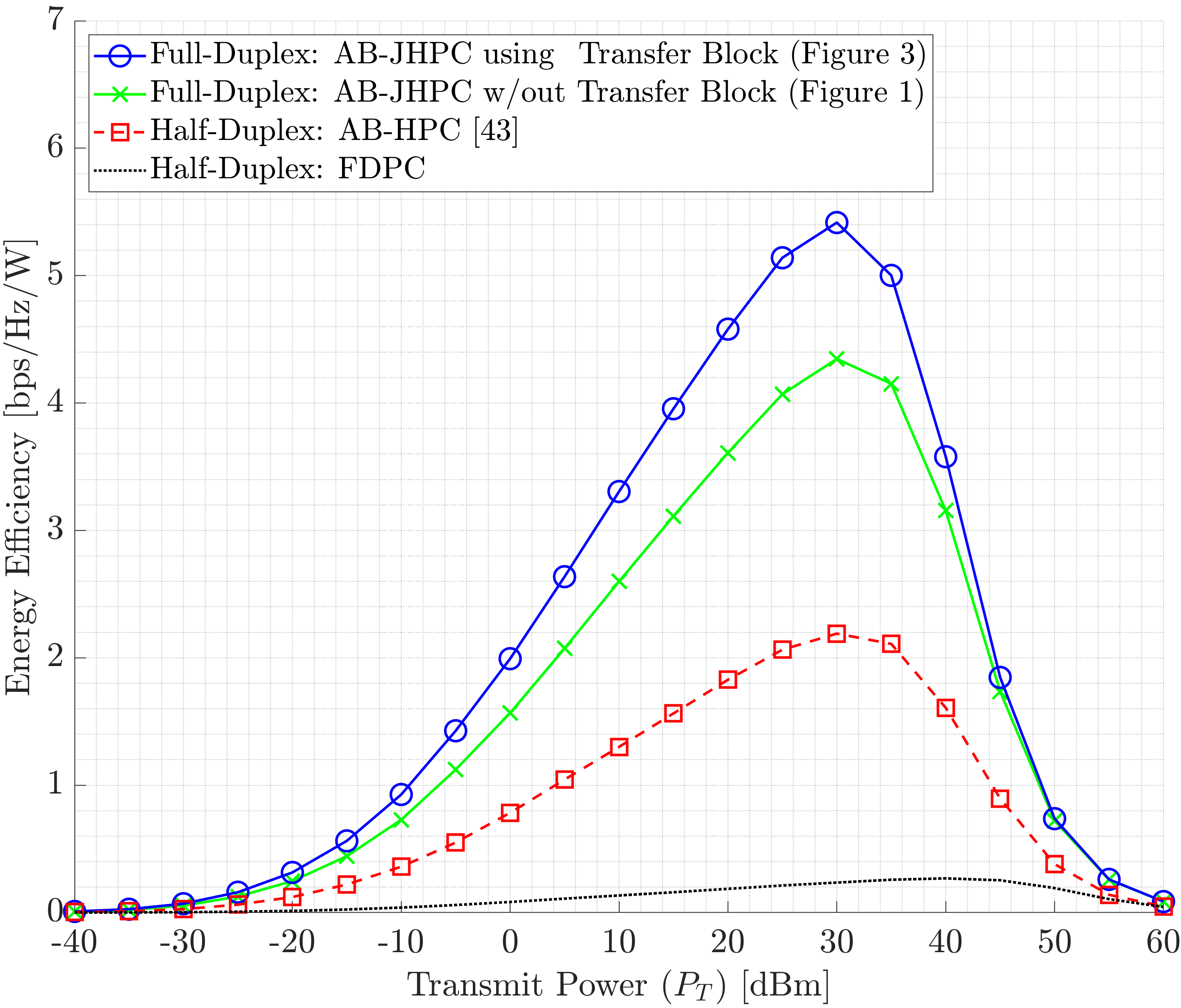}\label{fig_10a_DataStream}}
	\vspace{2ex}\\
	\subfigure[]
	{\includegraphics[width=\columnwidth]{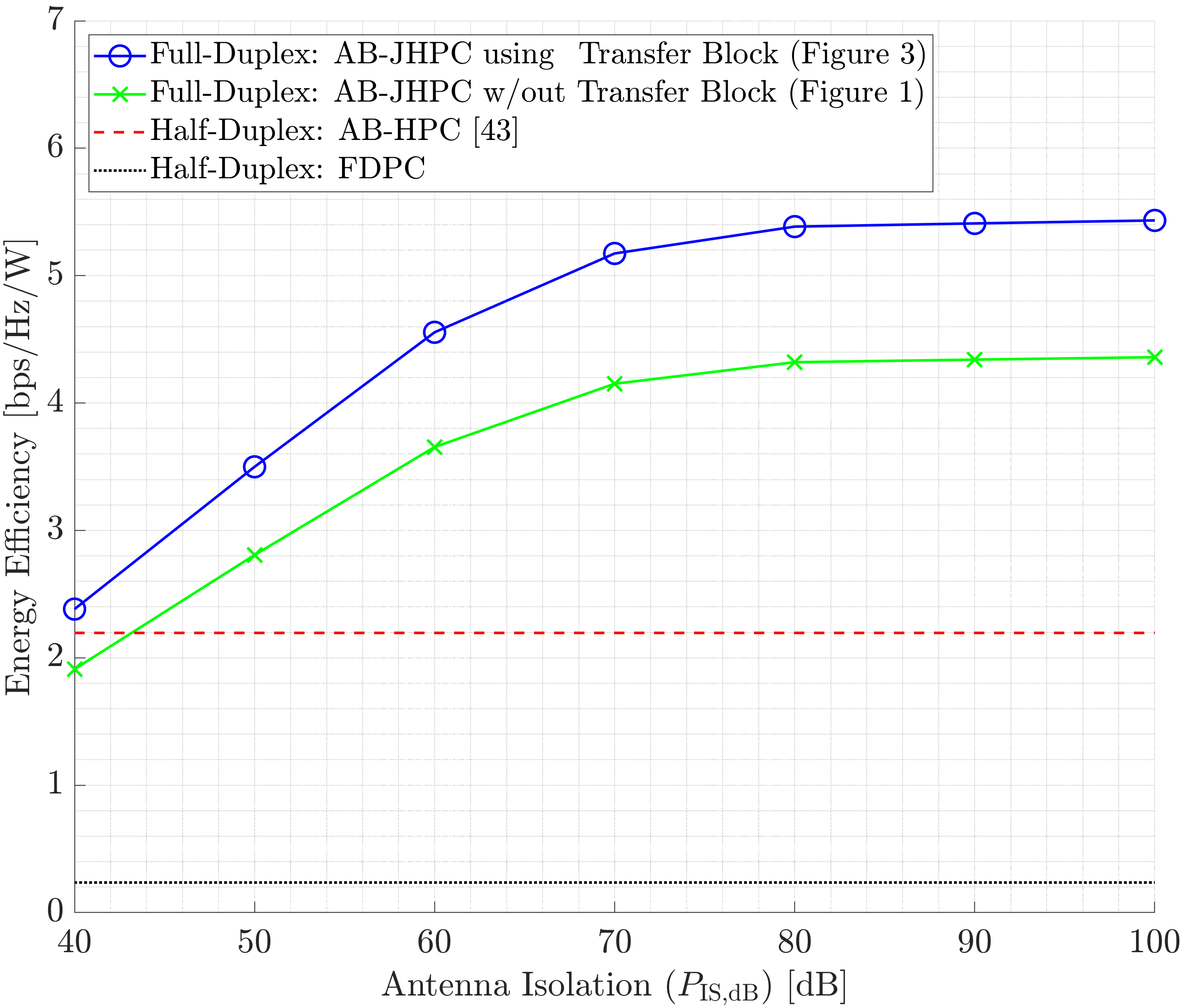}\label{fig_10b_DataStream}}
	\caption{Energy efficiency (a) versus the transmit power ($P_T$) for $P_{\textrm{IS,dB}}=60$ dB and (b) versus the antenna isolation ($P_{\textrm{IS,dB}}$) for $P_{T}=30$ dBm. Comparison of the proposed AB-JHPC with S-MMSE technique using full-duplex with respect to AB-HPC \cite{ASIL_VTC_DAC_ADC} and FDPC using half-duplex.}
	\label{fig_10_DataStream} 
	\vspace{-1ex}
\end{figure}

Fig. \ref{fig_10_DataStream} shows the energy efficiency results for both full-duplex and half-duplex schemes. Moreover, we consider the utilization of transfer block architecture to enhance the energy efficiency results by reducing the number of RF chains\footnote{AB-HPC using the half-duplex transmission does not consider the transfer block architecture \cite{ASIL_VTC_DAC_ADC}.}.
As proven in Section \ref{sec_RF_Reduction}, the total achievable rate performance remains the same whether the transfer block architecture is utilized or not. 
First, Fig. \ref{fig_10a_DataStream} plots the energy efficiency versus the transmit power $P_T$, where $P_{\textrm{IS,dB}}=60$ dB.
The numerical results show that using the transfer block improves the energy efficiency (e.g., at $P_T=30$ dB, we achieve $5.4$ bps/Hz/W by using the transfer block, whereas it reduces to $4.3$ bps/W/Hz without transfer block). 
On the other hand, the energy efficiency of FDPC is remarkably low compared to the hybrid schemes due to the utilization of power-hungry RF chains.
As explained in Table \ref{table_RF_Chain_CSI}, the proposed AB-JHPC technique requires significantly less number of RF chains than FDPC, which results in the reduced power consumption.
Second, Fig. \ref{fig_10b_DataStream} depicts the energy efficiency versus the antenna isolation ($P_{\textrm{IS,dB}}$) for $P_{T}=30$ dBm.
We observe that the energy efficiency improves for the higher values of $P_{\textrm{IS,dB}}$. 
However, the energy efficiency is almost saturated when $P_{\textrm{IS,dB}}$ is higher than 80 dB. 

\begin{figure}[!t]
	\centering
	\includegraphics[width=\columnwidth]{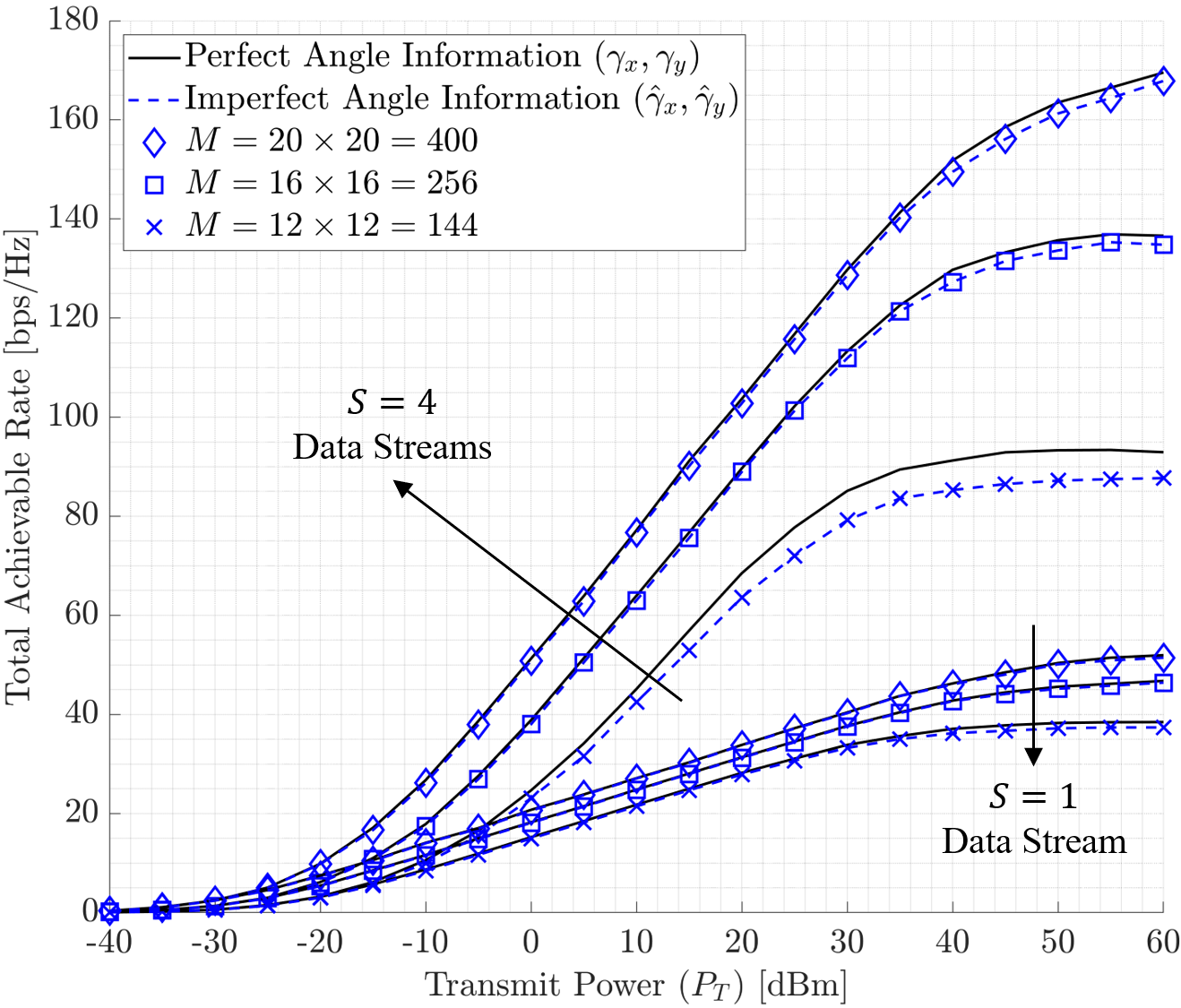}
	\caption{Achievable rate performance of the proposed AB-JHPC with S-MMSE technique versus the transmit power ($P_T$) for perfect and imperfect angle information, where the antenna isolation is $P_{\textrm{IS,dB}}=74$ dB \cite{FD_AntennaSep_74dB}.}
	\label{fig_11_imperfect_angle}
\end{figure}

{
\subsection{Imperfect Angle Information (AoD/AoA)}

As discussed earlier, the RF beamformer design is based on the slow time-varying AoD/AoA information (i.e., the mean of elevation angle ($\theta$) and azimuth angle ($\psi$), and their spread).
	Therefore, according to the perfect angle information, the transmit/receive RF beamformers are developed via $\gamma_x=\sin\left(\theta\right)\cos\left(\psi\right)$ and $\gamma_y=\sin\left(\theta\right)\sin\left(\psi\right)$ coefficients as illustrated in Fig. \ref{fig_4_RF_Design_AoA_AoD}.
However, considering the imperfect angle information, 
	the estimated coefficients are defined as
	$\hat{\gamma}_x = \gamma_x + \Delta\hat{\gamma}_{x}$
	and
	$\hat{\gamma}_y = \gamma_y + \Delta\hat{\gamma}_{y}$,
	where $\Delta\hat{\gamma}_{x}$ and $\Delta\hat{\gamma}_{y}$ are the estimation errors on $\gamma_x$ and $\gamma_y$ \cite{Imperfect_AoD, Imperfect_AoD_2,Imperfect_AoD_3}.
Here, 
$\Delta\hat{\gamma}_{x} \in \left[
-\frac{\epsilon_x}{2\pi},
+\frac{\epsilon_x}{2\pi} 
\right]$
and
$\Delta\hat{\gamma}_{y} \in \left[
-\frac{\epsilon_y}{2\pi},
+\frac{\epsilon_y}{2\pi} 
\right]$
are uniformly distributed random variables, where $\epsilon_x$ and $\epsilon_y$ denote the maximum angle estimation errors up to the beamwidth between the first nulls \cite{Imperfect_AoD}. 
	When URA has $M=M_x\times M_y$ antennas (i.e., $M_x$ and $M_y$ antennas along $x$-axis and $y$-axis as shown in Fig. \ref{fig_2_TxRxAntennas}), we have $\epsilon_x=\frac{2\pi}{M_x}$ and $\epsilon_y=\frac{2\pi}{M_y}$ \cite{Imperfect_AoD_3}.
	
Fig. \ref{fig_11_imperfect_angle} demonstrates the effect of imperfect angle information on the achievable rate performance of the proposed AB-JHPC with S-MMSE technique, where the antenna isolation is $P_{\textrm{IS,dB}}=74$ dB as in \cite{FD_AntennaSep_74dB}.
By assuming the same size transmit and receive URAs at each node ($M=M_{t,i}=M_{r,i}$), three square URA cases are considered as 
	${M=144}$, 
	${M=256}$, and
	${M=400}$.
Also, we assume either ${S=1}$ or ${S=4}$ data streams transmitted by each node for two-way communications.
	The numerical results imply that the proposed AB-JHPC technique for the full-duplex massive MIMO systems is robust to the imperfect angle information.
The achievable rate performance with the imperfect angle information closely approaches the perfectly estimated angle information case, as the URA size increases.
	The main reason for this behavior is the reduced angle estimation error with larger URAs (i.e., $\epsilon_x=\frac{2\pi}{M_x}$ and $\epsilon_y=\frac{2\pi}{M_y}$). 
		For instance, when the URA size is chosen as $M=144$ to support $S=4$ data streams, the performance gap between the perfect and imperfect angle information is observed as $4.7$ bps/Hz at $P_T=30$ dBm, where the perfect angle information provides $84.4$ bps/Hz achievable rate capacity. 
On the other hand, when the URA size is increased to $M=256$ ($M=400$), the performance gap reduces to only $1.2$ bps/Hz ($0.9$ bps/Hz) at $P_T=30$ dBm, while the total achievable rate capacity with perfectly estimated angle information is also enhanced to $113.2$ bps/Hz ($129.7$ bps/Hz).
	In other words, in the case of imperfect angle information utilized at the RF beamformer design, we achieve $99.1\%$ ($99.3\%$) of the capacity obtained with the perfect angle information, when the URA size is considered as $M=256$ ($M=400$).	
}

\section{Conclusions} \label{sec_Conc}
In this paper, a novel angular-based joint hybrid precoding/combining (AB-JHPC) has been proposed for the full-duplex mmWave massive-MIMO systems.
Our primary objectives in the proposed AB-JHPC technique are to 
(i) 	enhance the quality of self-interference $\textrm{cancellation (SIC)}$,
(ii) 	maximize the power of desired/intended signal,
(iii) 	reduce the channel estimation overhead size,
(iv) 	decrease the hardware cost/complexity via a low number of RF chains.
First, the RF-stage of AB-JHPC has been constructed via slow time-varying AoD/AoA information, where the transmit/receive RF beamformers aim to maximize the desired/intended signal power and cancel the strong SI power.
Therefore, a joint RF beamformer structure has been designed for generating orthogonal transmit/receive beams covering AoD/AoA support of intended channel and excluding AoD/AoA support of SI channel.
Then, the BB-stage of AB-JHPC has been built via the reduced-size effective intended channel.
Thus, the proposed AB-JHPC technique does not require any instantaneous SI channel knowledge neither in the RF-stage nor in the BB-stage.
We have primarily considered the well-known SVD approach to develop the BB precoder and BB combiner.
In addition to SVD, we have also proposed a new semi-blind minimum mean square error (S-MMSE) technique in the BB-stage design, which seeks to suppress the power of far-field SI component by using the slow time-varying channel characteristics. 
Afterwards, a transfer block architecture has been employed to minimize the RF chain utilization without sacrificing the achievable rate performance.
The promising numerical results show that the joint transmit/receive RF beamformer suppresses the power of far-field (near-field) SI channel by $81.5$ dB ($44.5$ dB).
On the other hand, the SI signal is considerably mitigated via the proposed AB-JHPC technique. 
We observe that $85.7$ dB SIC is achieved via only AB-JHPC and the total amount of SIC can be linearly increased via the advanced antenna isolation techniques.
In terms of the total achievable rate capacity, the proposed full-duplex mmWave massive MIMO systems can double the total achievable rate capacity compared to its half-duplex counterpart.
Also, the proposed S-MMSE algorithm remarkably outperforms the conventional SVD approach, especially in the high transmit power regime.
It is shown that the transfer block architecture utilizing less number of RF chains further improves the energy efficiency of the full-duplex mmWave massive MIMO systems.

\bibliographystyle{IEEEtran}
\bibliography{bibAsil_2103}

\begin{thebibliography}{10}
\providecommand{\url}[1]{#1}
\csname url@samestyle\endcsname
\providecommand{\newblock}{\relax}
\providecommand{\bibinfo}[2]{#2}
\providecommand{\BIBentrySTDinterwordspacing}{\spaceskip=0pt\relax}
\providecommand{\BIBentryALTinterwordstretchfactor}{4}
\providecommand{\BIBentryALTinterwordspacing}{\spaceskip=\fontdimen2\font plus
\BIBentryALTinterwordstretchfactor\fontdimen3\font minus
  \fontdimen4\font\relax}
\providecommand{\BIBforeignlanguage}[2]{{%
\expandafter\ifx\csname l@#1\endcsname\relax
\typeout{** WARNING: IEEEtran.bst: No hyphenation pattern has been}%
\typeout{** loaded for the language `#1'. Using the pattern for}%
\typeout{** the default language instead.}%
\else
\language=\csname l@#1\endcsname
\fi
#2}}
\providecommand{\BIBdecl}{\relax}
\BIBdecl

\bibitem{5G_Mas_MIMO_3}
M.~Agiwal, A.~Roy, and N.~Saxena, ``Next generation 5{G} wireless networks: A
  comprehensive survey,'' \emph{IEEE Commun. Surveys Tuts.}, vol.~18, no.~3,
  pp. 1617--1655, 3rd Quart. 2016.

\bibitem{Mass_mmWave_REV}
A.~L. Swindlehurst, E.~Ayanoglu, P.~Heydari, and F.~Capolino, ``Millimeter-wave
  massive {MIMO}: The next wireless revolution?'' \emph{IEEE Commun. Mag.},
  vol.~52, no.~9, pp. 56--62, Sept. 2014.

\bibitem{6G_prospective_look}
L.~{Bariah}, L.~{Mohjazi}, S.~{Muhaidat}, P.~C. {Sofotasios}, G.~K. {Kurt},
  H.~{Yanikomeroglu}, and O.~A. {Dobre}, ``A prospective look: Key enabling
  technologies, applications and open research topics in {6G} networks,''
  \emph{IEEE Access}, vol.~8, pp. 174\,792--174\,820, 2020.

\bibitem{Mass_mmWave_Survey}
S.~A. Busari, K.~M.~S. Huq, S.~Mumtaz, L.~Dai, and J.~Rodriguez,
  ``Millimeter-wave massive {MIMO} communication for future wireless systems: A
  survey,'' \emph{IEEE Commun. Surveys Tuts.}, vol.~20, no.~2, pp. 836--869,
  2018.

\bibitem{mmWave_Survey_2}
S.~{Kutty} and D.~{Sen}, ``Beamforming for millimeter wave communications: An
  inclusive survey,'' \emph{IEEE Commun. Surveys Tuts.}, vol.~18, no.~2, pp.
  949--973, 2nd Quart. 2016.

\bibitem{mmWave_Antenna_Size}
W.~{Hong}, K.~{Baek}, Y.~{Lee}, Y.~{Kim}, and S.~{Ko}, ``Study and prototyping
  of practically large-scale {mmWave} antenna systems for {5G} cellular
  devices,'' \emph{IEEE Commun. Mag.}, vol.~52, no.~9, pp. 63--69, 2014.

\bibitem{mmWave_Survey}
R.~W. {Heath}, N.~{González-Prelcic}, S.~{Rangan}, W.~{Roh}, and A.~M.
  {Sayeed}, ``An overview of signal processing techniques for millimeter wave
  {MIMO} systems,'' \emph{IEEE J. Sel. Topics Signal Process.}, vol.~10, no.~3,
  pp. 436--453, Apr. 2016.

\bibitem{MassiveMIMO_TDD}
E.~{Zeydan}, O.~{Dedeoglu}, and Y.~{Turk}, ``Experimental evaluations of
  {TDD}-based massive {MIMO} deployment for mobile network operators,''
  \emph{IEEE Access}, vol.~8, pp. 33\,202--33\,214, 2020.

\bibitem{MassiveMIMO_TDD_2}
H.~Q. {Ngo} and E.~G. {Larsson}, ``No downlink pilots are needed in {TDD}
  massive {MIMO},'' \emph{IEEE Trans. Wireless Commun.}, vol.~16, no.~5, pp.
  2921--2935, 2017.

\bibitem{MassiveMIMO_FDD}
J.~{Flordelis}, F.~{Rusek}, F.~{Tufvesson}, E.~G. {Larsson}, and O.~{Edfors},
  ``Massive mimo performance—{TDD} versus {FDD}: What do measurements say?''
  \emph{IEEE Trans. Wireless Commu.}, vol.~17, no.~4, pp. 2247--2261, 2018.

\bibitem{MassiveMIMO_FDD_2}
Y.~{Yang}, F.~{Gao}, Z.~{Zhong}, B.~{Ai}, and A.~{Alkhateeb}, ``Deep transfer
  learning-based downlink channel prediction for {FDD} massive {MIMO}
  systems,'' \emph{IEEE Trans. Commun.}, vol.~68, no.~12, pp. 7485--7497, 2020.

\bibitem{MassiveMIMO_FDD_3}
Z.~{Jiang}, A.~F. {Molisch}, G.~{Caire}, and Z.~{Niu}, ``Achievable rates of
  {FDD} massive {MIMO} systems with spatial channel correlation,'' \emph{IEEE
  Trans. Wireless Commun.}, vol.~14, no.~5, pp. 2868--2882, 2015.

\bibitem{FD_Survey}
Z.~Zhang, K.~Long, A.~V. Vasilakos, and L.~Hanzo, ``Full-duplex wireless
  communications: Challenges, solutions, and future research directions,''
  \emph{Proc. IEEE}, vol. 104, no.~7, pp. 1369--1409, July 2016.

\bibitem{FD_Survey_Secrecy}
D.~Kim, H.~Lee, and D.~Hong, ``A survey of in-band full-duplex transmission:
  From the perspective of {PHY and MAC} layers,'' \emph{IEEE Commun. Surveys
  Tuts.}, vol.~17, no.~4, pp. 2017--2046, 4th Quart. 2015.

\bibitem{FD_Survey_2019}
K.~E. {Kolodziej}, B.~T. {Perry}, and J.~S. {Herd}, ``In-band full-duplex
  technology: Techniques and systems survey,'' \emph{IEEE Trans. Microw. Theory
  Techn.}, vol.~67, no.~7, pp. 3025--3041, July 2019.

\bibitem{FD_In_Band_Survey}
G.~Liu, F.~R. Yu, H.~Ji, V.~C.~M. Leung, and X.~Li, ``In-band full-duplex
  relaying: A survey, research issues and challenges,'' \emph{IEEE Commun.
  Surveys Tuts.}, vol.~17, no.~2, pp. 500--524, 2nd Quart. 2015.

\bibitem{FD_AntennaSep_60dB}
M.~{Duarte}, A.~{Sabharwal}, V.~{Aggarwal}, R.~{Jana}, K.~K. {Ramakrishnan},
  C.~W. {Rice}, and N.~K. {Shankaranarayanan}, ``Design and characterization of
  a full-duplex multiantenna system for {WiFi} networks,'' \emph{IEEE Trans.
  Veh. Technol.}, vol.~63, no.~3, pp. 1160--1177, 2014.

\bibitem{FD_AntennaSep_74dB}
E.~{Everett}, A.~{Sahai}, and A.~{Sabharwal}, ``Passive self-interference
  suppression for full-duplex infrastructure nodes,'' \emph{IEEE Trans.
  Wireless Commun.}, vol.~13, no.~2, pp. 680--694, 2014.

\bibitem{FD_MIMO_SI_Canc_BF}
S.~{Huberman} and T.~{Le-Ngoc}, ``{MIMO} full-duplex precoding: A joint
  beamforming and self-interference cancellation structure,'' \emph{IEEE Trans.
  Wireless Commun.}, Apr. 2015.

\bibitem{FD_MIMO_mmWave}
X.~Liu, Z.~Xiao, L.~Bai, J.~Choi, P.~Xia, and X.-G. Xia, ``Beamforming based
  full-duplex for millimeter-wave communication,'' \emph{Sensors}, vol.~16,
  no.~7, p. 1130, July 2016.

\bibitem{FD_MIMO_Beam_Domain}
X.~{Xia}, K.~{Xu}, D.~{Zhang}, Y.~{Xu}, and Y.~{Wang}, ``Beam-domain
  full-duplex massive {MIMO}: Realizing co-time co-frequency uplink and
  downlink transmission in the cellular system,'' \emph{IEEE Trans. Veh.
  Technol.}, Oct. 2017.

\bibitem{FD_MIMO_ANALOG}
R.~{López-Valcarce} and N.~{González-Prelcic}, ``Analog beamforming for
  full-duplex millimeter wave communication,'' in \emph{2019 16th Int. Symp.
  Wireless Commun. Syst. (ISWCS)}, Aug. 2019, pp. 687--691.

\bibitem{Satyanarayana2019}
K.~{Satyanarayana}, M.~{El-Hajjar}, P.~{Kuo}, A.~{Mourad}, and L.~{Hanzo},
  ``Hybrid beamforming design for full-duplex millimeter wave communication,''
  \emph{IEEE Trans. Veh. Technol.}, vol.~68, no.~2, pp. 1394--1404, Feb. 2019.

\bibitem{FD_MIMO_MU_Learning}
K.~{Satyanarayana}, M.~{El-Hajjar}, A.~A.~M. {Mourad}, and L.~{Hanzo},
  ``Multi-user full duplex transceiver design for mmwave systems using
  learning-aided channel prediction,'' \emph{IEEE Access}, May 2019.

\bibitem{FD_MIMO_Relay_mmWave_2}
Y.~{Zhang}, M.~{Xiao}, S.~{Han}, M.~{Skoglund}, and W.~{Meng}, ``On precoding
  and energy efficiency of full-duplex millimeter-wave relays,'' \emph{IEEE
  Trans. Wireless Commun.}, vol.~18, no.~3, pp. 1943--1956, 2019.

\bibitem{FD_MIMO_Relay_mmWave}
Y.~{Cai}, Y.~{Xu}, Q.~{Shi}, B.~{Champagne}, and L.~{Hanzo}, ``Robust joint
  hybrid transceiver design for millimeter wave full-duplex {MIMO} relay
  systems,'' \emph{IEEE Trans. Wireless Commun.}, Feb. 2019.

\bibitem{FD_MIMO_1_bit_HPC}
J.~M.~B. {da Silva}, A.~{Sabharwal}, G.~{Fodor}, and C.~{Fischione}, ``1-bit
  phase shifters for large-antenna full-duplex {mmWave} communications,''
  \emph{IEEE Trans. Wireless Commun.}, vol.~19, no.~10, pp. 6916--6931, 2020.

\bibitem{Mass_MIMO_Precoding_Survey}
N.~Fatema, G.~Hua, Y.~Xiang, D.~Peng, and I.~Natgunanathan, ``Massive {MIMO}
  linear precoding: A survey,'' \emph{IEEE Syst. J.}, vol.~12, no.~4, pp.
  3920--3931, Dec. 2017.

\bibitem{ANALOG_BF}
Y.~{Wang}, W.~{Zou}, and Y.~{Tao}, ``Analog precoding designs for millimeter
  wave communication systems,'' \emph{IEEE Trans. Veh. Technol.}, vol.~67,
  no.~12, pp. 11\,733--11\,745, Dec. 2018.

\bibitem{ANALOG_BF_Heath}
X.~{Gao}, L.~{Dai}, S.~{Han}, C.~{I}, and R.~W. {Heath}, ``Energy-efficient
  hybrid analog and digital precoding for mmwave {MIMO} systems with large
  antenna arrays,'' \emph{IEEE J. Sel. Areas Commun.}, vol.~34, no.~4, pp.
  998--1009, Apr. 2016.

\bibitem{Mass_MIMO_Hyb_Survey}
I.~{Ahmed}, H.~{Khammari}, A.~{Shahid}, A.~{Musa}, K.~S. {Kim}, E.~{De
  Poorter}, and I.~{Moerman}, ``A survey on hybrid beamforming techniques in
  5{G}: Architecture and system model perspectives,'' \emph{IEEE Commun.
  Surveys Tuts.}, vol.~20, no.~4, pp. 3060--3097, 4th Quart. 2018.

\bibitem{Mass_MIMO_Hybrid_Survey}
A.~F. Molisch, V.~V. Ratnam, S.~Han, Z.~Li, S.~L.~H. Nguyen, L.~Li, and
  K.~Haneda, ``Hybrid beamforming for massive {MIMO}: A survey,'' \emph{IEEE
  Commun. Mag.}, vol.~55, no.~9, pp. 134--141, Sept. 2017.

\bibitem{Mass_MIMO_Hybrid_Survey_2}
M.~{Rihan}, T.~{Abed Soliman}, C.~{Xu}, L.~{Huang}, and M.~I. {Dessouky},
  ``Taxonomy and performance evaluation of hybrid beamforming for {5G} and
  beyond systems,'' \emph{IEEE Access}, vol.~8, pp. 74\,605--74\,626, Mar.
  2020.

\bibitem{Mass_MIMO_Hyb_Low_Comp}
L.~Liang, W.~Xu, and X.~Dong, ``Low-complexity hybrid precoding in massive
  multiuser {MIMO} systems,'' \emph{IEEE Wireless Commun. Lett.}, vol.~3,
  no.~6, pp. 653--656, Dec. 2014.

\bibitem{MassMIMO_hybrid_NO_ADMA}
H.~{Lin}, F.~{Gao}, S.~{Jin}, and G.~Y. {Li}, ``A new view of multi-user hybrid
  massive {MIMO}: Non-orthogonal angle division multiple access,'' \emph{IEEE
  J. Sel. Areas Commun.}, vol.~35, no.~10, pp. 2268--2280, Oct. 2017.

\bibitem{SU_mMIMO_OFDM}
F.~{Sohrabi} and W.~{Yu}, ``Hybrid analog and digital beamforming for {mmWave
  OFDM} large-scale antenna arrays,'' \emph{IEEE J. Sel. Areas Commun.},
  vol.~35, no.~7, pp. 1432--1443, July 2017.

\bibitem{OMP_Full_CSI}
O.~E. {Ayach}, S.~{Rajagopal}, S.~{Abu-Surra}, Z.~{Pi}, and R.~W. {Heath},
  ``Spatially sparse precoding in millimeter wave {MIMO} systems,'' \emph{IEEE
  Trans. Wireless Commun.}, vol.~13, no.~3, pp. 1499--1513, Mar. 2014.

\bibitem{SU_mMIMO_Dynamic_SubArray}
S.~{Park}, A.~{Alkhateeb}, and R.~W. {Heath}, ``Dynamic subarrays for hybrid
  precoding in wideband {mmWave MIMO} systems,'' \emph{IEEE Trans. Wireless
  Commun.}, vol.~16, no.~5, Mar. 2017.

\bibitem{JSDM_LargeArray}
A.~Adhikary, J.~Nam, J.~Y. Ahn, and G.~Caire, ``Joint spatial division and
  multiplexing--the large-scale array regime,'' \emph{IEEE Trans. Inf. Theory},
  vol.~59, no.~10, pp. 6441--6463, Oct. 2013.

\bibitem{MassMIMO_hybrid_JSDM_FA}
T.~Ketseoglou and E.~Ayanoglu, ``Downlink precoding for massive {MIMO} systems
  exploiting virtual channel model sparsity,'' \emph{IEEE Trans. Commun.},
  vol.~66, no.~5, pp. 1925--1939, 2018.

\bibitem{ASIL_ABHP_Access}
A.~{Koc}, A.~{Masmoudi}, and T.~{Le-Ngoc}, ``{3D} angular-based hybrid
  precoding and user grouping for uniform rectangular arrays in massive
  {MU-MIMO} systems,'' \emph{IEEE Access}, vol.~8, pp. 84\,689--84\,712, May
  2020.

\bibitem{ASIL_Subconnected_GC}
W.~{Zheng}, A.~{Koc}, and T.~{Le-Ngoc}, ``Sub-connected hybrid precoding
  architectures in massive {MIMO} systems,'' in \emph{2020 IEEE Global Commun.
  Conf. (GLOBECOM)}, Dec. 2020, pp. 1--6.

\bibitem{ASIL_PSO_PA_WCNC}
A.~{Koc} and T.~{Le-Ngoc}, ``Swarm intelligence based power allocation in
  hybrid massive {MIMO} systems,'' in \emph{2021 IEEE Wireless Commun. and
  Netw. Conf. (WCNC)}, Mar. 2021, pp. 1--7.

\bibitem{ASIL_MC_PIMRC}
A.~{Koc}, A.~{Masmoudi}, and T.~{Le-Ngoc}, ``{3D} angular-based hybrid
  precoding for multi-cell {MU}-massive-{MIMO} systems in {C-RAN}
  architecture,'' in \emph{2020 IEEE 31th Annu. Int. Symp. Pers. Indoor Mobile
  Radio Commun. (PIMRC)}, Sept. 2020, pp. 1--6.

\bibitem{ASIL_VTC_DAC_ADC}
A.~{Koc} and T.~{Le-Ngoc}, ``Hybrid millimeter-wave massive {MIMO} systems with
  low {CSI} overhead and few-bit {DACs/ADCs},'' in \emph{2020 IEEE 92th Veh.
  Technol. Conf. (VTC2020-Fall)}, Dec. 2020, pp. 1--5.

\bibitem{HPC_Wireless_Backhaul}
S.~{Ni}, J.~{Zhao}, H.~H. {Yang}, and Y.~{Gong}, ``Enhancing downlink
  transmission in {MIMO HetNet} with wireless backhaul,'' \emph{IEEE Trans.
  Veh. Technol.}, July 2019.

\bibitem{MassiveMIMO_Backhaul}
A.~{Bonfante}, L.~{Galati Giordano}, D.~{López-Pérez}, A.~{Garcia-Rodriguez},
  G.~{Geraci}, P.~{Baracca}, M.~M. {Butt}, and N.~{Marchetti}, ``5g massive
  mimo architectures: Self-backhauled small cells versus direct access,''
  \emph{IEEE Transactions on Vehicular Technology}, vol.~68, no.~10, pp.
  10\,003--10\,017, 2019.

\bibitem{dahlman20164g}
E.~Dahlman, S.~Parkvall, and J.~Skold, \emph{{4G}, {LTE}-advanced {Pro} and the
  Road to {5G}}.\hskip 1em plus 0.5em minus 0.4em\relax Academic Press, 2016.

\bibitem{ASIL_Mobeen_2D_Array}
M.~{Mahmoud}, A.~{Koc}, and T.~{Le-Ngoc}, ``{2D} antenna array structures for
  hybrid massive {MIMO} precoding,'' in \emph{2020 IEEE Global Commun. Conf.
  (GLOBECOM)}, Dec. 2020, pp. 1--6.

\bibitem{3D_Beamforming_ElevAzim}
T.~\hspace{-0.5ex}{Wang}, B.~{Ai}, R.~{He}, and Z.~{Zhong}, ``Two-dimension
  \hspace{-0.25ex}direction-of-arrival \hspace{-.25ex}estimation
  \hspace{-0.25ex}for \hspace{-0.25ex}massive {MIMO} systems,'' \emph{IEEE
  Access}, vol.~3, pp. 2122--2128, Nov. 2015.

\bibitem{ChannelModels}
X.~{Cheng}, B.~{Yu}, L.~{Yang}, J.~{Zhang}, G.~{Liu}, Y.~{Wu}, and L.~{Wan},
  ``Communicating in the real world: {3D MIMO},'' \emph{IEEE Wireless Commun.},
  vol.~21, no.~4, pp. 136--144, Aug. 2014.

\bibitem{balanis2015antenna}
C.~Balanis, \emph{Antenna Theory: Analysis and Design}.\hskip 1em plus 0.5em
  minus 0.4em\relax Wiley, 2015.

\bibitem{AoD_Est_2_Decades}
H.~{Krim} and M.~{Viberg}, ``Two decades of array signal processing research,''
  \emph{IEEE Signal Process. Mag.}, vol.~13, no.~4, pp. 67--94, July 1996.

\bibitem{ASIL_Xiaoyi_Channel_Estimation}
X.~{Zhu}, A.~{Koc}, R.~{Morawski}, and T.~{Le-Ngoc}, ``A semi-deterministic
  channel estimation approach based on geospatial data and fuzzy c-means,'' in
  \emph{2021 IEEE Wireless Commun. and Netw. Conf. (WCNC)}, June 2021, pp.
  1--6.

\bibitem{AHMED_FD}
A.~{Masmoudi} and T.~{Le-Ngoc}, ``Channel estimation and self-interference
  cancelation in full-duplex communication systems,'' \emph{IEEE Trans. Veh.
  Technol.}, vol.~66, no.~1, pp. 321--334, 2017.

\bibitem{Report_5G_UMi_UMa_Rel16}
{3GPP TR 38.901}, ``5{G}: Study on channel model for frequencies from 0.5 to
  100 {GHz},'' Tech. Rep. Ver. 16.1.0, Nov. 2020.

\bibitem{Report_5G_Macro_PL_Rel_16}
{3GPP TR 36.931}, ``{LTE}; evolved universal terrestrial radio access
  ({E-UTRA}); radio frequency ({RF}) requirements for {LTE} pico node {B},''
  Tech. Rep. Ver. 16.0.0, July 2020.

\bibitem{EnergyEff_1}
X.~{Gao}, L.~{Dai}, S.~{Han}, C.~{I}, and R.~W. {Heath}, ``Energy-efficient
  hybrid analog and digital precoding for mmwave {MIMO} systems with large
  antenna arrays,'' \emph{IEEE J. Sel. Areas Commun.}, vol.~34, no.~4, pp.
  998--1009, Apr. 2016.

\bibitem{EnergyEff_2}
Y.~{Wang}, W.~{Zou}, and Y.~{Tao}, ``Analog precoding designs for millimeter
  wave communication systems,'' \emph{IEEE Trans. Veh. Technol.}, vol.~67,
  no.~12, pp. 11\,733--11\,745, Dec. 2018.

\bibitem{EnergyEff_3}
{Shuguang Cui}, A.~J. {Goldsmith}, and A.~{Bahai}, ``Energy-constrained
  modulation optimization,'' \emph{IEEE Trans. Wireless Commun.}, vol.~4,
  no.~5, pp. 2349--2360, Sept. 2005.

\bibitem{Imperfect_AoD}
J.~{Hu}, F.~{Shu}, and J.~{Li}, ``Robust synthesis method for secure
  directional modulation with imperfect direction angle,'' \emph{IEEE Commun.
  Lett.}, vol.~20, no.~6, pp. 1084--1087, 2016.

\bibitem{Imperfect_AoD_2}
H.~{Zhang}, Y.~{Xiao}, Y.~{Xiao}, and W.~{Xiang}, ``Impact of imperfect angle
  estimation on spatial and directional modulation,'' \emph{IEEE Access},
  vol.~8, pp. 7081--7092, 2020.

\bibitem{Imperfect_AoD_3}
D.~{Zhu}, J.~{Choi}, and R.~W. {Heath}, ``Two-dimensional {AoD and AoA}
  acquisition for wideband millimeter-wave systems with dual-polarized
  {MIMO},'' \emph{IEEE Trans. Wireless Commun.}, vol.~16, no.~12, pp.
  7890--7905, 2017.

\end{thebibliography}

\begin{IEEEbiography}[{\includegraphics[width=1in,height=1.25in,clip,keepaspectratio]{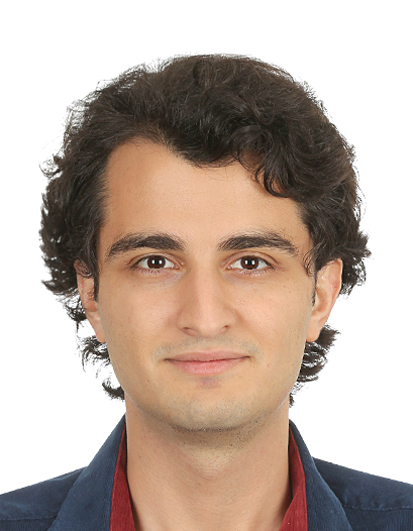}}]{Asil Koc}
	(Student Member, IEEE) received the B.Sc. degree (Hons.) in electronics and communication engineering and the M.Sc. degree (Hons.) in telecommunication engineering from Istanbul Technical University, Istanbul, Turkey, in 2015 and 2017, respectively. He is currently pursuing the Ph.D. degree in electrical engineering with McGill University, Montreal, QC, Canada. 
	
	From 2015 to 2017, he was a Research and a Teaching Assistant with the Electronics and Communication Engineering Department, Istanbul Technical University. Since 2017, he has been a Teaching Assistant with the Electrical and Computer Engineering Department, McGill University. His research interests include, but not limited to wireless communications, massive MIMO, full-duplex, spatial modulation, energy harvesting, and cooperative networks. He was a recipient of the Erasmus Scholarship Award funded by the European Union, the McGill Engineering Doctoral Award, the STARaCom Collaborative Grant funded by the FRQNT, and the Graduate Research Enhancement and Travel Award funded by McGill University.
\end{IEEEbiography}

\begin{IEEEbiography}[{\includegraphics[width=1in,height=1.25in,clip,keepaspectratio]{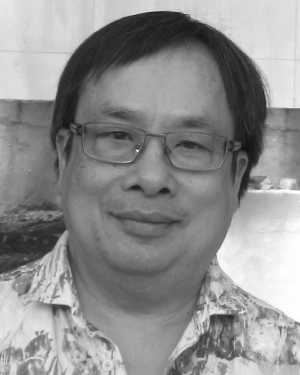}}]{Tho Le-Ngoc}
	(Life Fellow, IEEE) received the B.Eng. degree in electrical engineering, in 1976, the M.Eng. degree in microprocessor applications from McGill University, Montreal, in 1978, and the Ph.D. degree in digital communications from the University of Ottawa, Canada, in 1983. 
	
	From 1977 to 1982, he was with Spar Aerospace Ltd., Sainte-Anne-de-Bellevue, QC, Canada, involved in the development and design of satellite communications systems. From 1982 to 1985, he was with SRTelecom Inc., Saint-Laurent, QC, Canada, where he developed the new point-tomultipointDA- TDMA/TDM Subscriber Radio System SR500. From 1985 to 2000, he was a Professor with the Department of Electrical and Computer Engineering, Concordia University, Montreal. Since 2000, he has been with the Department of Electrical and Computer Engineering, McGill University. His research interest includes broadband digital communications. He is a Fellow of the Engineering Institute of Canada, the Canadian Academy of Engineering, and the Royal Society of Canada. He was a recipient of the 2004 Canadian Award in Telecommunications Research and the IEEE Canada Fessenden Award, in 2005. He holds the Canada Research Chair (Tier I) on Broadband Access Communications.
\end{IEEEbiography}

\end{document}